\documentclass{article}
\usepackage{graphicx}

\begin{document}
\newcommand{\w}[1]{{\bf #1}}
\newcommand{\be}{\begin{equation}}
\newcommand{\ee}{\end{equation}}
\newcommand{\de}{{\rm d}}
\newcommand{\ie}{{\rm i}}
\newcommand{\bea}{\begin{eqnarray}}
\newcommand{\eea}{\end{eqnarray}}
\newcommand{\dpr}{^{\prime\prime}}
\newcommand{\nn}{\nonumber \\}
\newcommand{\ex}[1]{{\rm e}^{#1}}
\newcommand{\nicht}[1]{ }
\newcommand{\hrho}{\hat\rho}

\title{From Elastica to Floating Bodies of Equilibrium}

\author{Franz Wegner\\
Institute for Theoretical Physics\\
Ruprecht-Karls-University Heidelberg, Germany}

\maketitle

\begin{abstract}
A short historical account of the curves related
to the two-di\-men\-sio\-nal floating bodies of equilibrium
and the bicycle problem is given.
Bor, Levi, Perline and Tabachnikov found, quite a
number had already been described as {\it Elastica} 
by Bernoulli and Euler and as
{\it Elastica under Pressure} or {\it Buckled Rings}
by Levy and Halphen.
Auerbach already realized that Zindler had described
curves for the floating bodies problem. An even larger class
of curves solves the bicycle problem.

The subsequent sections deal with some supplemental details:
Several derivations of the equations for the 
elastica and elastica under pressure are given.
Properties of Zindler curves and some
work on the problem of floating bodies
of equilibrium by other mathematicians are considered.
Special cases of elastica under pressure reduce to algebraic curves,
as shown by Greenhill.
Since most of the curves considered here are bicycle curves,
a few remarks concerning them are added.
\end{abstract}

\tableofcontents

\section{Introduction}

Mikl\'os R\'edei introduces in his paper
"On the tension between mathematics and physics"\cite{Redei}
the "supermarket picture" of the relation of mathematics 
and physics: that mathematics is like a supermarket and
physics its customer.

When in 1938 Auerbach solved  the problem of floating
bodies of equilibrium\cite{Auerbach} for the density
$\rho=1/2$ he could go to the
supermarket 'Mathematics' and found the solution in form
of the Zindler curves\cite{Zindler21}.

When I thought about this problem for densities $\rho\not=1/2$
I found a differential equation, went to the
supermarket and found elliptic functions as ingredients
for the solution. But in the huge supermarket I did not
find the finished product. Later Bor, Levi, Perline, and Tabachnikov\cite{BLPT}
showed me the shelf, where the boundary
curves as Elastica under Pressure had been put already in
the 19th century. The linear limit, which I had also
considered had been put there as Elastica already in the 17th century
by James Bernoulli\cite{BernoulliJ1,BernoulliJ2}
and in the 18th century by Leonhard Euler.\cite{Euler}
Fortunately good mathematics has no date of expiry.

Section \ref{hist} presents
a short historical survey of the
curves and their applications, called {\it Elastica} and
{\it Elastica under  Pressure} or {\it Buckled Rings}.
These curves, known since the seventeenth and nineteenth century,
first as solutions of elastic problems, have shown up as
solutions of quite a number of other problems, in particular
as boundaries of two-dimensional bodies which can float in
equilibrium in all orientations; this later problem is also
solved by {\it Zindler curves}.

These classes of curves yield solutions to the bicycle problem.
One asks in this problem
for front and rear traces of a bicycle, which do not allow
to conclude, in which direction the bicycle went,
that is the traces are identical in both directions.
The curves that give the traces of the front wheel and
the traces of the rear wheels are the boundary and the envelope of the
water lines, resp., of the floating bodies of equilibrium.

The subsequent sections deal with some supplemental details:
In section 
\ref{Deriv} several derivations of the equations for the 
elastica and elastica under pressure are given.
Section \ref{1/2Zind} deals with the
Zindler curves and work on the problem of floating bodies
of equilibrium by other mathematicians, including criticism.
Section \ref{Green} is devoted to work by Greenhill, who found
that special cases of the elastica under pressure can be 
represented by algebraic curves.
Since most of the curves considered here are special
cases of bicycle curves, section \ref{bicycle} brings some 
remarks on these curves.

\section{Historical survey} \label{hist}

In this paper we consider a class of planar curves, which surprisingly
show up in quite a number of different physical and mathematical problems.
These curves are not generally known, since they are represented
by elliptic functions, although special cases can be represented
by more elementary functions.

\subsection{The Curves} \label{cur}

These curves appear in two flavors; they obey in the linear form
the equation
\be
\frac 1{\sqrt{1+y^{\prime 2}}} = ay^2 + b \label{lin}
\ee
in Cartesian coordinates $(x,y=y(x))$, $y'=\de y/\de x$.
The circular form is described by
\be
\frac 1{\sqrt{r^2+r^{\prime 2}}} = ar^2 +b +\frac c{r^2}, \label{cir}
\ee
in polar coordinates $(r,\phi)$ with $r=r(\phi)$ and $r'=\de r/\de\phi$.
Eq. (\ref{lin}) yields the curvature $\kappa$,
\be
\frac 1{y'} \frac{\de}{\de x}\frac 1{\sqrt{1+y^{\prime 2}}}
= -\frac{y\dpr}{(1+y^{\prime 2})^{3/2}} = \kappa = 2ay.
\label{lin2}
\ee
From (\ref{cir}) we obtain the curvature
\be
\frac 1{rr'} \frac{\de}{\de\phi}
\frac{r^2}{\sqrt{r^2+r^{\prime 2}}}=
\frac{r^2+2r^{\prime 2}-rr\dpr}{(r^2+r^{\prime 2})^{3/2}}
= \kappa = 4ar^2 + 2b. \label{cir2}
\ee
We relate the polar coordinates to Cartesian coordinates
and shift $y$ by $r_0$,
\be
r\cos\phi = r_0+y, \quad r\sin\phi = x.
\ee
Then (\ref{cir2}) reads
\be
\kappa = 4ar_0^2 + 8ar_0y + 4a(x^2+y^2) +b. \label{cirlin}
\ee
If we now choose
\be
b=-2ar_0^2, \quad a = \tilde a/(4r_0),
\ee
and perform the limit $r_0\rightarrow \infty$, then
eq. (\ref{cirlin}) reads
\be
\kappa = 2\tilde a y,
\ee
which is the linear form (\ref{lin2}). Thus the linear form
is a limit of the circular form, where the radius $r_0$
goes to infinity.

The equation of the curves can be formulated
coordinate-independent,
\be
2\kappa\dpr +\kappa^3 -\mu\kappa -\sigma=0, \label{EuLa}
\ee
where the prime now indicates the derivative with respect to the
arc length. Multiplication by $\kappa'$ allows integration,
\be
\kappa^{\prime 2} + \frac{\kappa^4}4 - \mu\frac{\kappa^2}2
-\sigma\kappa = 2\hat E. \label{EuLa2}
\ee
The coefficient $\sigma$ vanishes in the linear case.
The derivation will become apparent, when we formulate
the physical and/or mathematical problems solved
by the curves. But the relation between eqs. (\ref{lin}, \ref{cir})
and eqs. (\ref{EuLa}, \ref{EuLa2}) can also be given directly.\cite{BLPT}

\subsection{Linear Elastica}\label{IntroElast}

The linear curves show up in {\it Elastica}. The question is: How
does an elastic beam (or wire or rod) of given length bend?
It may be fixed at two ends and the directions of
the wire at both ends are given, or it is fixed at one end
and loaded at the other end.
Bending under load was already considered in the 13th century by Jordanus de
Nemore,  around 1493 by Leonardo da Vinci\cite{Balarini},
in 1638 by Galileo Galilei, and in 1673 by
Ignace-Gaston Pardies.\cite{Levien,Truesdell1},
although they could not give correct results.\cite{Levien,Todhunter}
James Bernoulli following Hooke's ideas obtained a correct
differential equation assuming that the curvature is proportional
to the bending moment, and partially solved it in 1691-1692.
\cite{BernoulliJ1},
\be
\frac{\de y}{\de x} = \frac{x^2}{\sqrt{a^4-x^4}}. \label{D1}
\ee
Huygens\cite{Huygens} argued in 1694 that the problem had a larger variety
of solutions and sketched several of them.
The more general differential equation
was given by Bernoulli\cite{BernoulliJ2} in 1694-1695,
\be
\frac{\de y}{\de x} = \frac{x^2\pm ab}{\sqrt{a^4-(x^2\pm ab)^2}},
\label{D2}
\ee
This equation yields
\be
\frac 1{\sqrt{1+(\frac{\de x}{\de y})^2}} = \frac{x^2\pm ab}{a^2},
\ee
which is equation (\ref{lin}) with $x$ and $y$ exchanged and 
different notation for the constants.
James Bernoulli also realized "... among all curves of a given
length drawn over a straight line the elastic curve is the one
such that the center of gravity of the included area is the
furthest distant from the line, just as the catenary is the one
such that the center of gravity of the curve is the furthest
distant ..."\cite{BernoulliJ2}. Thus he found that the cross section 
of a volume of water contained in a cloth sheet is bounded
(below the water line) by the elastica curve.

In 1742 Daniel Bernoulli (nephew of James) proposed in a letter to
Leonhard Euler
that the potential energy of a bent beam is proportional to the
integral of the square of the curvature $\kappa$ integrated over the
arclength $s$ of the beam $\int\de s \kappa^2$. In a 1744 paper
Euler\cite{Euler} used variation techniques to solve the problem.
He found eq. (\ref{D2}), classified the solutions, discussed
the stability, and realized that the curvature is proportional
to $x$,
\be
\kappa = \frac{y\dpr}{(1+y^{\prime 2})^{3/2}} = \frac{2x}{a^2}.
\label{D3}
\ee

This property plays a role in at least two other physical phenomena:

In 1807 Pierre Simon Laplace\cite{Laplace} investigated the shape of
the capillary. The surface of a fluid trapped between two parallel
vertical plates obeys also (\ref{D3}), since the difference of
pressure inside and outside the fluid is proportional to the curvature
of the surface.

Charges in a linearly increasing magnetic field move due to the
Lorentz force on trajectories with curvature proportional to the
magnetic field and thus again on trajectories given by the
curves of elastica. Without being aware of this equivalence
Evers, Mirlin, Polyakov, and W\"olfle considered them in their
paper on the semiclassical theory of transport in a random magnetic
field.\cite{Evers}

\begin{figure}[h!]
\includegraphics[scale=0.5,angle=-90]{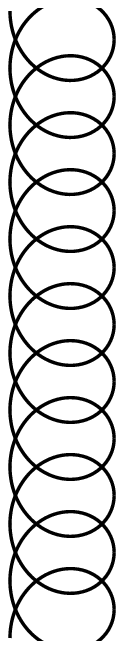}\hspace{5mm}
\includegraphics[scale=0.5,angle=-90]{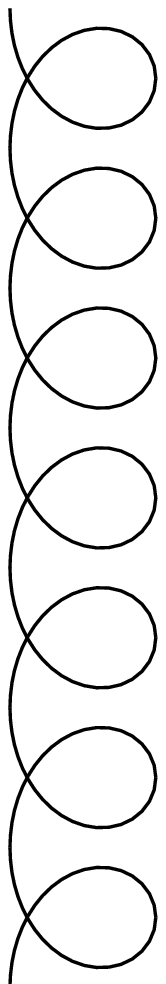}\vspace{2mm}

\includegraphics[scale=0.5,angle=-90]{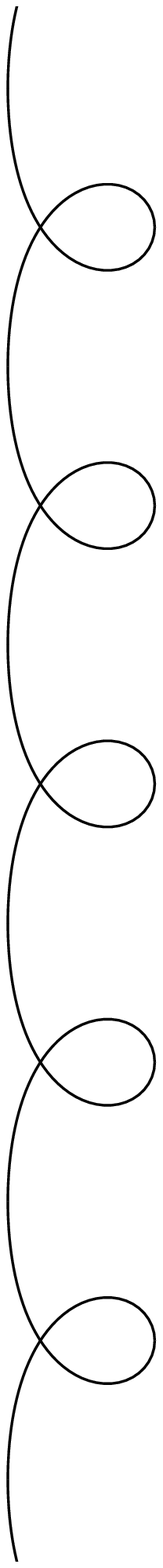}\vspace{2mm}

\includegraphics[scale=0.5,angle=-90]{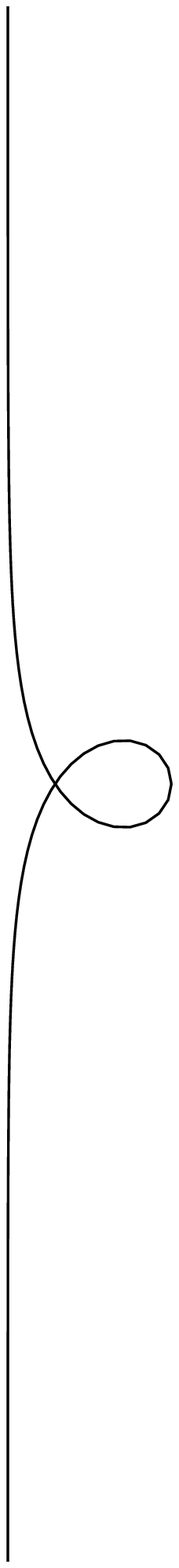}\vspace{2mm}

\includegraphics[scale=0.5,angle=-90]{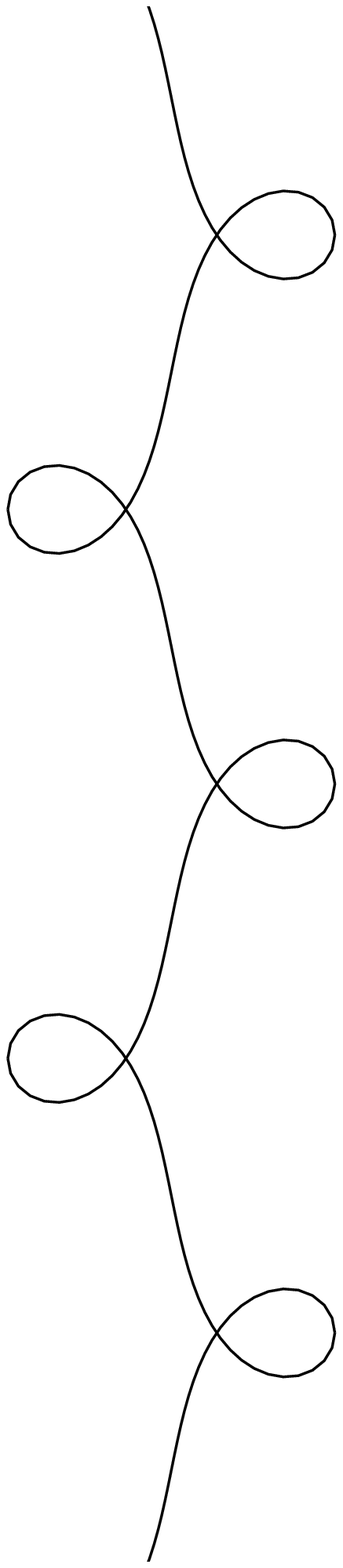}
\caption{Examples of linear Elastica (i)}
\label{fEl1}
\end{figure}

\begin{figure}[h!]
\includegraphics[scale=0.5,angle=90]{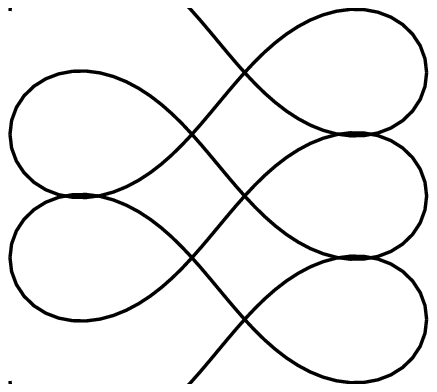}
\hspace{5mm}
\includegraphics[scale=0.5,angle=90]{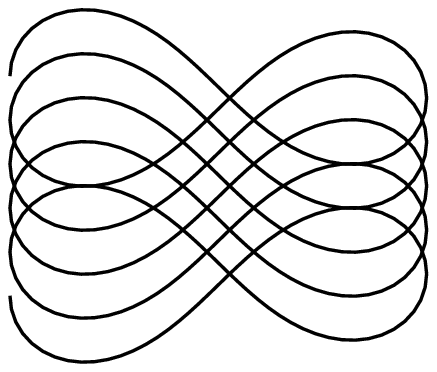}
\hspace{5mm}
\includegraphics[scale=0.5,angle=90]{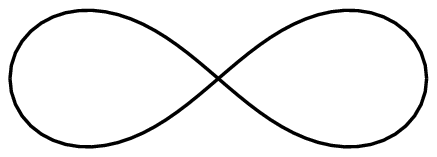}
\hspace{5mm}
\includegraphics[scale=0.5,angle=90]{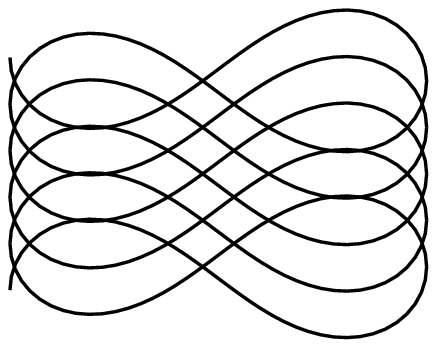}
\hspace{5mm}
\includegraphics[scale=0.5,angle=90]{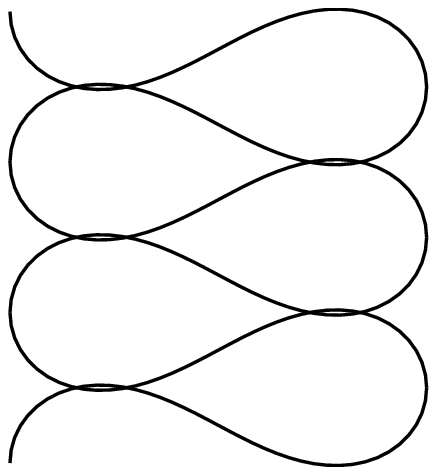}
\caption{Examples of linear Elastica (ii)}
\label{fEl2}
\end{figure}

\begin{figure}[h!]
\includegraphics[scale=0.5,angle=90]{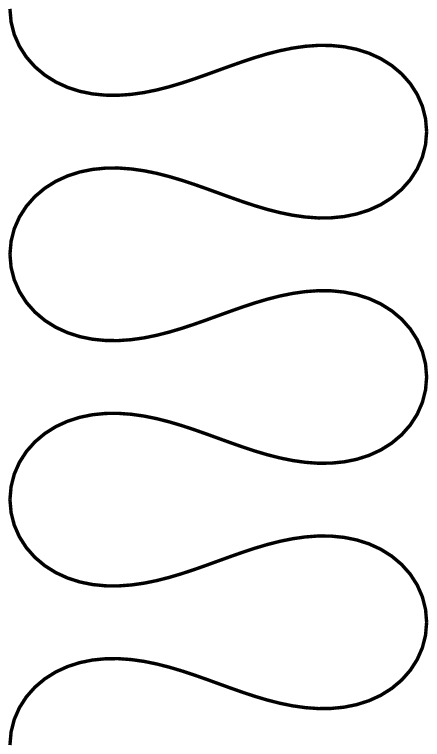}
\hspace{2mm}
\includegraphics[scale=0.5,angle=90]{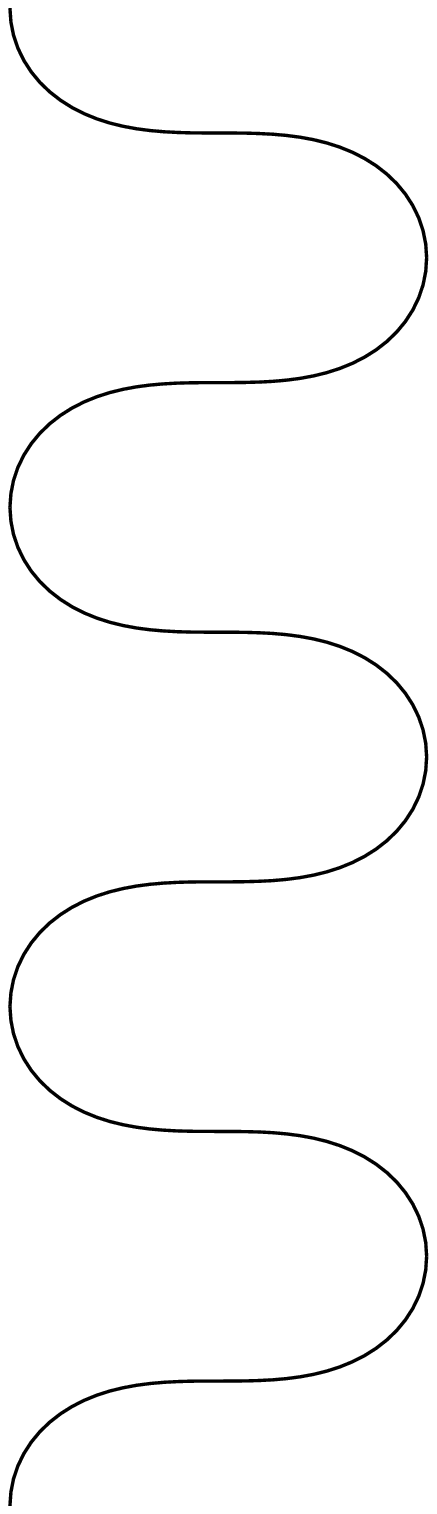}
\vspace{2mm}

\includegraphics[scale=0.5,angle=90]{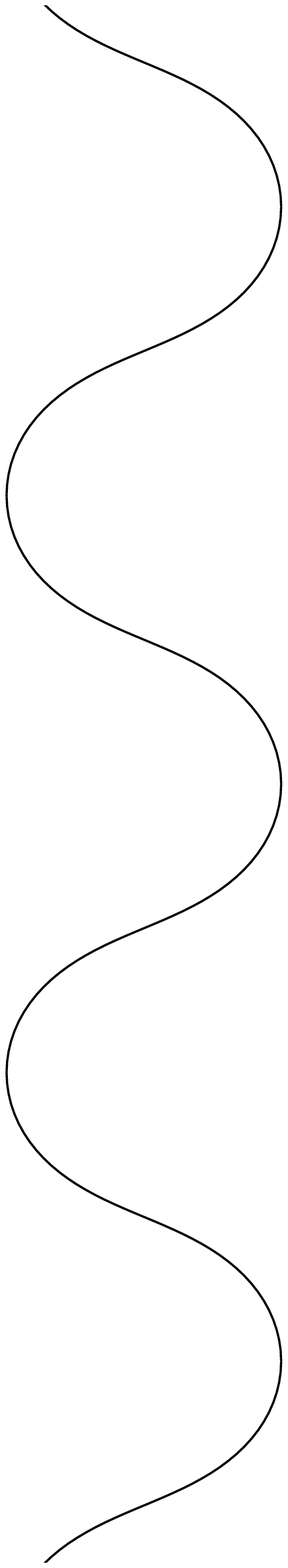}
\vspace{2mm}

\includegraphics[scale=0.5,angle=90]{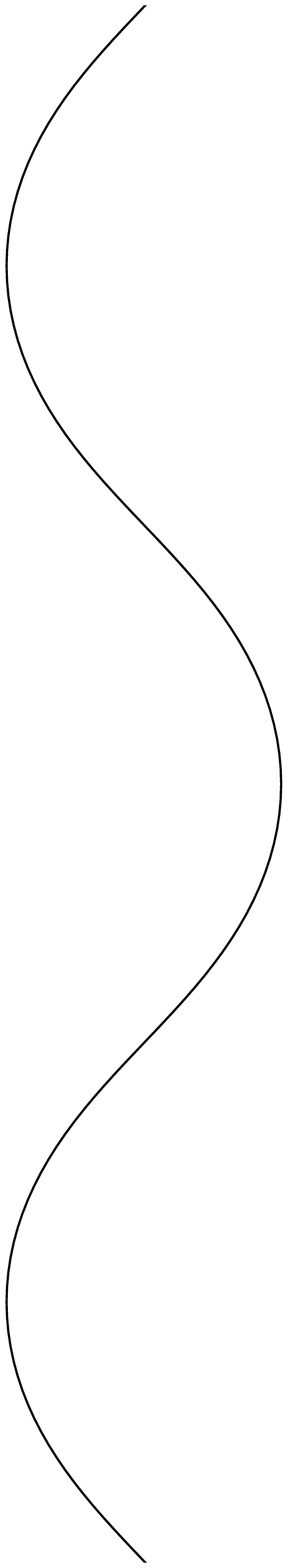}
\caption{Examples of linear Elastica (iii)}
\label{fEl3}
\end{figure}

In 1859 Kirchhoff\cite{Kirchhoff} introduced a kinetic analogue by
showing that the problem of elastica is related to the movement
of a pendulum. Set
\be
\dot x = \cos\theta, \quad \dot y = \sin\theta,
\ee
where the dot indicates the derivative with respect to the
arc $s$. Then one obtains
\be
\dot\theta = \kappa, \quad y' = \tan\theta.
\ee
Thus using eqs. (\ref{lin2}) and (\ref{lin}) one obtains
\be
\dot\theta^2 =4a^2y^2 =4a (\cos\theta-b).
\ee
Multiplication by $mr^2/2$ (with $m$ for the mass and $r$ for
the length of the pendulum) and a corresponding choice
of the constants $a$ and $b$, yields
\be
\frac{mr^2}2 \dot\theta^2 -mgr\cos\theta = E.
\ee
This is the energy of a pendulum, if we substitute time $t$
for the arc $s$ in the derivative. Thus $\theta(t)$ is the
time dependence of the angle of the pendulum against its
lowest position. The periodic movement is obvious.
Suppose $a$ is positive. Then for $-1<b<1$ the pendulum
will swing in a finite interval 
$-\theta_0\le\theta\le+\theta_0$. These are the inflectional
solutions.
If $b<-1$ then the pendulum will move across the highest
point $\theta=\pi$. These are the non-inflectional solutions.
The limit case $b=-1$ yields a
non-periodic solution (infinite period).

Some examples of elastica are shown in figures \ref{fEl1} to 
\ref{fEl3}. The first two rows of figure \ref{fEl1} show 
non-inflectional cases where the pendulum moves across
the highest point.
The third row shows the aperiodic limit case $b=-1$.
It is called {\it syntractrix of Poloni} (1729).
The last row of figure \ref{fEl1} and figures \ref{fEl2} and \ref{fEl3}
show inflectional cases corresponding to 
periodic oscillations without reaching
the highest point. This yields a large variety of shapes
including the {\it Eight} in the middle of figure \ref{fEl2}
called {\it lemnoid}.
The second drawing in figure \ref{fEl3} corresponds to
the case where the pendulum moves up to a horizontal position.
It is the {\it rectangular elastica} or {\it right lintearia}.
The pendulum swings below the horizontal position
in the last two rows of figure \ref{fEl3}.
In all cases the curves show
equilibrium positions of the elastic beam. However, only
sufficiently short pieces of the curves correspond to
a stable equilibrium or even the absolute minimum of the 
potential energy.

\begin{figure}[h!]
\includegraphics[scale=0.6]{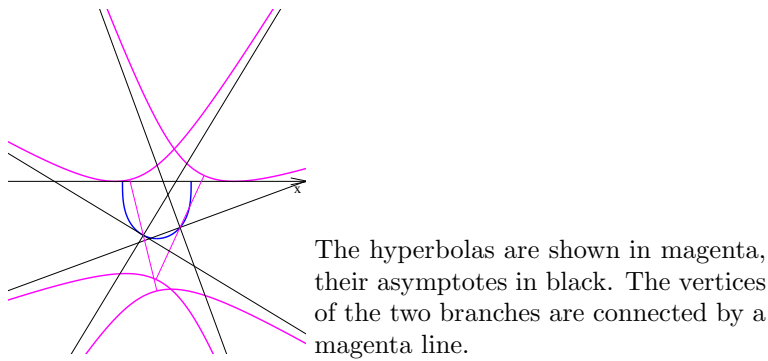}
\parbox[b]{6cm}{The hyperbolas are shown in magenta, their asymptotes in black.
The vertices of the two branches are connected by a magenta line.}
\caption{Rectangular elastica as roulette of hyperbola}
\label{fhyp1}
\end{figure}
{\bf Rectangular elastica as roulette of hyperbola} 
The rectangular elastic curve is the locus of the center of a
rectangular hyperbola rolling without slipping on a straight
line. The upper branch of the hyperbola is shown in figure \ref{fhyp1}
in two different orientations. The midpoint between the branches
lies on the blue rectangular linteria. (Sturm 1841, see \cite{Ferreol},
Greenhill 1892 \cite{Greenhill92}).

An excellent survey with many figures on the history of the elastica
has been given by Raph Levien.\cite{Levien}
Also Todhunter\cite{Todhunter} and
Truesdell\cite{Truesdell1,Truesdell2} give
reviews of the history of elasticity. Many details are found in the
treatise by Love\cite{Love} on the mathematical theory of 
elasticity. In his PhD thesis\cite{Born06} (1906) Max Born
investigated elastic wires theoretically and experimentally 
in the plane and also in three dimensional space. 
The solution of eq. (\ref{lin}) or equivalently eqs. (\ref{D1}, \ref{D2})
was given by Saalsch\"utz\cite{Saal80} in 1880 by means of
elliptic functions. The elliptic functions were developed
by Abel and Jacobi mainly in the years 1826 -- 1829 in several
articles in Crelles Journal.\cite{Crelle}. Abel died in 1829,
Jacobi published his fundamental work in the same
year.\cite{Jacobi29}
The explicit solutions are not given here.
They can be found, e.g., in sect. 263 of \cite{Love}, in sect. 13
of \cite{Levien} and in ref. \cite{We4a,We4}.
Engineers often call 'Bernoulli-Euler beam theory'
approximations, in which the beam is only slightly bent.\cite{Wiki}

\subsection{Elastica under Pressure (buckled rings)}\label{IntroPress}

\begin{figure}[th!]
\includegraphics[scale=0.5]{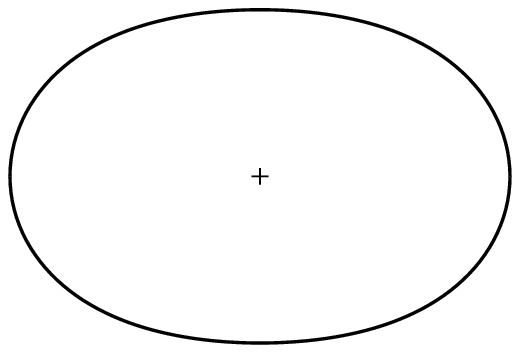}
\hspace{2mm}
\includegraphics[scale=0.5]{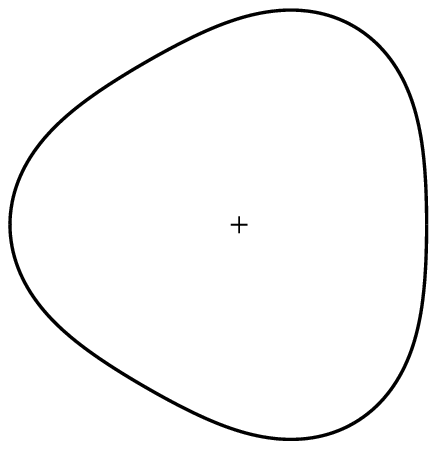}
\hspace{2mm}
\includegraphics[scale=0.5]{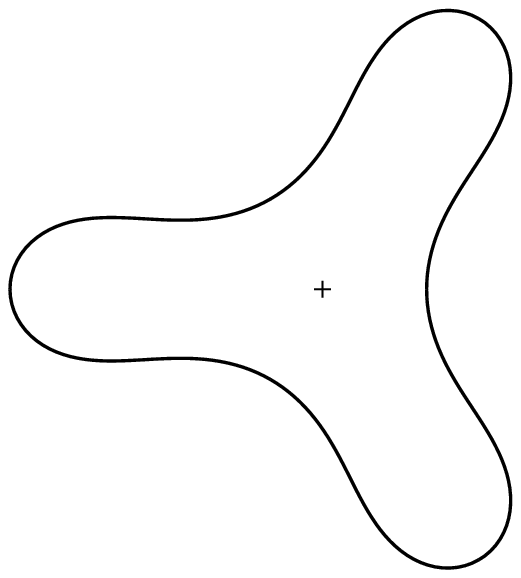}
\vspace{2mm}

\includegraphics[scale=0.5]{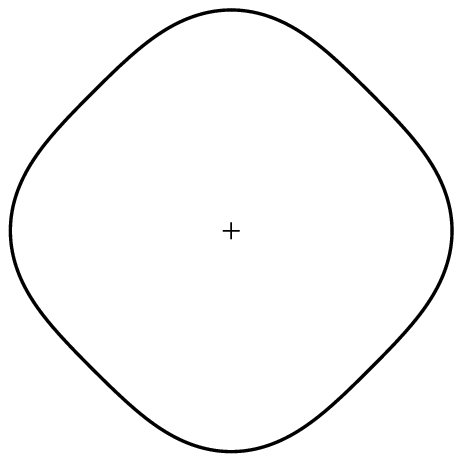}
\hspace{2mm}
\includegraphics[scale=0.5]{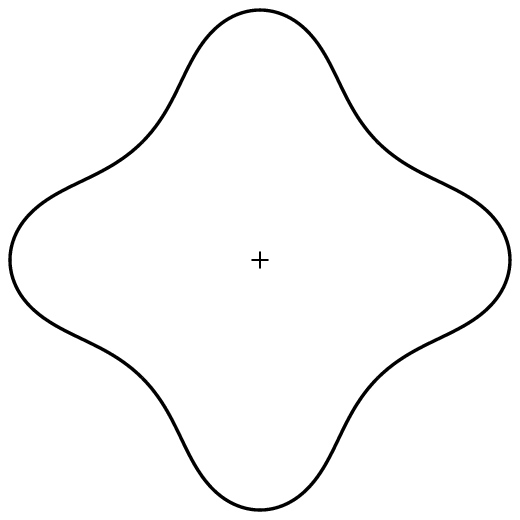}
\hspace{2mm}
\includegraphics[scale=0.5,angle=90]{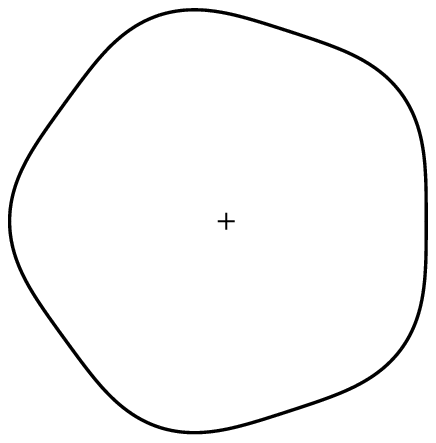}
\hspace{2mm}
\includegraphics[scale=0.5,angle=90]{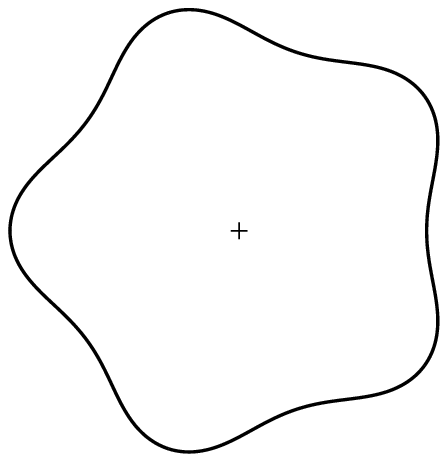}
\vspace{2mm}

\includegraphics[scale=0.5]{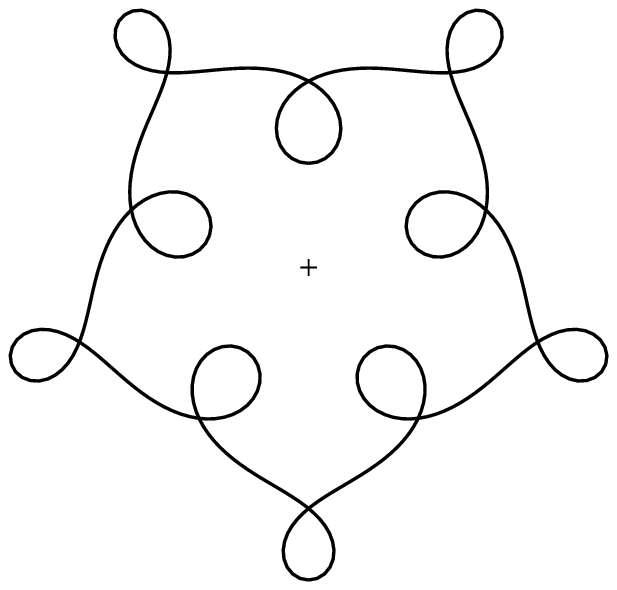}
\hspace{2mm}
\includegraphics[scale=0.3,angle=90]{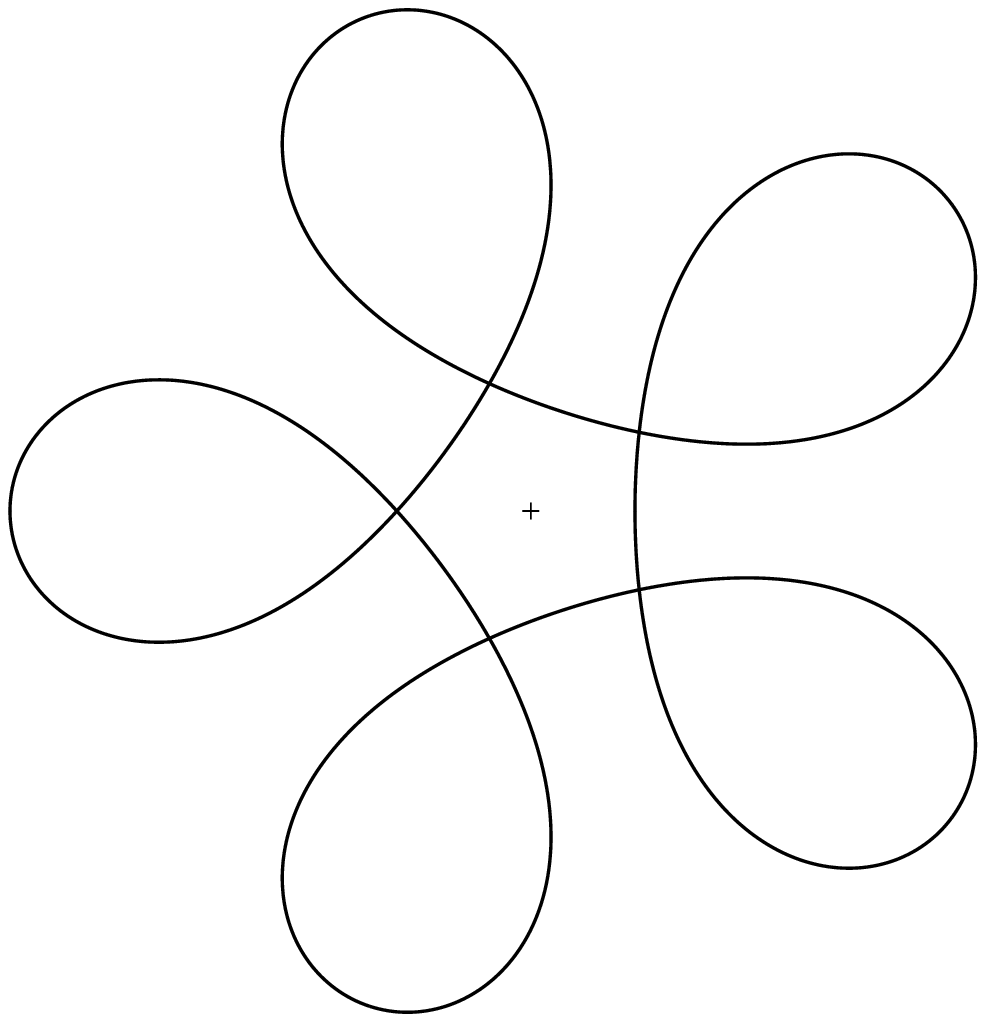}
\hspace{2mm}
\includegraphics[scale=0.5]{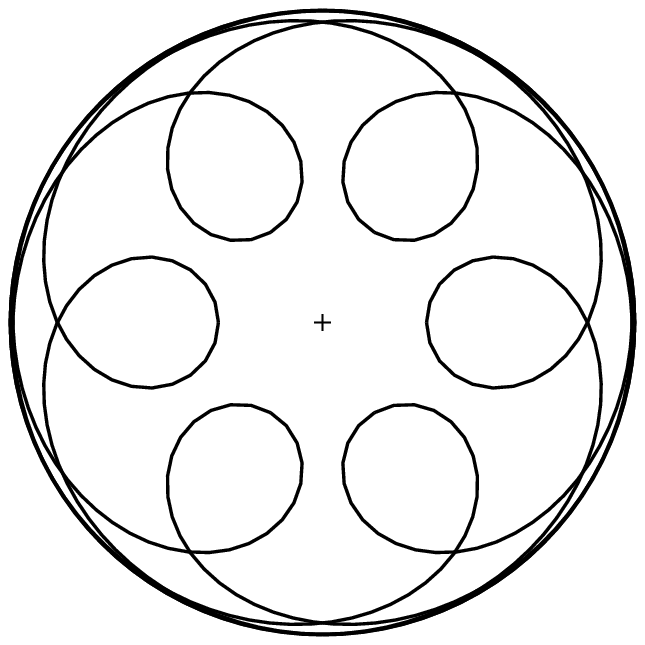}

\caption{Examples of Elastica under Pressure (buckled rings)}
\label{fElP}
\end{figure}

By now we considered elastica to which only forces acted at
the ends. A more general problem considers elastic
wires, on which forces act along the arc. Maurice Levy realized in
1884\cite{Levy} that the case, where a constant force $P$
per arc length acts perpendicularly on the wire, yields the
differential equation (\ref{cir}). He showed that this
problem could be solved by elliptic functions and found
two types of solutions.
They are called buckled rings, if the wire is closed.
The constraints are equivalent if instead the perimeter and the area
inside the ring are given.
Halphen worked out the results in some detail in
the same year\cite{Halphen84}
and included it in his 'Trait\'e des fonctions elliptic et
de leurs applications'.\cite{Halphen88}
Some elastica under pressure are shown in figure \ref{fElP}.
Their symmetry is given by the dieder groups $D_p$ with $p=2,3,3$
in the first row, $p=4,4,5,5$ in the second row and $p=5,5,6$ in
the third row. All of them are solutions of equation (\ref{cir}).
Whereas those in the first and second row can be considered
as deformed circles, those in the third row show double points.
Similarly as for the elastica the curves show equilibrium
configurations. But often only small parts of them are in stable
equilibrium.

Greenhill\cite{Greenhill99} considered the same problem in 1899 and
looked particularly for curves that can be expressed by pseudo-elliptic
functions. Thus some of the solutions are algebraic curves.
The simplest example besides the circle is given by
\be 
r^3 = a^3 \cos(3\phi)
\ee
with the curvature
\be 
\kappa = 4r^2/a^3
\ee
in polar coordinates $(r,\phi)$, which may be written
\be 
(x^2+y^2)^3 = a^3x(x^2-3y^2)
\ee
in Cartesian coordinates $(x,y)$. This Kiepert curve (W. Roberts and L. Kiepert
1870, see \cite{Ferreol}) looking like a cloverleaf is
shown in fig. \ref{fgp2p3c}.

\begin{figure}[h!]
\includegraphics[scale=0.5, angle=-90]{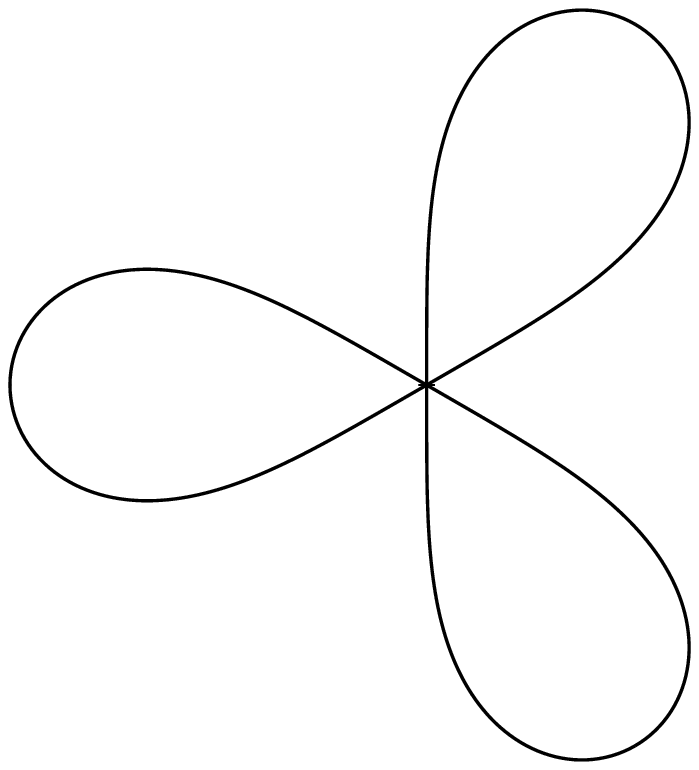}
\caption{Kiepert curve $r^3=a^3\cos(3\phi)$}
\label{fgp2p3c}
\end{figure}

{\bf Area-constrained planar elastica in biophysics}\\
Cells in biology have usually nearly constant volume and 
constant surface area. Their shape is to a large extend
determined by the minimum of the membrane bending energy,
see e.g. Helfrich\cite{Helf73} and Svetina and
Zeks\cite{SveZek89}. The two-dimensional analogue was considered
by Arreaga, Capovilla et al.\cite{Arreaga02,CapChrGuv02}
and by Goldin et al.\cite{Goldin10}. Since pressure and
area are
conjugate quantities, the shapes are also given by those
of elastica under pressure. Now of course, constant area and
constant length of the bounding loop are given. The authors
refer for the determination of the shape to the Lagrange
equations (\ref{EuLa}, \ref{EuLa2}) as reported by Langer and 
Singer for elastica\cite{LanSin84} and for buckled rings
\cite{Lan99}, see also the references \cite{BLPT,Singer}.
The equations for the elastica under pressure can also be
obtained by considering the forces and momenta in the
rods.\cite{TadOdeh67}
The solution of the equation  (\ref{cir}) for buckled rings
in terms of elliptic functions can be found in
\cite{Levy,Halphen84,Halphen88,Greenhill99}
and in \cite{We4a,We4}. Reference \cite{We4} contains
further figures. 

\subsection{Floating Bodies of Equilibrium} \label{IntroFloat}

\subsubsection{Ulam's Problem in two dimensions} \label{IntroFloatUlam}

The curves mentioned in the previous subsections
appear in two other problems: the problem of
{\it Floating Bodies of Equilibrium} and the {\it Bicycle Problem}.
The first of these problems is
related to the problem 19 in the Scottish Book by Ulam\cite{Ulam}:
"Is a solid of uniform density which will float in water
in every position a sphere?" The two-dimensional version
of the problem concerns a cylinder of uniform density
$\rho$ which floats in water in equilibrium in every position
with its axis parallel to the water surface. Sought is
the curve different from a circle confining the cross section of the cylinder
perpendicular to its axis.

The density of the log be $\rho$ (more precisely $\rho$ is the
ratio of the density of the log over that of the liquid).
The area of the cross section
be $A$, the part above and below the water-line are denoted by
$A_1$ and $A_2$. Then Archimedes' law requires
\be 
A_1 = (1-\rho) A, \quad A_2 =\rho A.
\ee
The distance of the center of gravity of the cross section
above the water-line be $h_1$, that below the water-line $h_2$,
the length of the log $L$. Then the potential energy is
\be 
\rho(1-\rho)ALg(h_1+h_2).
\ee
Thus $h_1+h_2$ has to be constant. The line connecting the two centers of
gravity has to be perpendicular to the water-line.
Rotation by an infinitesimal angle yields that the length $2l$
of the water-line obeys
\be 
\frac 23 l^3 \left( \frac 1{A_1} + \frac 1{A_2} \right)
= h_1+h_2.
\ee
Thus the length of the water-line does not depend on the orientation
of the log.
The conditions that the area below the water-line and the length
of the water-line are constant implies that the part of the perimeter
of the cross-section below the water-line is constant. It also
implies that the envelope of the water-lines is given by the 
midpoints of the water-lines.

This two-dimensional problem has attracted many mathematicians.

\begin{figure}[h!]
\includegraphics{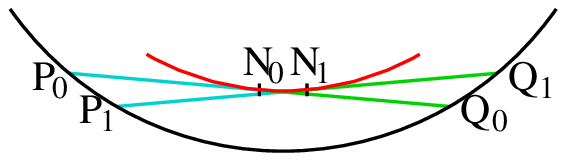}
\caption{Lower part of the boundary (black) of the floating
body}
\label{watbic}
Two waterlines $P_0Q_0$ and $P_1Q_1$ are in blue and cyan.
The midpoints $N_0$ and $N_1$ lie on the envelope (red) of 
the water-lines.
\end{figure}

\subsubsection{Density $\rho=1/2$} \label{IntroFloat=1/2}

There is a large class of solutions for $\rho=1/2$.
The solutions are not related to the elastica, but it
seems worthwhile to mention them.
Basically the solutions were found by Zindler\cite{Zindler21}, although
he did not consider this physical problem, but found convex
curves which have the property that chords between two points
on the boundary which bisect the perimeter have constant
length $2l$ and simultaneously cut the enclosed area in two
halves. They can be parameterized by
\bea
x(\alpha) = l \cos(\alpha) + \xi(\alpha), &&
y(\alpha) = l \sin(\alpha) + \eta(\alpha), \label{Zin1} \\
\xi(\alpha) = \int^{\alpha} \de\beta \cos(\beta) \hrho(\beta), &&
\eta(\alpha) = \int^{\alpha} \de\beta \sin(\beta) \hrho(\beta),
\label{Zin2}
\eea
with parameter $\alpha$,
where $\xi$ and $\eta$ obey
\be
\xi(\alpha+\pi) = \xi(\alpha), \quad \eta(\alpha+\pi) =\eta(\alpha).
\label{Zin3}
\ee
$(\xi,\eta)$ are the coordinates of the  envelope of the water lines,
and $|\hrho(\alpha)|$ is the radius of curvature of the envelope.
Typically the envelope has an odd number of cusps.

\begin{figure}[h!]
\parbox[b]{6cm}
{\includegraphics{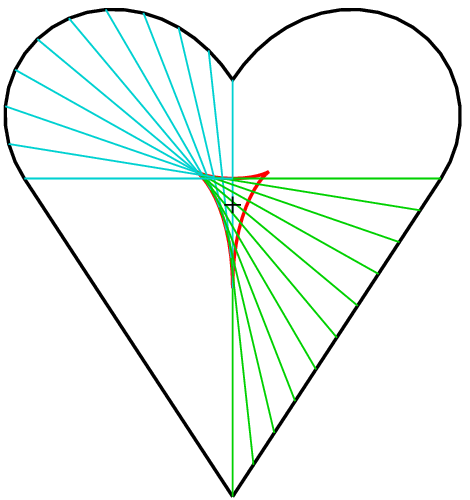}
%{\epsfig{file=fhau2.eps}
\caption{Heart-shaped Zindler curve, see Auerbach\cite{Auerbach}}
\label{fhau2}}
\parbox[b]{6cm}
{\includegraphics{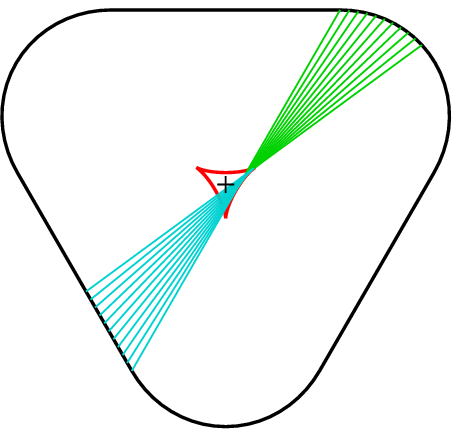}
%{\epsfig{file=fhau3.eps}
\caption{Zindler curve, see Auer\-bach\cite{Auerbach}}
\label{fhau3}}
\end{figure}

\begin{figure}[h!]
\parbox[b]{6cm}
{\includegraphics{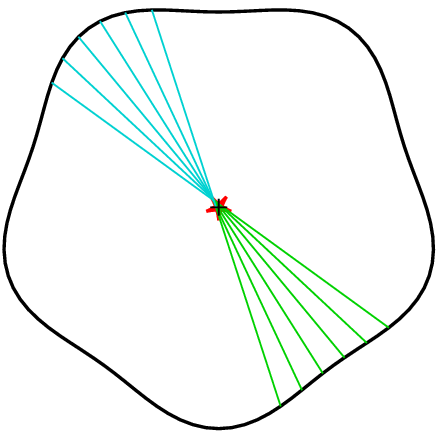}
%{\epsfig{file=fhrc5.eps}
\caption{Zindler curve, five-fold symmetry}
\label{fhrc5}}
\parbox[b]{6cm}
{\includegraphics[angle=-90]{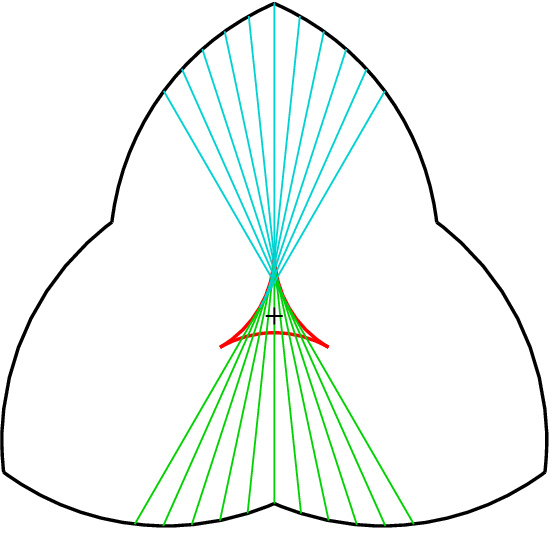}
%{\epsfig{file=fhrk3.eps,angle=-90}
\caption{Zindler curve, see Sal\-gal\-ler and
Kostelianetz\cite{SalKos},
similarly by Zindler\cite{Zindler21}}
\label{fhrk3}}
\end{figure}

Condition (\ref{Zin3}) implies $\hrho(\beta+\pi)=-\hrho(\beta)$.
The chords run from
$(x(\alpha),y(\alpha))$ to $(x(\alpha+\pi),y(\alpha+\pi))$.
Zindler did not consider the centers of gravity of both
halfs of the area. Otherwise he would have realized that their distance
does not depend on the angle $\alpha$
and the line between them is always perpendicular to the chord. This class of
curves was also found by
Auerbach\cite{Auerbach} in 1938 and by
Geppert\cite{Geppert} in 1940.
Special cases were given by Salkowski\cite{Salkow} in
1934 and by Salgaller and Kostelianetz\cite{SalKos}
in 1939.
Examples of Zindler curves are shown in  figures \ref{fhau2} to \ref{fhrk3}.
They are due to Auerbach\cite{Auerbach}, Zindler\cite{Zindler21},
and Salgaller and Kostelianetz\cite{SalKos}. 
The envelopes of the water lines are shown in red, the water lines
in blue and cyan.

\subsubsection{Density $\rho\not=$ 1/2} \label{IntroFloatNot1/2}

For a long time it was not clear, whether solutions for
$\rho\not=1/2$ exist.
Gilbert\cite{Gilbert} in his nice article
'How things float' claims in section 3
{\it 'Different heart-shaped cross sections work for other densities (he means
densities different from $1/2$) and there are other solutions that are not
heart-shaped.'}
Indeed, there are cross-sections that
are not heart-shaped for density 1/2 and densities different from
1/2. But I do not know a heart-shaped solution for density different
from 1/2 and I doubt that at the time he wrote the paper a solution
for densities different from 1/2 was known. At least he does not
give reference to such a solution.

Attempts to find solutions for
$\rho\not=1/2$ by
Salkowski\cite{Salkow} in 1934,
Gericke\cite{Gericke} in 1936,
and Ruban\cite{Ruban} in 1939 failed. As I will explain later,
it seems that Ruban was close to a solution.
It was proven by several authors that chords, which form
a triangle or a quadrangle yield only circles.

Bracho, Montejano, and
Oliveros\cite{BraMonOli01,BraMonOli04,OliMont99} were
probably the first to find solutions for densities different from 
$1/2$. They consider a carousel, which is a dynamical
equilateral polygon in which the midpoint of each edge travels
parallel to it. The trace of the vertices describe the boundary
and the midpoints outline the envelope of the waterline.
In this way they found solutions where the chords form
an equilateral pentagon.
However, their solutions were not sufficiently 
convex, since the water line cuts the cross section in 
some positions several times.

For special densities $\rho$ one can deform the circular cross-section into one
with $p$-fold symmetry axis and mirror symmetry.
\be 
r(\phi) = r_0(1+2\epsilon\cos(p\phi)
+2\sum_{n=2}^{\infty} c_n\cos(pn\phi)), \label{rphi}
\ee
where the coefficients $c_n$ are functions of $\epsilon$ and $p$ with
$c_n=O(\epsilon^n)$. The corresponding $p-2$ densities depend on $\epsilon$.
Surprisingly, the perturbation expansion in $\epsilon$
yielded (up to order $\epsilon^7$) one and the same solution
for all $p-2$ densities, although it had to be expected only for
pairs with density $\rho$ and $1-\rho$. The present author
reported this result in \cite{We2}.
This result was unexpected. It was probably true to all 
orders in $\epsilon$ and thus deserved further investigation.
(In eq. (83) of \cite{We2}v3 $c_{\delta_0}$ should be replaced 
by $s_{\delta_0}$).

In a first step the limit $p\rightarrow\infty$ was considered with
$r_0\propto p$ and $\epsilon\propto 1/p$.
This corresponds to the transition
from the circular case to the linear case mentioned in
subsection \ref{cur}. In this limit only terms
$\sum_{n,k} c_{n,k}p^{n+2k-1}\epsilon^{n+2k}\cos(pn\phi)$ in the
expansion (\ref{rphi}) contribute (with odd $n$).

{\bf Property of constant distance}
These curves have a remarkable property, which I call the
property of constant distance:

Consider two copies of the curves. Choose an arbitrary point
on each curve. Then in the linear case there exists always a 
length $\delta u$ by which the curves can be shifted against
each other, and in the circular case there exists an
angle $\delta\phi$ by which the two curves can be rotated against each
other, so that the distance $2l$ between the two points stays
constant, if they move on both curves by the same arc
distance $s$.

Considering this procedure the other way round, we may
shift curves in the linear case continuously against each other
and watch how the distance $2l$ increases with $\delta u$,
or we may rotate the curves in the circular case continuously
against each other and watch how $2l$ varies with
$\delta\phi$. When $\delta\phi$ is increased by $2\pi/p$,
then both curves fall unto themselves, and a solution for the
floating bodies is found, provided the curve is sufficiently
convex, so that the chord between the two
points does not intersect the curve at another point.

\begin{figure}[h!]
\parbox[b]{3cm}
{\includegraphics[scale=0.5]{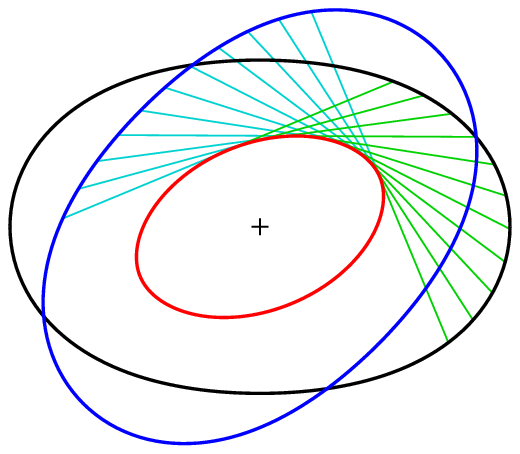}}
%{\epsfig{file=fgp1p2c.eps,scale=0.5}}
\parbox[b]{4.3cm}
{\includegraphics[scale=0.5]{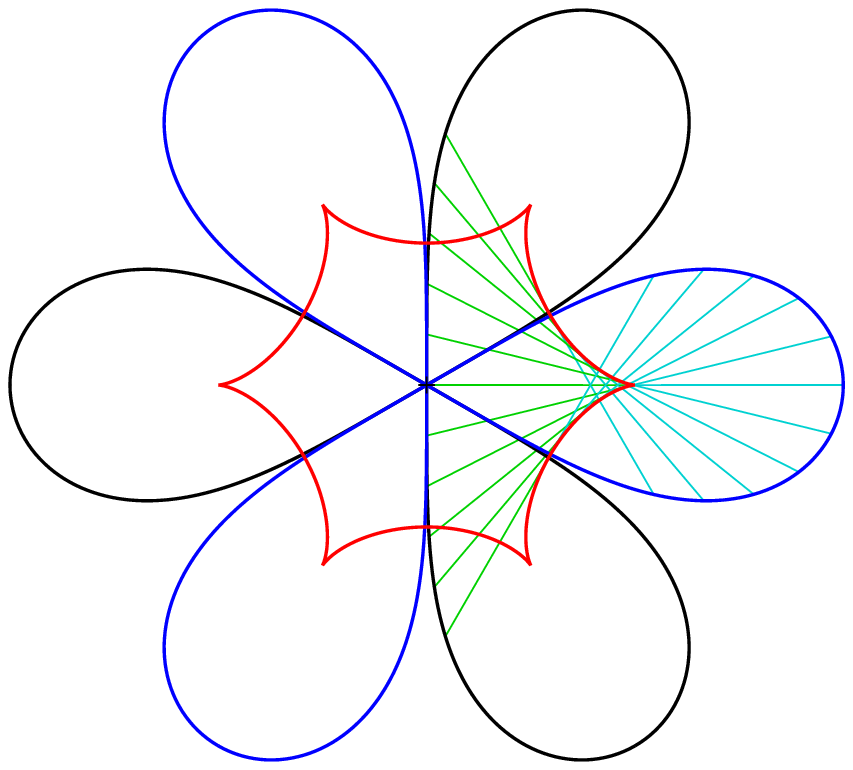}}
%{\epsfig{file=fgp2p3d.eps,scale=0.5}}
\parbox[b]{4.3cm}
{\includegraphics[scale=0.5]{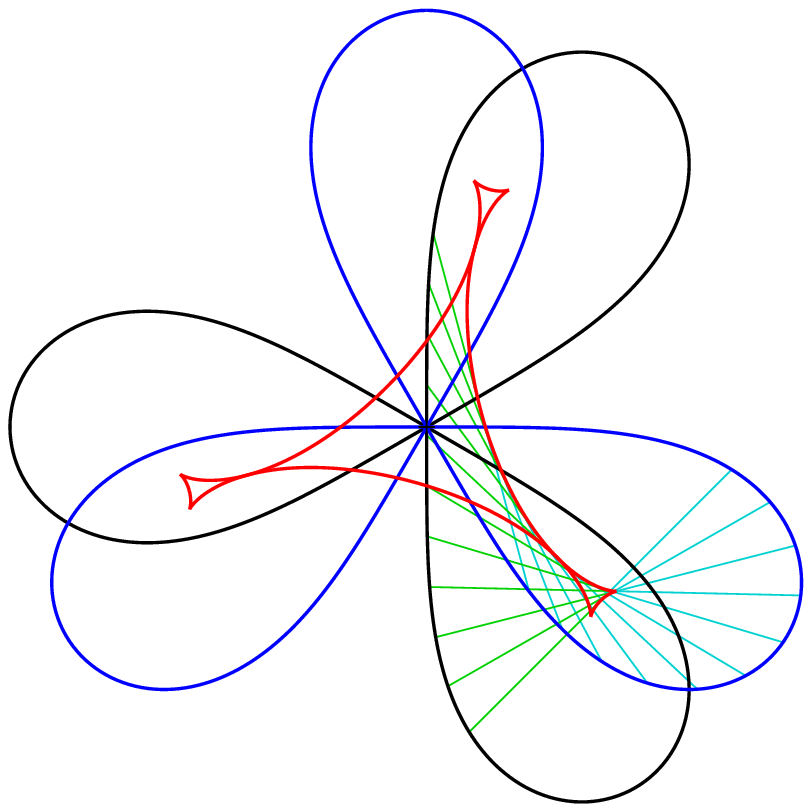}}
%{\epsfig{file=fgp2p3e.eps,scale=0.5}}
\caption{Property of constant distance}\label{fFB1}
The first figure of \ref{fElP} and figure \ref{fgp2p3c} are shown
in two copies rotated against each other by $45^0$,
$60^0$ and $30^0$.
\end{figure}

In figure \ref{fFB1} three examples for the property of
constant distance are shown. Two copies of the first of the
figures \ref{fElP} are shown in black and blue. The lines
of constant distance $2l$ are drawn in cyan and green
switching color in the middle, where they touch the red
envelope. Similarly these lines are shown for copies of the
Kiepert curve, figure \ref{fgp2p3c}, rotated against each other
by $\delta\phi=60^0$ and $\delta\phi=30^0$, resp.
The length $2l$ of the chord for the Kiepert curve is given by
$2l=a|\sin(3\delta\phi/2)|^{1/3}$.

\begin{figure}
\parbox[b]{2.8cm}
{\includegraphics[scale=0.5]{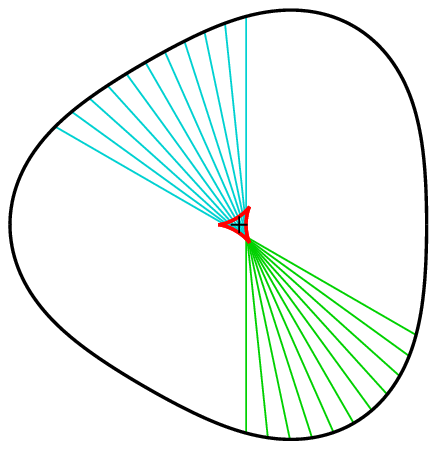}}
%{\epsfig{file=fgp1p3a.eps,scale=0.5}}
\parbox[b]{2.8cm}
{\includegraphics[scale=0.5]{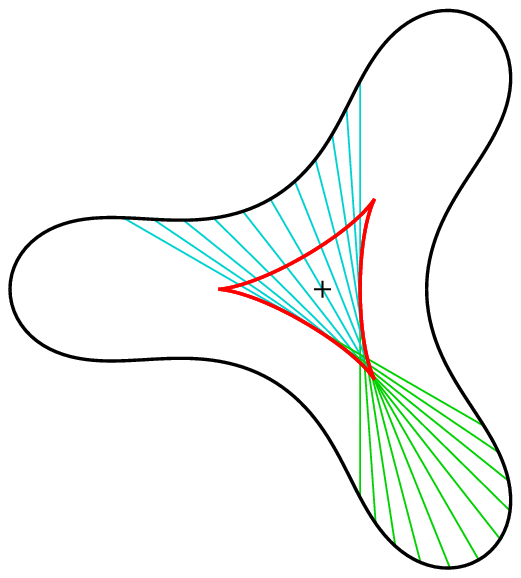}}
%{\epsfig{file=fgp1p3c.eps,scale=0.5}}
\parbox[b]{2.8cm}
{\includegraphics[scale=0.5]{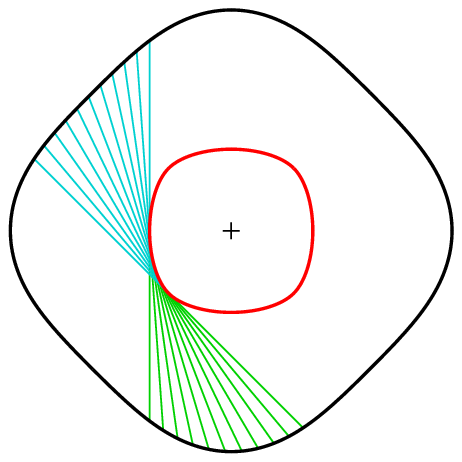}}
%{\epsfig{file=fgp1p4k.eps,scale=0.5}}
\parbox[b]{2.8cm}
{\includegraphics[scale=0.5]{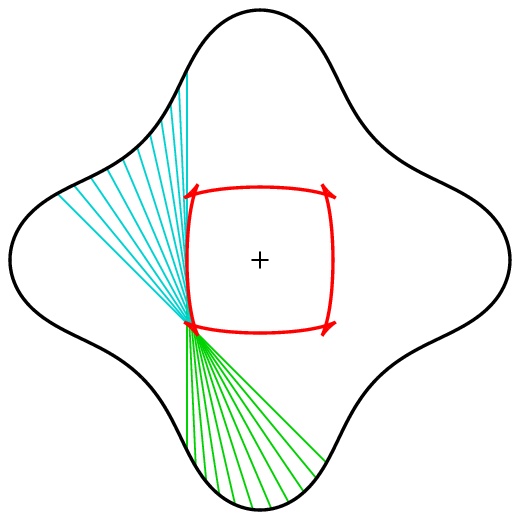}}
%{\epsfig{file=fgp1p4c.eps,scale=0.5}}
\caption{Floating bodies of equilibrium, $p=3$
and $p=4$}\label{fFB2}
\end{figure}

\begin{figure}[h!]
\parbox[b]{3.8cm}
{\includegraphics[scale=0.5]{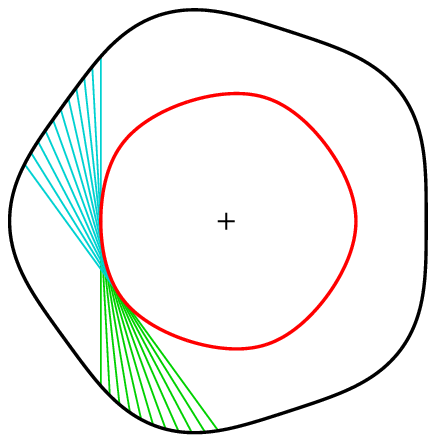}}
%{\epsfig{file=fgp1p5j.eps,scale=0.5}}
\parbox[b]{3.8cm}
{\includegraphics[scale=0.5]{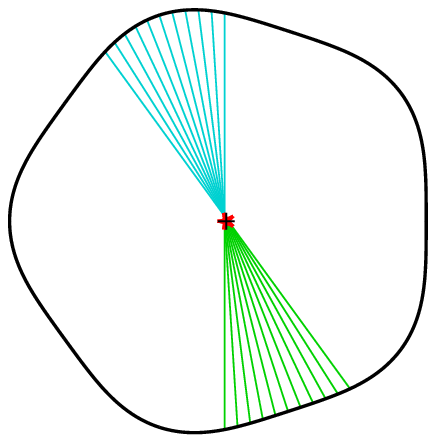}}
%{\epsfig{file=fgp1p5k.eps,scale=0.5}}
\parbox[b]{3.8cm}
{\includegraphics[scale=0.5]{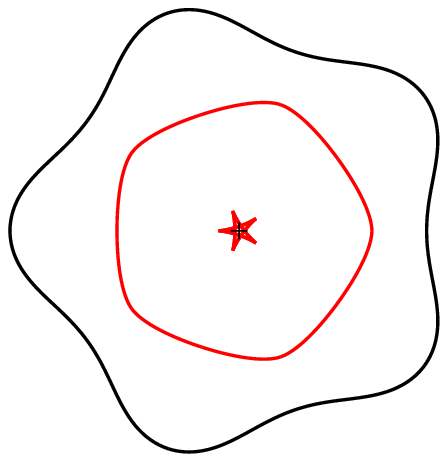}}
%{\epsfig{file=fgp1p5a.eps,scale=0.5}}
\caption{Floating bodies of equilibrium, $p=5$}\label{fFB3}
\end{figure}

The distance $2l$ shrinks for the buckled ring (first
figure of figure \ref{fFB1}) after rotation by $2\pi/p=180^0$
to zero. Therefore it cannot serve as floating
body of equilibrium. The buckled rings in figure \ref{fFB2} of
symmetry $D_3$ and $D_4$ and those in figure \ref{fFB3}
of symmetry $D_5$
are boundaries of floating bodies of equilibrium.
Rotation by $2\pi/p$ yields a non-zero distance $l$.
The waterlines are shown in green and cyan,
the envelope of the waterlines in red. The figures with odd $p$
are also solutions for $\rho=1/2$, thus special Zindler curves.
The figures with $p=5$ are besides solutions for $\rho=1/2$
also solutions for a density $\rho>1/2$ and for a density
$\rho<1/2$.

\begin{figure}[h!]
\includegraphics[scale=0.5,angle=-90]{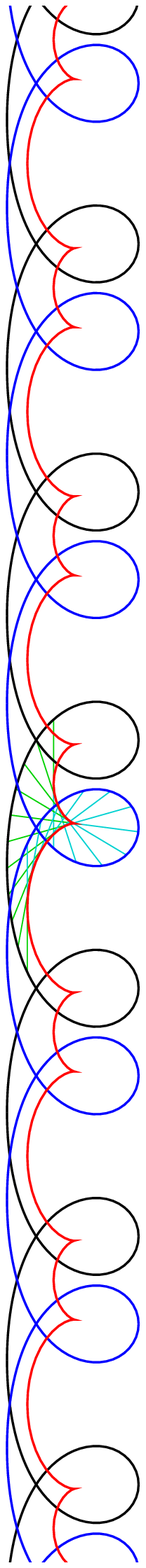}
\vspace{2mm}

\includegraphics[scale=0.5,angle=90]{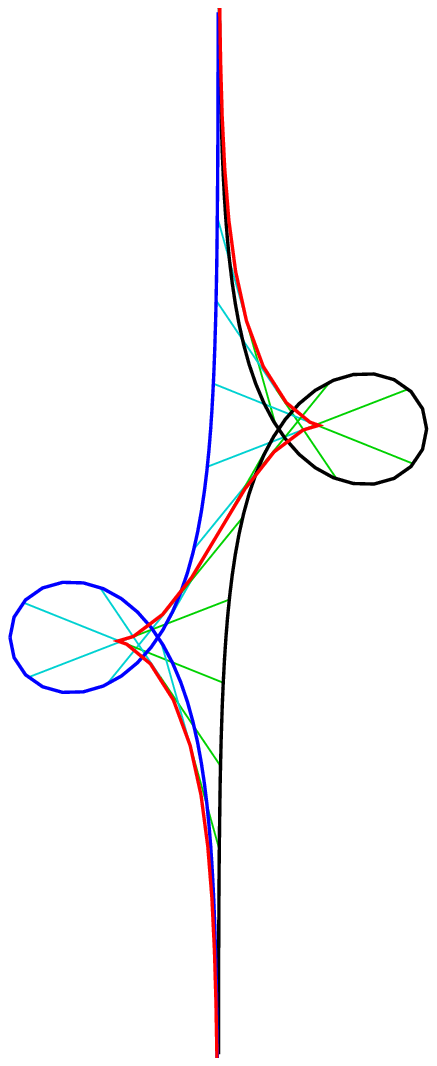}
\hspace{2mm}
\includegraphics[scale=0.5,angle=-90]{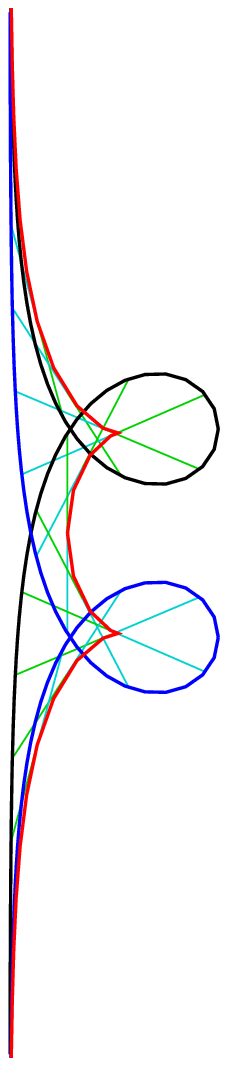}
\vspace{2mm}

\includegraphics[scale=0.5,angle=-90]{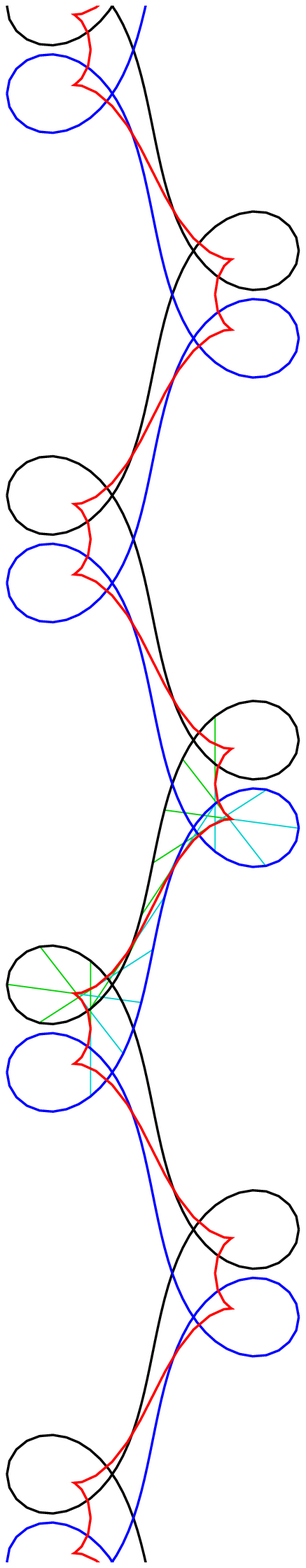}
\vspace{2mm}

\includegraphics[scale=0.5,angle=90]{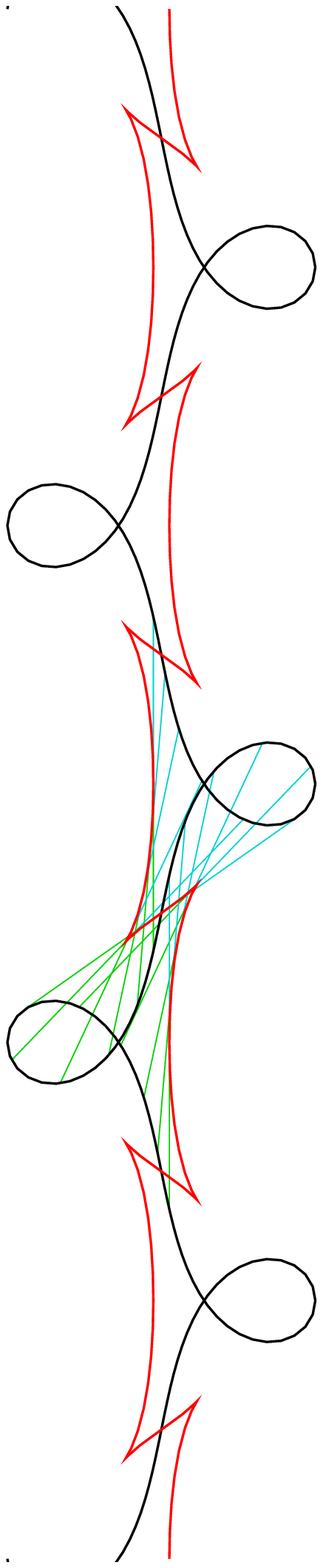}
\vspace{2mm}

\includegraphics[scale=0.5,angle=90]{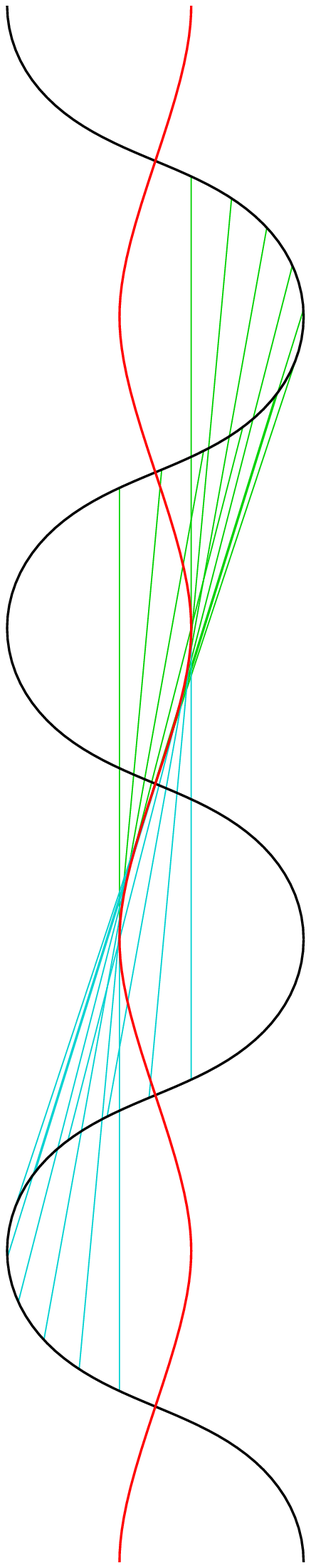}
\caption{Property of constant distance for the linear case}\label{flin}
\end{figure}

We turn to the linear case with examples in figure \ref{flin}.
In the first to third row examples of figures
from the second to fourth row of figure \ref{fEl1} are shown.
They are shifted by a distance $\delta u$ and in one case one curve
is reflected. This reflected curve is solution of eq.
\ref{lin} with the same constants $a$ and $b$.
In the fourth and fifth row two examples are shown, where
the figures were shifted so far that they fall on each
other, together with lines of length $2l$ and the envelopes.

The derivation of the
differential equations (\ref{cir},\ref{lin}) for the curves
are contained in \cite{We1} based on \cite{We2,We3}.
First the linear case was dealt with,
where first large distances, then infinitesimal distances,
and finally arbitrary distances were considered
(section 2 of \cite{We3}). It yields eq. (\ref{lin})
(Eq. (17) of \cite{We3} and Eq. (27) of \cite{We1}).
The circular case is considered in
section 3 of \cite{We3} and in section 3.2 of \cite{We1}.
In deriving this equation the author assumed that also for
non-integer periodicity $p$ such chords (of infinitesimal
length) exist between
the curves rotated against each other by nearly $2\pi$.
This assumption
yields a differential equation of order 3, (43) of \cite{We3} and (33) of
\cite{We1}.
It can be integrated easily to eq. (\ref{cir})
(Eq. (47) of \cite{We3} and Eq. (37) of \cite{We1}).
Explicit solution
of these equations showed that the property of constant distance
holds.\cite{We4a,We4}

The problem is originally non-local, since it connects end-points
of the chords generally without the necessecity of closing them
to a polygon of chords. The eqs. (\ref{lin}, \ref{cir}), however,
reduce it to a local problem: The equations connect only locus and
direction of the curve at the same point.

It came to my big surprise, when Bor, Levine, Perline, and
Tabachnikov\cite{BLPT}
pointed out, that the problem of elastica under pressure and
the problem of floating bodies of equilibrium in two dimensions
are governed by the same differential equation (\ref{cir}).

{\bf Charges in magnetic fields}
The curvature of the boundary curves is quadratic in the radius $r$,
$\kappa = 4ar^2 +2b$ according to eq. (\ref{cir2}). Thus charges
moving in a perpendicular magnetic field of such an r-dependence,
will move along these curves.

\begin{figure}[h!]
\includegraphics[scale=0.6]{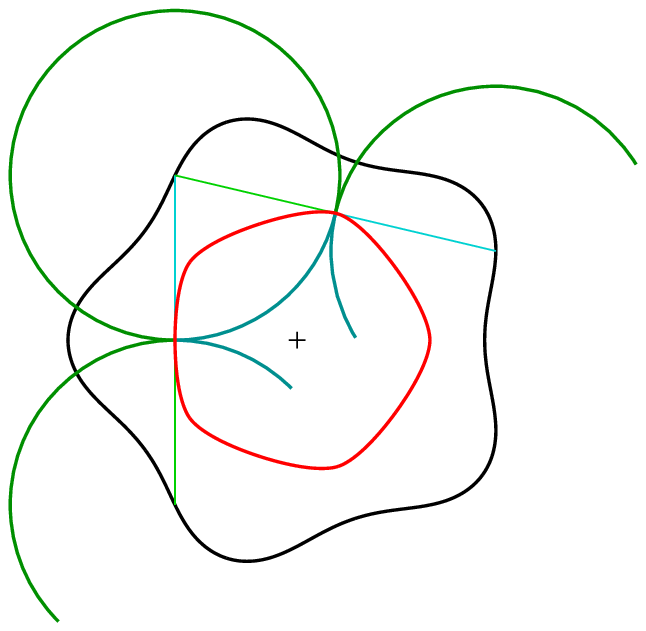}
\caption{Magnetic billiard for perpendicular incidence}\label{fgmaga}
The red boundary is given by the envelope of the chords.
%\end{figure}
\vspace{2mm}

%\begin{figure}[h!]
\includegraphics[scale=0.3,angle=-90]{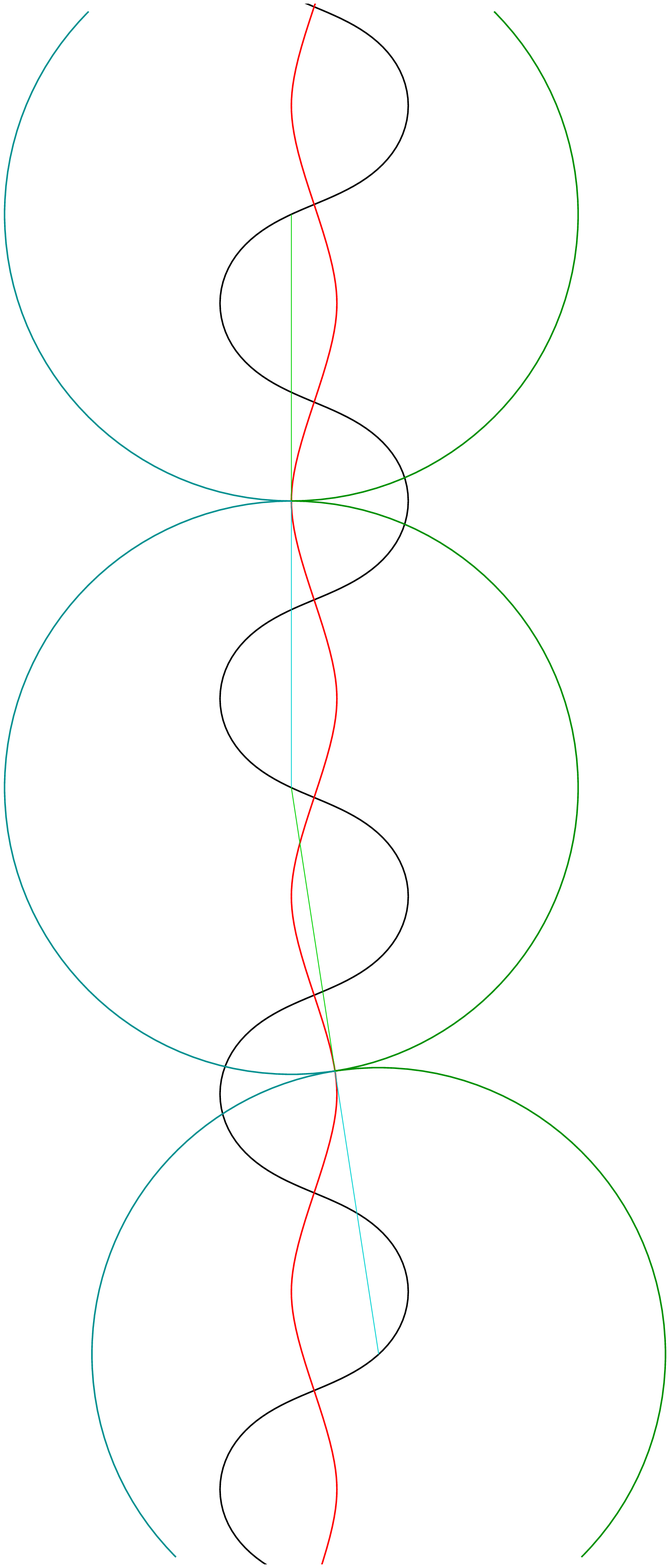}
\caption{Magnetic billiard for perpendicular incidence}\label{flmagb}
The red boundary is given by the envelope of the chords.
\end{figure}

A different system is a dynamical billiard. There a particle alternates
between free motion and specular reflections at a boundary
(angle of incidence equals angle of reflection).
Circles are boundaries with the property that the particle leaves
and arrives at the boundary at the same angle $\delta$.
Gutkin\cite{Gutkin11} found boundaries of billiards with this property,
but different from circles.

In a magnetic billiard the particle is charged and subject to a constant
perpendicular magnetic field. Thus it does not move on straight lines,
but on Larmor circles of radius $R$ given by its charge and mass, and by the
strength of the magnetic field. If the angle $\delta$ with the boundary
of the billiard is a right angle, then billiards bounded
by the (red) envelopes of the chords have this Gutkin $\delta$ property, if the
radius equals half the length of the chords, $R=l$.
(Bialy, Mironov, and Shalom\cite{BiMiSh01}). These Larmor arcs
can be inside (shown in dark blue), but also outside the boundary,
(shown in dark green) in figure \ref{fgmaga}. Such Larmor arcs
are also possible at the envelopes of the chords of linear elastica,
see figure \ref{flmagb}.

\begin{figure}[h!]
\includegraphics[scale=0.8]{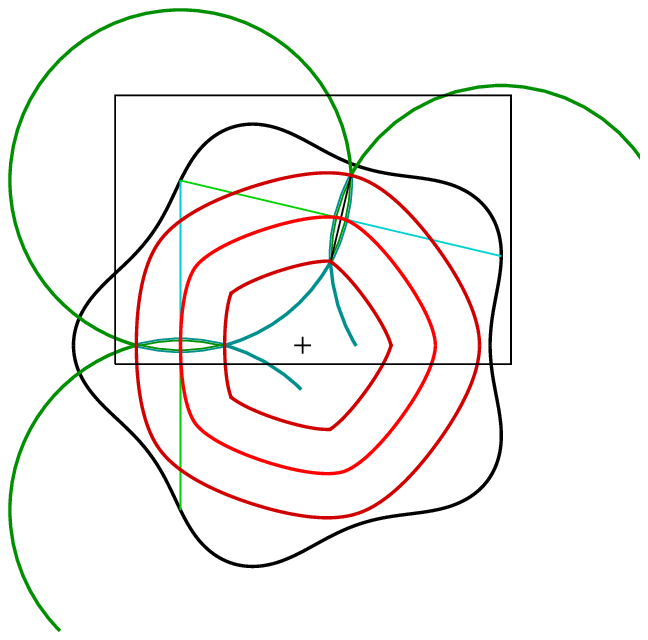} \vspace{2mm}
\includegraphics{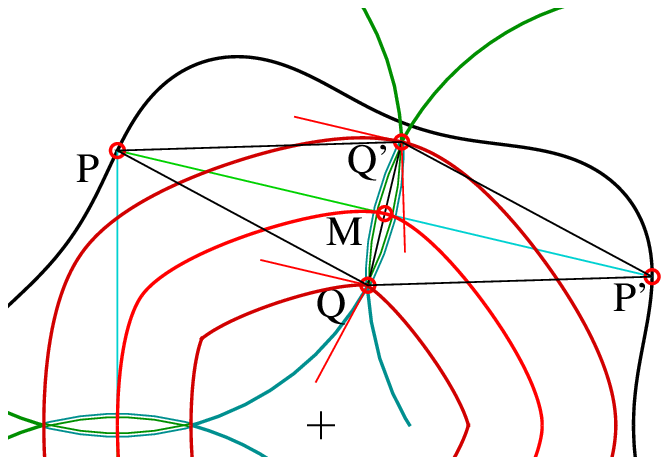}
\caption{Magnetic billiards for $\delta\not=90^0$}\label{fgmagbc}
The dark red parallel curves are the boundaries of the billiards.
Right figure: Cut-out.
\end{figure}

If the particle does not meet the boundary at right angle, then
the boundary is described by a parallel curve to the envelope.
These parallel curves are shown in figure \ref{fgmagbc} in dark red.
The chord $PP'$ touches the red envelope at the midpoint $M$.
The parallel curves and the Larmor circles go through $Q$ and $Q'$.
The tangents at these curves at $Q$ and $Q'$ are indicated by red
lines. The angle between the tangents at $Q$ be $\delta$, that at
$Q'$ is $\delta'$. Both angles add up to $180^0$. The angles in the
rhomb $PQP'Q'$ with midpoint $M$ obey
$\angle Q'PM = \angle QPM = \angle Q'P'M = \angle QP'M = 90^0-\delta$.
It is now obvious that the radius of the Larmor arcs obey $l=R\sin\delta$
and the distance between the red envelope and the boundaries of the
billiards is $R\cos\delta$.

This construction is restricted to angles $\delta$ sufficiently
close to $90^0$ and envelopes sufficiently convex.\cite{BiMiSh01}
Obviously the boundaries are not allowed to have double points.

\subsection{The Bicycle Problem}

The bicycle problem is closely related to the problem of
finding floating bodies of equilibrium.
It was addressed by Finn\cite{Finn,Finn04,Tabach06}.
The problem goes back to a criticism  of the discussion between
Sherlock Holmes and Watson in {\it The
Adventure of the Priory School}\cite{Doyle} on which way a
bicycle went whose
tires' traces are observed. Let the distance between the front 
and the rear
wheel of the bicycle be $l$. The end points of the tangent lines of length
$l$ to the trace of the rear wheel in the direction the bicycle went yields
the points of the traces of the front wheel. Thus if the tangent lines in both
directions end at the trace of the front wheel, it is open which way the
bicycle went. Thus curves $\gamma$ for the rear wheel 
(in red in Fig. \ref{watbic}) and $\Gamma$ for the
front wheel (in black in fig. \ref{watbic}) are solutions for such an ambiguous
direction of the bicycle.
The tire track problem consists in finding such curves $\Gamma$ and $\gamma$
different from the trivial solutions of circles and straight lines.

Obviously the solutions of the two-dimensional floating body
problem solve the bicycle problem, but also the linear
elastica and the Zindler curves solve the problem.
There are more solutions to the bicycle curves.
Finn argues that the variety of bicycle curves is
much larger: Draw from one point ($N_0$) of the
rear tire track the tangent to the front wheel in both 
directions and give an arbitrary smooth tire track between 
these two points ($P_0$) and ($Q_0$) in figure \ref{watbic}.
Then one can continue tire track curves in 
both directions.\cite{Finn,Finn04}.

We shortly explain why Zindler curves are bicycle curves.
Let the bicyclist go in one direction so that the
front wheel is at $(x(\alpha),y(\alpha))$ given in eq.
(\ref{Zin1}) and with the rear wheels at
$(\xi(\alpha).\eta(\alpha))$. Then the bicyclist going in the 
opposite direction is with its front wheels at
\be 
x_-(\alpha) = -l \cos(\alpha) + \xi(\alpha), \quad
y_-(\alpha) = -l \sin(\alpha) + \eta(\alpha). \label{Zin4}
\ee
Since
$(x_-(\alpha),y_-(\alpha)) = (x(\alpha+\pi),y(\alpha+\pi))$
and the traces of the rear wheels agree due to (\ref{Zin3}),
both bicyclists use the same traces for their wheels
and one cannot determine, which way the bicyclist went.

Zindler curves, but also a number of curves from elastica
and buckled rings yield envelopes, that is traces of the
rear wheels with cusps. Then the rear wheel has to go back
and forth. For these curves the front wheel has to be
turned by more than the right angle. Driving along these
traces requires artistic skills and a suitable bicycle.
Apart from the Zindler
curves (figs. \ref{fhau2} to \ref{fhrk3}) this is the case for
the first, second, and fourth buckled ring of figure \ref{fFB2},
the inner rear trace of figure \ref{fFB3}, and the elastica
of the fourth row of fig. \ref{flin}.
However, the third traces of fig. \ref{fFB2}, the outer
trace of fig. \ref{fFB3}, and the traces of the fifth row
of fig. \ref{flin} can be easily traversed. They constitute good
solutions of the bicycle problem.

\section{Derivation of the differential equations}\label{Deriv}

The equations governing the elastic beam (wire) have been given in
different ways. They are developed in this section. These differential
equations have been given in numerous papers apart from their
original derivation in numerous papers. I mention only
\cite{BLPT,Lan99,LanSin84,Levien,Love,Singer,SinHan19,TadOdeh67}
and references therein.

\subsection{Bernoulli - Huygens solution}\label{BerHuy}

\begin{figure}[h!]
\includegraphics{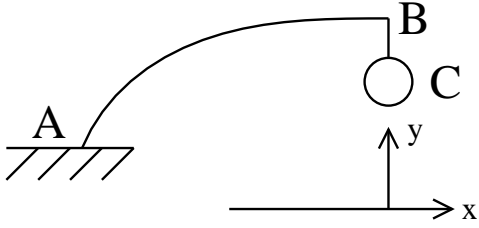}
\caption{Beam AB with load C}
\label{fBH}
\end{figure}

James Bernoulli considered initially a beam AB loaded by a weight C
at the end assuming the beam to be horizontally at the point
of the load, see figure \ref{fBH}.
He assumed the curvature $\kappa$ to be a function $f$ of the 
moment. Hence,
\be
\kappa = \frac{\de\phi}{\de s} 
= \frac{\de}{\de x} \frac{y'}{\sqrt{1+y^{\prime 2}}} = f(x),
\ee
where $\phi$ is the angle of the tangent at the curve against the x-axis.
Integration yields
\be
y' = \frac{S(x)}{\sqrt{1-S^2(x)}}, \quad S(x) = \int_0^x \de\xi f(\xi).
\ee
Assuming a linear relation between the curvature and the moment gives
\be
f(x) = \frac{2x}{a^2}, \quad S(x) = \frac{x^2\pm ab}{a^2}
\ee
and thus equation (\ref{D2}). The {\it rectangular elastica}
primarily considered by James Bernoulli is obtained for $b=0$.

\subsection{Forces and Pressure}\label{DerivForce}

\begin{figure}[h!]
\includegraphics{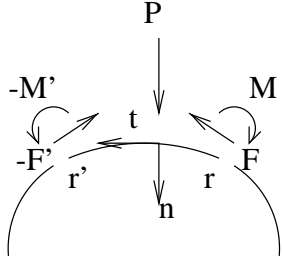}
\caption{Wire, forces, torques, pressure, and tangent and normal vectors}
\label{fpiece}
\end{figure}

Here we consider the force and torque acting in the elastic wire
similar to that given by Levy\cite{Levy} and derive eqs. (\ref{lin2},
\ref{cir2}).
Let us cut out a piece from $\w r$
to $\w r'$ (figure \ref{fpiece}). At the ends act forces $\w F$ and $-\w F'$.
In addition a force per length $P$ perpendicular to the
wire exerts the force
\be
\w F_P = P \w e_3\times(\w r'-\w r)
\ee
on the piece of wire, where $\w e_3$ is the unit vector perpendicular
to the plane.

The total force vanishes in the static case,
\be
\w F-\w F' + P \w e_3\times(\w r'-\w r) = 0.
\ee
Integration yields the force acting on the wire,
\be
\w F(\w r) = \w F_0 + P \w e_3 \times \w r.
\ee
Next we consider the torque acting on the piece of wire.
Due to the curvature of the wires there are torques $\w M$ and
$-\w M'$ at the ends of the wires. Moreover $\w r\times \w F$
and $-\w r'\times \w F'$ are exerted by the forces at the ends
and the force on the piece exerts
\bea
\w M_P=\int\de\w M_P = \int(-\w r\times\de\w F_P)
= -P\int\w r\times(\w e_3\times\de\w r) \nn
= -P\int\w e_3(\w r\de\w r)
=-\frac12 P\w e_3(r^{\prime2}-r^2).
\eea
The total torque vanishes,
\be
\w M-\w M' + \w r\times\w F -\w r'\times\w F' +\w M_P =0.
\ee
This yields
\be
\w M(\w r) = -\w r\times\w F_0 -\frac12 P\w e_3 r^2 + \w M_0.
\label{M}
\ee
Let us introduce
\be
\w M=\w e_3 M, \quad \w M_0=\w e_3 M_0, \quad
\w e_3\times\w F_0 = \w F_n.
\ee
Multiplication of (\ref{M}) by $\w e_3$ yields
\be
M(\w r) = \w r\cdot\w F_n -\frac 12 P r^2 + M_0
=-\frac 12 P(\w r-\frac{\w F_n}P)^2  +\frac{\w F_n^2}{2P} + M_0.
\label{Mr}
\ee
The torque is proportional to the curvature $\kappa$
\be
M = - EI\kappa,
\ee
where $E$ is the elastic modulus and $I$ the second moment
with respect to the axis in $e_3$-direction
through the center of gravity of the cross section of the wire.
Love calls this the {\it ordinary approximate theory} and
discusses it in sections 255 --  258 of his book.\cite{Love}

If there is no external force, $P=0$, then the first
equation (\ref{Mr}) yields
\be
\kappa =  -\frac 1{EI}(\w r\cdot\w F_n + M_0)
\ee
in agreement with eq. (\ref{lin2}).
Hence $\kappa$ increases linearly with $\w r$ parallel to $\w F_n$.

If $P\not=0$, then
we replace $\w r-\w F_n/P \rightarrow \w r$ and obtain
\be
\kappa = 4ar^2 + 2b, \quad a=P/(8EI), \quad
b=-(\w F^2_n/(2P)+M_0)/(2EI).
\ee
Hence the curvature $\kappa$ increases quadratically with $\w r$
in accordance with eq. (\ref{cir2}).

\subsection{Equation for the curvature}\label{DerivCurve}

We start with the Frenet-Serret formula for the tangential vector $\w t$ and
the normal vector $\w n$ of the wire
\be
\frac{\de\w t}{\de s} = \kappa \w n, \quad
\frac{\de\w n}{\de s} = -\kappa \w t.
\label{K1}
\ee
We express the force as
\be
\w F = F_t \w t + F_n \w n. \label{K2}
\ee
Then going along the wire (beam) we obtain
\be
\frac{\de}{\de s} (F_t \w t + F_n \w n) = P \w n. \label{K3}
\ee
Thus
\bea
\frac{\de F_t}{\de s} - \kappa F_n = 0, \label{K4} \\
\frac{\de F_n}{\de s} + \kappa F_t = P. \label{K5}
\eea
The torque obeys
\be
\frac{\de M}{\de s} = - F_n. \label{K6}
\ee
Finally the torque is assumed to be proportional to the curvature
\be
M = -EI \kappa. \label{K7}
\ee
This yields
\be
EI \frac{\de\kappa}{\de s} = F_n. \label{K8}
\ee
We insert this relation in eq. (\ref{K4}) and obtain
\be
\frac{\de F_t}{\de s} - EI \kappa \frac{\de\kappa}{\de s} = 0,
\ee
which can be integrated to
\be
F_t - \frac{EI}2 \kappa^2 = c'.
\ee
Equations (\ref{K8}) and (\ref{K5}) yield
\be
EI\frac{\de^2\kappa}{\de s^2} = \frac{\de F_n}{\de s}
= -\kappa F_t + P
=\kappa(-c' - \frac{EI}2 \kappa^2) + P.
\ee
Hence
\be
\frac{\de^2\kappa}{\de s^2} + \frac 12 \kappa^3 + \frac{c'}{EI} \kappa
- \frac P{EI} = 0. \label{eqkap}
\ee
If we multiply this equation by $\de\kappa/\de s$ and integrate,
then we obtain
\be
\frac 12 \left(\frac{\de\kappa}{\de s}\right)^2
+ \frac 18\kappa^4 + \frac{c'}{2EI}\kappa^2 - \frac P{EI} \kappa = \hat E.
\label{eqkap2}
\ee
These are the equations (\ref{EuLa}) and (\ref{EuLa2}).

\subsection{Geometrical derivation}\label{DerivGeom}

In this subsection we will derive this equation requiring the extreme
of the integral over the square of the curvature with appropriate
side conditions.
We perform a purely geometrical derivation. As a function of the
arc parameter $s$ we introduce the angle $\phi$ of the tangent
against the x-axis and the Cartesian coordinates. The origin
is at $s=0$. The curve starts with the angle $\phi_0$.
\bea
\phi(s) &=& \phi_0 + \int_0^s \de s' \kappa(s'), \\
x(s) &=& \int_0^s \de s' \cos(\phi(s')), \\
y(s) &=& \int_0^s \de s' \sin(\phi(s')).
\eea
The length of the arc be $s_0$. The area between the arc and the
straight line from the origin to the endpoint of the arc
$(x(s_0),y(s_0))$ is given by
\be
A = \frac 12 \int_0^{s_0} \de s
(x(s) \sin(\phi(s)) - y(s) \cos(\phi(s))).
\ee
We may have several side conditions on the curve:
the angle at the end point $\phi(s_0)$,
the coordinates of the end point  and the area fixed.
Thus the corresponding quantities have to be subtracted from the
integral over $\kappa^2/2$ by Lagrange multipliers
$\lambda_1 ... \lambda_5$,
\be
I= \frac 12 \int_0^{s_0} \de s \kappa^2(s) 
-\lambda_1 \phi(s_0) -\lambda_2 x(s_0) - \lambda_3 y(s_0)
-\lambda_4 A - \frac{\lambda_5}2(x^2(s_0)+y^2(s_0)). \label{LagI}
\ee
In total these may be too many conditions. But for those
we do not take into account, we set $\lambda_i=0$.
The variation of $I$ has to vanish,
\bea
\delta I &=& \int_0^{s_0} \de s \delta\kappa(s) F(s), \\
F(s) &=& \kappa(s) -\lambda_1
+(\lambda_2+\lambda_5 x(s_0)) \int_s^{s_0} \de s' \sin(\phi(s')) \nn
&-& (\lambda_3+\lambda_5 y(s_0)) \int_s^{s_0} \de s' \cos(\phi(s')) \nn
&-& \frac{\lambda_4}2 \int_s^{s_0} \de s' (x(s')\cos(\phi(s'))
+y(s')\sin(\phi(s'))) \nn
&+& \frac{\lambda_4}4 \left( \int_s^{s_0} \de s' \sin(\phi(s'))
\right)^2 +\frac{\lambda_4}4 \left( \int_s^{s_0}
\de s' \cos(\phi(s')) \right)^2 \\
&=& \kappa(s) - \lambda_1 + \lambda_2 (y(s_0)-y(s)) - \lambda_3
(x(s_0)-x(s)) \nn
&+& \frac{\lambda_4}2 [x(s)(x(s)-x(s_0)) + y(s)(y(s)-y(s_0))] \nn
&+& \lambda_5(x(s_0)y(s)-y(s_0)x(s)). \label{F0}
\eea
The equation $F(s)=0$ has to be solved.One immediately sees
from eq. (\ref{F0})
that $\kappa$ depends (for $\lambda_4\not=0$) quadratically
on the distance from some point.
The derivatives of $F$ with respect to $s$, indicated by a dot
vanish,
\newcommand{\dddot}[1]{\stackrel{\mbox{...}}{#1}}
\bea
\dot F(s) &=& \dot\kappa(s) +K_1 = 0, \\
K_1 &=& 
\cos(\phi(s))[\lambda_3+\lambda_4 x(s) - \frac{\lambda_4}2 x(s_0)
-\lambda_5 y(s_0)]\nn
&+& \sin(\phi(s))[-\lambda_2+\lambda_4 y(s) - \frac{\lambda_4}2 y(s_0)
+\lambda_5 x(s_0)], \\
\ddot F(s) &=& \ddot\kappa(s) +K_2\kappa +\lambda_4 = 0, \\
K_2 &=& 
-\sin(\phi(s))[\lambda_3+\lambda_4 x(s) - \frac{\lambda_4}2 x(s_0)
-\lambda_5 y(s_0)] \nn
&+& \cos(\phi(s))[-\lambda_2+\lambda_4 y(s) - \frac{\lambda_4}2 y(s_0)
+\lambda_5 x(s_0)], \\
\dddot F(s) &=& \dddot\kappa(s) +K_2\dot\kappa(s) -K_1\kappa(s)^2
= 0.
\eea
Elimination of $K_1$ and $K_2$ yields
\be
\kappa \dot F(s) -\frac 1{\kappa^2}\dot\kappa \ddot F(s)
+\frac 1{\kappa} \dddot F(s)
= \kappa\dot\kappa
- \frac 1{\kappa^2} \dot\kappa \ddot\kappa
-\frac{\lambda_4}{\kappa^2} \dot\kappa
+\frac 1{\kappa} \dddot\kappa = 0. \label{F123}
\ee
This expression is a complete derivative. Integration and
multiplication by $\kappa$ gives
\be
\frac 12 \kappa^3 + \ddot\kappa +\lambda_4 +C_1\kappa = 0.
\label{F123a}
\ee
This is the Lagrange equation for the relative extrema of
the bending energy.
Multiplication of this expression by $\dot\kappa$ and another
integration yields
\be
\frac 12 \dot\kappa^2 + \frac 18 \kappa^4
+\lambda_4\kappa + \frac{C_1}2 \kappa^2 = C_2, \label{F123b}
\ee
which agrees with eqs. (\ref{eqkap}) and (\ref{eqkap2}) by renaming the
constants.
Apart from eq. (\ref{F123}) also boundary conditions for
$F=0$, $\dot F=0$, and $\ddot F(0)$ at $s=0$ or $s=s_0$
have to be fulfilled. Thus $\kappa$, $\dot\kappa$, and
$\ddot\kappa$ at $s=0$ or $s=s_0$ are expressed by $\lambda_i$
and $x(s_0)$, $y(s_0)$, $\phi$ at $s=0$ or $s=s_0$.
Insertion in (\ref{F123a}) and (\ref{F123b}) yields the corresponding
constants $C_1$ and $C_2$.

\subsection{Water contained in a cloth sheet}

James Bernoulli showed that water contained in a long
cloth sheet (in y-direction) is bound in the xz-direction
by an elastic curve. (See Levien\cite{Levien}, footnotes 3 and 5,
and Truesdell\cite{Truesdell1}, p. 201).
This problem goes under the name of {\it lintearia}.

\begin{figure}[h!]
\includegraphics[angle=270]{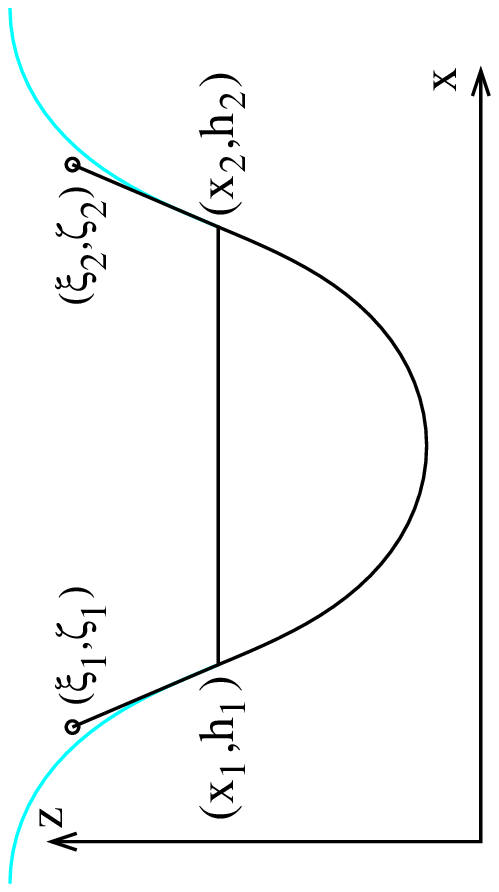}
\caption{Cross section of the cloth filled with water 
for $x_1>\xi_1$ and $x_2<\xi_2$.}
\label{flma2}

\includegraphics[angle=270]{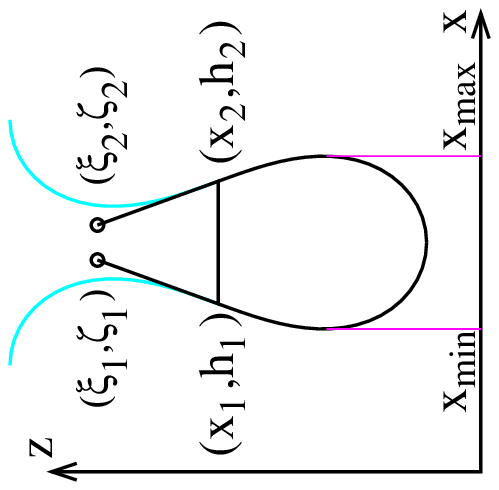}
\caption{Cross section of the cloth filled with water
for $x_1<\xi_1$ and $x_2>\xi_2$.}
\label{flmd2}
The cyan curve indicates in both cases
the continuation of the elastica curve.
\end{figure}

The cloth ends should be fixed at $(\xi_1,\zeta_1)$ and
$(\xi_2,\zeta_2)$. The height of the water-line is $h$.
The water surface ranges from $x_1$ to $x_2$ as in figures
\ref{flma2} and \ref{flmd2}.

We give two derivations, a longer one using variational techniques,
and a shorter one, considering forces and pressure as in
subsection \ref{DerivCurve}.

\subsubsection{Variation of the potential energy}

We denote the area of the cross section by $A$,
the breadth of the cloth by $s$,
the potential energy by $V$,
the length of the cloth in y-direction $L_y$,
the gravitational constant $g$,
and the density of the water $\rho$.
Then the area of the cross section $A$, the potential energy $V$
and the breadth $s$ of the cloth are given by

\newcommand{\intx}{\int_{x_1}^{x_2}\de x}

\bea
A &=& \intx (h(x)-z(x)), \\
V &=& \frac{L_y g \rho}2 \intx (h^2(x)-z^2(x)), \\
s &=& s_1+s_2+s_3, \\
s_1 &:=& \sqrt{(\xi_1-x_1)^2+(\zeta_1-h_1)^2}, \\
s_2 &:=& \sqrt{(\xi_2-x_2)^2+(\zeta_2-h_2)^2}, \\
s_3 &:=& \intx \sqrt{1+z^{\prime 2}}, \\
z' &:=& \frac{\de z}{\de x}, \quad h':=\frac{\de h}{\de x}.
\eea
Of course we know that $h(x)$ does not depend on $x$. However,
it is of advantage to have $h(x_1)$ and $h(x_2)$ as two
variables, which can be varied independently.

We look for the extreme of $V-\alpha A-\sigma S$. The variations are
\bea
\delta A &=& \intx(\delta h(x)-\delta z(x)), \\
\delta V &=&  L_yg\rho \intx (h\delta h -z \delta z), \\
\delta s_1 &=& \frac{\delta x_1(x_1-\xi_1)
+ \delta h_1 (h_1-\zeta_1)}{s_1}, \\
\delta s_2 &=& \frac{\delta x_2(x_2-\xi_2)
+ \delta h_2 (h_2-\zeta_2)}{s_2}, \\
\delta s_3 &=& \delta x_2 \sqrt{1+z_2^{\prime 2}}
-\delta x_1 \sqrt{1+z_1^{\prime 2}}
+\intx \frac{z'}{\sqrt{1+z^{\prime 2}}}
\frac{\de \delta z}{\de x}, \\
z'_i &:=& z'(x=x_i), \quad h'_i := h'(x=x_i).
\eea
We have not included contributions proportional to $\delta x_i$ in
$\delta A$ and $\delta V$, since $h_i=z_i$ at $x_i$.
Partial integration transforms the integral to
\be
\intx \frac{z'}{\sqrt{1+z^{\prime 2}}}
\frac{\de \delta z}{\de x} =
\delta z_2 \frac{z'_2}{\sqrt{1+z_2^{\prime 2}}}
-\delta z_1 \frac{z'_1}{\sqrt{1+z_1^{\prime 2}}}
-\intx \delta z \frac{\de}{\de x}\left(\frac{z'}
{\sqrt{1+z^{\prime 2}}}\right).
\ee
The variation of $V-\alpha A-\sigma S$ contains contributions
proportional to $\delta x_i,\delta h_i$, and $\delta z_i$
and an integral over $x$,
\bea
&& \intx (L_y g \rho h(x)-\alpha) \delta h(x) \\
&-& \intx \left(L_y g \rho z(x) -\alpha
- \sigma\frac{\de}{\de x}\left(\frac{z'}
{\sqrt{1+z^{\prime 2}}}\right)\right) \delta z(x)
\eea
The variation of $\delta h(x)$ yields constant $h$ as expected,
\be
\alpha = L_y g \rho h.
\ee
The variation of $\delta z(x)$ yields
\be
L_y g \rho (z(x)-h) = \sigma\frac{\de}{\de x}\left(\frac{z'}
{\sqrt{1+z^{\prime 2}}}\right)
= \sigma \frac{z\dpr}{(1+z^{\prime 2})^{3/2}}
= \sigma \kappa.
\ee
Thus the curvature increases proportional to the depth
measured from the water-surface.

Since the variation of $h(x_i)=z(x_i)$ yields
\be
\delta h_i +h'_i \delta x_i = \delta z_i + z'_i \delta x_i, \quad
\delta z_i = \delta h_i - z'_i \delta x_i,
\ee
we obtain the contributions
\bea
&& \delta x_1 \left(\frac{x_1-\xi_1}{s_1}
\mp \frac 1{\sqrt{1+z_1^{\prime 2}}}\right)
+\delta h_1 \left(\frac{h-\zeta_1}{s_1}
\mp \frac{z_1'}{\sqrt{1+z_1^{\prime 2}}}\right) \nn
&+& \delta x_2 \left( \frac{x_2-\xi_2}{s_2} 
\pm \frac{z_2'}{\sqrt{1+z_2^{\prime 2}}}\right)
+ \delta h_2 \left( \frac{h-\zeta_2}{s_2}
\pm \frac{z_2'}{\sqrt{1+z_2^{\prime 2}}}\right). \label{dir}
\eea
The factors of $\delta x_i$ and $\delta h_i$ have to vanish.
They yield the direction of the lines from the lines of
the suspension to the lines where the cloth touches the waterline.
If $x_1>\xi_1$ and $\xi_2>x_2$ as in fig. \ref{flma2},
then the upper signs in (\ref{dir}) apply, and
the slope is con\-ti\-nu\-ous across the waterline as expected.

If instead $x_1<\xi_1$ and $\xi_2<x_2$ as in fig. \ref{flmd2},
then $z(x)$ is double-valued with values $z_-(x)$ and $z_+(x)$.
Then the $x$-integral of $s_3$ reads
\be
s_3 = \int_{x_{\rm min}}^{x_{\rm max}} \de x \sqrt{1+z_-^{\prime 2}(x)}
+\int_{x_{\rm min}}^{x_1} \de x \sqrt{1+z_+^{\prime 2}(x)}
+\int_{x_2}^{x_{\rm max}} \de x \sqrt{1+z_-^{\prime 2}(x)}.
\ee
Accordingly the contributions from $s_3$ change sign and the lower
signs in (\ref{dir}) apply. Again the slope is continuous across the
waterline. The expression for the area and similarly for
the potential have different signs in front of the integrals,
\be
A = \int_{x_{\rm min}}^{x_{\rm max}} \de x (h(x)-z_-(x))
-\int_{x_{\rm min}}^{x_1} \de x (h(x)-z_+(x))
-\int_{x_2}^{x_{\rm max}} \de x (h(x)-z_+(x)).
\ee

\subsubsection{Considering forces and pressure}

A simpler derivation can be given by considering the forces
and the pressure as in subsection \ref{DerivCurve}.

Since the cloth can be bent without exerting any forces
or torques, one has $F_n=0$, $M=0$, $c=0$. Thus eqs.
(\ref{K4}) and (\ref{K5}) read
\be
\frac{\de F_t}{\de s}=0, \quad \kappa \frac{F_t}{L_y} = P,
\ee
where $F_t$ is the total force over the length $L_y$.
The pressure acts from inside and depends on $z$,
\be
P=-\rho g (h-z).
\ee
This yields
\be
F_t = {\rm const}, \quad \kappa = \frac{L_y\rho g (z-h)}{F_t}
\ee
in agreement with equation (\ref{lin2}).

\subsection{Elastica as roulette of Hyperbola}

We determine the differential equation for the roulette of
the hyperbola. We parametrize the coordinates $(x_h,y_h)$
of the hyperbola by a parameter $p$,
\bea
x_h = a\cosh(p), && y_h = b \sinh(p), \\
\frac{\de x_h}{\de p} = a\sinh(p), &&
\frac{\de y_h}{\de p} = b \cosh(p).
\eea
$a$ and $b$ are the semi axes of the hyperbola.
The arc length $s$ of the hyperbola reads
\be
\de s = -\sqrt{q} \de p, \quad
q=a^2\sinh^2(p)+b^2\cosh^2(p)
= \frac{a^2+b^2}2 \cosh(2p) + \frac{b^2-a^2}2.
\ee
The coordinates $(x,y)$ of the midpoint between the branches of
the hyperbola rolling on the x-axis without slipping obey
\be
\left(\begin{array}c x \\ y \end{array}\right)
= \left(\begin{array}c s \\ 0 \end{array}\right)
+\frac 1{\sqrt q} \left(\begin{array}{cc} a\sinh(p) & b\cosh(p) \\
-b\cosh(p) & a\sinh(p) \end{array}\right)
\left(\begin{array}c x_h \\ y_h \end{array}\right),
\ee
which yields
\be
x = s+\frac{a^2+b^2}{2\sqrt{q}}\sinh(2p), \quad
y = -\frac{ab}{\sqrt{q}}.
\ee
The derivatives
\be
\frac{\de x}{\de p} = \frac{a^2b^2}{q^{3/2}}, \quad
\frac{\de y}{\de p} = \frac{ab(a^2+b^2)}{2q^{3/2}}\sinh(2p)
\ee
yield
\be
\frac{\de y}{\de x} = \frac{a^2+b^2}{2ab}\sinh(2p).
\ee
We express $\sinh(2p)$ by $q$ and $q$ by $y$,
\be
q=\left(\frac{ab}y\right)^2, \quad
\left(q+\frac{a^2-b^2}2\right)^2
= \left(\frac{a^2+b^2}2\sinh(2p)\right)^2
+\left(\frac{a^2+b^2}2\right)^2.
\ee
This yields the differential equation for the Sturm roulette
for the hyperbola rolling on a straight line
\bea
\frac{\de y}{\de x} &=& \frac 1{ab}\sqrt{\left(q+\frac{a^2-b^2}2\right)^2
-\left(\frac{a^2+b^2}2\right)^2}
= \frac 1{ab}\sqrt{(q+a^2)(q-b^2)} \nn
&=& \frac 1{ab}\sqrt{\left(\frac{a^2b^2}{y^2}+a^2\right)
\left(\frac{a^2b^2}{y^2}-b^2\right)}
=\frac 1{y^2}\sqrt{(b^2+y^2)(a^2-y^2)}.
\eea
For a rectangular hyperbola one has $b=a$ and this equation
reduces to
\be
\frac{\de y}{\de x} = \frac{\sqrt{a^4-y^4}}{y^2},
\ee
which is eq. (\ref{D1}) for the rectangular elastica,
only the coordinates $x$ and $y$ exchanged.

\begin{figure}[h!]
\includegraphics{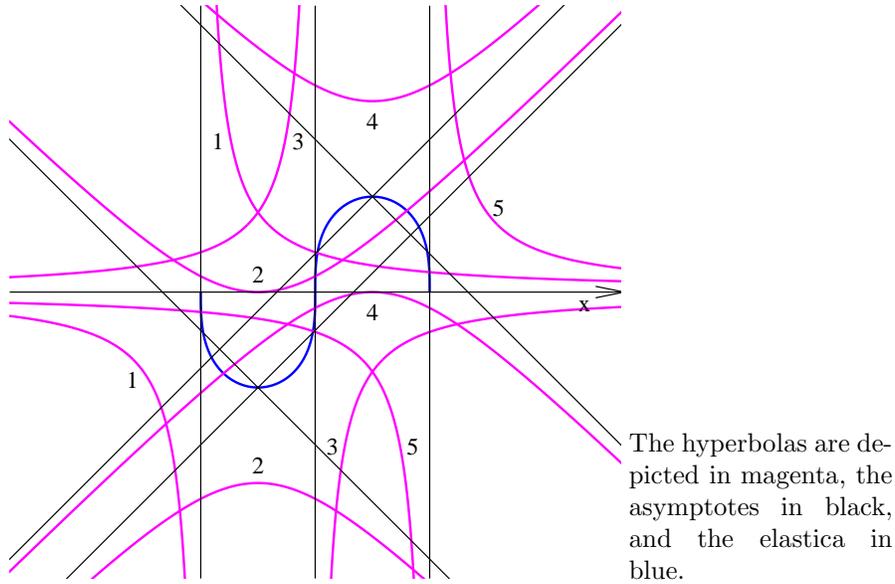}
\parbox[b]{35mm}{The hyperbolas are depicted in magenta, the
asymptotes in black, and the elastica in blue.}
\caption{Full period of the rectangular elastica as roulette
of hyperbola}
\label{fhyp2}
\end{figure}

The roulette shown in figure \ref{fhyp1} shows only half a period
of the rectangular elastica. One obtains the full period, if
one first rolls the upper branch of the hyperbola, as shown
in figure \ref{fhyp2} starting with hyperbola 1 over hyp. 2
to hyp. 3, where for 1 and 3 one asymptote is along the x-axis.
Now we roll the lower branch from hyp. 3 to 4 and 5. This
yields the second half of the elastica. The hyperbola 5 has
the same orientation as number 1, only shifted by one period
along the x-axis. Now one may continue for another period,
and so on.

\section{The case $\rho=1/2$}\label{1/2}

The boundaries of the two-dimensional floating bodies of equilibrium
with density $1/2$ are Zindler curves. We describe these closed curves
in the following subsection. Chords of the curves bisect both the
boundary and the enclosed area. The centers of gravity of these
halves have constant distance and their connecting line is perpendicular
to the chord (subsect. \ref{1/2Grav}).
In the following subsection \ref{1/2Rem} I comment
on some papers which investigate plain regions with the
property that chords of constant length cut the region in
two pieces of constant areas. Some of them deal with the problem
of floating bodies of equilibrium, others are purely
geometrical.
 
\subsection{Zindler curves}\label{1/2Zind}

Zindler considers mainly in sects. 6, 7 and 10 of
\cite{Zindler21} convex plain areas with the property
that any chord between two points bisecting the perimeter has
constant length and bisects the area.

The envelope of the chords is defined by Equation (\ref{Zin2})
with the constraint (\ref{Zin3}) 
which implies $\xi(\pi)=\xi(0)$, $\eta(\pi)=\eta(0)$.
The boundary can be parameterized by equation (\ref{Zin1}).
Hence
\be
x(\alpha+\pi)=\xi(\alpha)-l\cos(\alpha), \quad
y(\alpha+\pi)=\eta(\alpha)-l\sin(\alpha).
\ee
The diameters $D(\alpha)$ ranges from
$(x(\alpha),y(\alpha))$ to
$(x(\alpha+\pi),y(\alpha+\pi))$. It
bisects the perimeter and the area enclosed by the boundary, provided $l$ is
sufficiently large, so that
the diameters cut the boundary only at the end points
of the diameter.
\begin{figure}
\includegraphics[angle=-90]{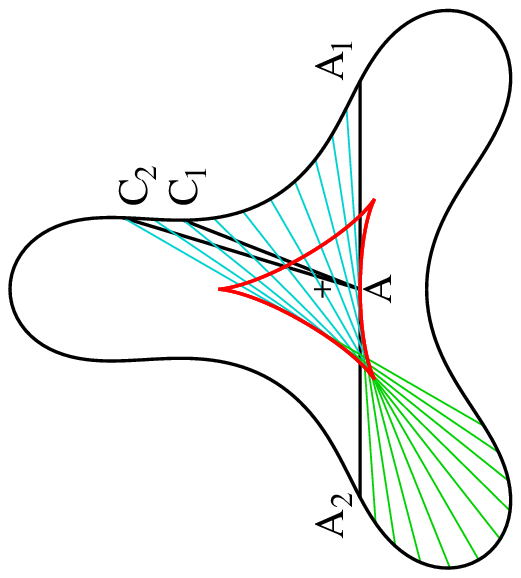}
 \caption{Two halves of the area}
 \label{halfArea}
 The area is cut in halves by the diameter $D(\alpha)=A_1AA_2$.
 The coordinates of $A$ are $\xi(\alpha),\eta(\alpha)$, and those of
 $A_1,A_2,C_1,C_2$ are (x,y) at $\alpha,\alpha+\pi,\gamma,\gamma+\de\gamma$,
resp.
 The infinitesimal triangle of eq. (\ref{inftri}) is $AC_1C_2$.
\end{figure}

We consider now the two regions cut by the diameter $D(\alpha)$.
They consist of infinitesimal triangles $AC_1C_2$, (fig. \ref{halfArea})
\be
A=(\xi(\alpha),\eta(\alpha)),\quad C_1=(x(\gamma),y(\gamma)),
\quad C_2=(x(\gamma+\de\gamma),y(\gamma+\de\gamma)), \label{inftri}
\ee
with $\gamma=\alpha_1=\alpha...\alpha+\pi$ for region 1 and
$\gamma=\alpha_2=\alpha-\pi...\alpha$ for region 2.

Then the perimeter is given by
\be
u_{1,2}=\int_{\alpha_{1,2}}^{\alpha_{1,2}+\pi} \de\gamma
\sqrt{x'(\gamma)^2+y'(\gamma)^2}
=\int_{\alpha_{1,2}}^{\alpha_{1,2}+\pi} \de\gamma
\sqrt{\hrho^2(\gamma)+l^2}.
\ee
Since $\hrho(\gamma-\pi)=-\hrho(\gamma)$, one obtains
the same integral. Hence $u_2=u_1$. The diameter bisects
the perimeter.

The area of the infinitesimal triangle (\ref{inftri})
is given by
\be
\de A = \frac12 \de\gamma
[(x(\gamma)-\xi(\alpha))y'(\gamma)
-(y(\gamma)-\eta(\alpha))x'(\gamma)].
\ee
\newcommand{\xiga}{\xi_{\gamma,\alpha}}
\newcommand{\etaga}{\eta_{\gamma,\alpha}}
With the abbreviations
\be
\xiga:=\xi(\gamma)-\xi(\alpha), \quad
\etaga:=\eta(\gamma)-\eta(\alpha)
\ee
we obtain
\bea
x(\gamma)-\xi(\alpha) &=& l\cos(\gamma)+\xiga, \nn
y(\gamma)-\eta(\alpha) &=& l\sin(\gamma)+\etaga, \nn
x'(\gamma) &=& -l\sin(\gamma)+\hrho(\gamma)\cos(\gamma), \nn
y'(\gamma) &=& l\cos(\gamma)+\hrho(\gamma)\sin(\gamma)
\eea
and
\bea
\de A &=& \frac12 \de\gamma [l^2
+l\xiga\cos(\gamma)
+l\etaga\sin(\gamma) \nn
&+& \xiga\hrho(\gamma)\sin(\gamma)
-\etaga\hrho(\gamma)\cos(\gamma)].
\eea
One finds that both areas $A_1$ and $A_2$ are equal
\bea
A_1 = A_2 &=& \frac{\pi}2 l^2 - \Delta, \nn
\Delta &=& \int_{\alpha}^{\alpha+\pi} \de\gamma
\eta(\gamma) \hrho(\gamma) \cos(\gamma)\nn
&=& -\int_{\alpha}^{\alpha+\pi} \de\gamma
\xi(\gamma) \hrho(\gamma) \sin(\gamma).
\eea
They are independent of $\alpha$, since the integrand
is a periodic function of $\gamma$ with period $\pi$.
The term proportional to $l$ vanishes, since it is a total
derivative of a function, which vanishes at the limits,
\bea
&&\int_{\alpha_{1,2}}^{\alpha_{1,2}+\pi}
\de\gamma[\xiga\cos(\gamma)+\etaga\sin(\gamma)]\nn
&=& \int_{\alpha_{1,2}}^{\alpha_{1,2}+\pi}
\de\gamma \frac{\de}{\de\gamma}
[\xiga\sin(\gamma)-\etaga\cos(\gamma)].
\eea

\subsection{Centers of Gravity}\label{1/2Grav}

Zindler did not consider the centers of gravity of the
two half areas. It is important for the floating body problem
that their distance is constant and that the straight line
between the two centers is normal to the chord.

The centers of gravity $(\bar x(\gamma),\bar y(\gamma))$
of the infinitesimal triangles (\ref{inftri}) are
given by
\bea
\bar x(\gamma) &=& \frac13 \xi(\alpha) +\frac23 (l\cos(\gamma)
+\xi(\gamma))
= \xi(\alpha) +\frac23 (l\cos(\gamma)+\xiga), \nn
\bar y(\gamma) &=& \frac13 \eta(\alpha) +\frac23 (l\sin(\gamma)
+\eta(\gamma))
= \eta(\alpha) +\frac23 (l\sin(\gamma)+\etaga).
\eea
Thus the centers of gravity $(\bar X,\bar Y)$ of the
halves of the area are given by the integrals
\bea
A_{1,2}\bar X_{1,2}(\alpha) &=& \int\de A \bar x = \int\de A \xi(\alpha)
+\frac13 \int\de\gamma [l^3 \cos(\gamma) \nn
&+& l^2\xiga(1+\cos^2(\gamma))
+l^2\etaga\cos(\gamma)\sin(\gamma) \nn
&+& l\xiga^2\cos(\gamma)+l\xiga\etaga\sin(\gamma) \nn
&+& l\xiga\hrho(\gamma)\sin(\gamma)\cos(\gamma)
-l\etaga\hrho(\gamma)\cos^2(\gamma) \nn
&+& \xiga^2\hrho(\gamma)\sin(\gamma)
-\xiga\etaga\hrho(\gamma)\cos(\gamma)], \\
A_{1,2}\bar Y_{1,2}(\alpha) &=& \int\de A \bar y = \int\de A \eta(\alpha)
+\frac 13 \int\de\gamma [l^3 \sin(\gamma) \nn
&+& l^2\xiga\cos(\gamma)\sin(\gamma)
+l^2\etaga(1+\sin^2(\gamma)) \nn
&+& l\xiga\etaga\cos(\gamma)
+l\etaga^2\sin(\gamma) \nn
&+& l\xiga\hrho(\gamma)\sin^2(\gamma)
-l\etaga\hrho(\gamma)\sin(\gamma)\cos(\gamma) \nn
&+& \xiga\etaga\hrho(\gamma)\sin\gamma
-\etaga^2\hrho(\gamma)\cos(\gamma)].
\eea
The result can be written
\bea
A_{1,2}\bar X_{1,2}(\alpha) &=& \pm (l^3 \hat x_3 +l \hat x_1)
+l^2 \hat x_2 + \hat x_0, \nn
A_{1,2}\bar Y_{1,2}(\alpha) &=& \pm (l^3 \hat y_3 +l \hat y_1)
+l^2 \hat y_2 + \hat y_0,
\eea
which yields
\be
\hat x_3 = - \frac23 \sin(\alpha), \quad
\hat y_3 = \frac23 \cos(\alpha).
\ee

The integral for $\hat x_1$ and $\hat y_1$ can be written
\bea
\hat x_1 &=& \frac23 \int\de\gamma \frac{\de}{\de \gamma}
[\xiga(\xiga\sin(\gamma) -\etaga\cos(\gamma))], \nn
\hat y_1 &=& \frac23 \int\de\gamma \frac{\de}{\de \gamma}
[\etaga(\xiga\sin(\gamma) -\etaga\cos(\gamma))].
\eea
They vanish at the limits. Thus $\hat x_1=\hat y_1=0$.
Thus the distance $h$ between both centers of gravity
obeys $A_{1,2}h=\frac43 l^3$ and the line between both centers
is perpendicular to the chord between the two areas.

\subsection{Remarks on other papers} \label{1/2Rem}

At least seven papers\cite{Auerbach,Geppert,Gericke,Hir33,%
Ruban,SalKos,Salkow} appeared from 1933 to 1940, which
discuss (i) characteristic properties of the circle, and (ii)
which (convex) plain regions have the property that
chords between two points of the boundary of constant
length cut the bounded region in two pieces of constant area.
I shortly report on them. The first one by
Hirakawa\cite{Hir33} stated two theorems:\\
Theorem I. A closed convex plane curve with the property that
all chords of fixed length span arcs of equal length, is a circle.\\
Theorem II. A plane oval, in which the areas cut off by chords of
equal length have the same content, is a circle.\\
Apparently it is meant that this should hold for all lengths of
chords. Salkowski emphasizes that considerable weakening of the
conditions yields similar results.

\subsubsection{Salkowski 1934} \label{Salk}

Salkowski\cite{Salkow} started in 1934 with these two theorems
and sharpened them. First he introduces what is now known as
Darboux butterfly:
\begin{figure}
\includegraphics{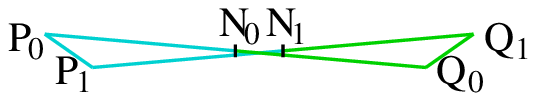}
 \caption{Darboux butterfly}
 \label{wb1} Usually $P_0,P_1,Q_0$ are given and $Q_1$ is constructed.
 Compare this figure with fig. \ref{watbic}.
\end{figure}

Consider a polygonal line $P_0,P_1,P_2,...P_n$ with constant
side length $P_iP_{i+1}=s$. Then one connects $P_0$ with $Q_0=P_n$
by the line $P_0Q_0=2d$. Then a point $P_{n+1}=Q_1$ is determined
so that $Q_0Q_1=s,\quad P_1Q_1=P_0Q_0=2d$. $Q_1$ is the point which
lies on the parallel to $P_1P_n$ through $P_0$.
The arcs $P_0P_1$ and $Q_0Q_1$ in fig. \ref{watbic} are equal.
They are replaced by straight lines of equal length in the Darboux butterfly
fig.\ref{wb1}.
In the limit considered here, where $P_0P_1=s$ tends to zero, the ratio of arc
and distance tends to one.
He argues that then Theorem I is equivalent
to Theorem II and that this remains true when $s$ tends to 0
(and correspondingly $n$ to $\infty$). He restricts the corresponding
curve $c$ to a curve without turning point. He considers the
isosceles trapezoid $P_iP_{i+1}Q_iQ_{i+1}$ with circumcircles
with centers $M_i$. This center is intersection point of the
middle normals on $P_iP_{i+1}$, $Q_iQ_{i+1}$, but also on
$P_iQ_i$, $P_{i+1}Q_{i+1}$. Denote the midpoint of $P_iQ_i$ by
$N_i$ and that of $P_{i+1}Q_{i+1}$ by $N_{i+1}$. In the limit
$s\rightarrow0$ the points $N_i$ yield the curve $(N)$. The
point $M_i$ yield the evolute $(M)$ of $(N)$. (I think, here
is a misprint in the paper: Instead of "die Punkte $N$ liegen
auf einer Evolute der Kurve" it should read "die Punkte $M$
liegen auf einer Evolute der Kurve". A little bit later seems
to be another misprint: Instead of "Mittelpunkt der Sehne $PN$"
one should read "Mittelpunkte $N$ der Sehne $PQ$".)

Salkowski continues: It may happen that the  trapezoid degenerates to a
rectangle. In this case the
curve $(N)$ has a cusp.
The tangents to the oval at the end points $P$ and
$Q$ are parallel and perpendicular to $PQ$. If the arc
$PQ$ is less than half of the circumference of the total
circumference, then there exists an arc $P'Q'$ of the same length with $P'Q'$
parallel $PQ$, but shorter chord
$P'Q'<PQ$. Thus the cusp of $(N)$ is only possible,
if $PQ$ bisects the circumference. Such an example for
$(N)$ is Steiner's hypocycloid with three cusps.

He shows now\\
Theorem III. If a plane regular piece of curve has the property that
three sets of chords of constant length $2d_1,2d_2,2d_3$ cut off constant
lengths of curve and form a triangle, then the curve is a circle.\\
(Gericke\cite{Gericke} gave in 1936 another proof of this theorem).\\
Theorem IV. If all chords over constant arcs of length $2s$ of a curve
have the same length $2d$ and the chords over arcs of length $s$ have
the same length $2d'$, then the curve is a circle.\\
Theorem IV'. If a set of quadrangles $PQRS$ with constant side lengths
$2d_1,2d_2$, $2d_3,2d_4$ can be inscribed in an oval with corresponding
constant arcs of curve, then the oval is a circle. In particular one
finds\\
Theorem V. If all chords of an oval, which cut off one fourth of the
circumference, have the same length, then the curve is a circle.

Finally Salkowski asks the general question:
Are there ovals $c$, in which
an $n$-gon with equal edges $2d$ can be shifted
so that the corner points divide the perimeter
in equal parts? One realizes that the area of the
$n$-gon has to have constant size, further at
least one angle is larger than a right angle.
Consider three consecutive corners $P,Q,R$ of
the figure with an obtuse angle at $Q$.
Shift the chord $PQ$ continuously to $QR$,
then the midpoint $N$ describes a piece of
an oval $(N)$ and its evolute describes
the curve $(M)$ of the midpoints of the
circles, which touch the oval in the end points
of the chords. Denote the midpoint of $PQ$ by
$N_0$, the midpoint of $QR$ by $N_1$.
$M_0$, $M_1$ are the intersections of the
mid-normals on $PQ$ and $QR$, resp. $M$
is the intersection of the two mid-normals.
Then the points $M_0,M,M_1$ constitute a
triangle with obtuse angle at $M$. The curve
$(M)$ touches the mid-normals at $M_0$ and
$M_1$.

So far I agree with the construction.
But now Salkowski continues: Thus the piece of
curve $MM_0$ (to my opinion it should read curve $M_1M_0$,
since $M$ is generally not in the curve)
is longer than the chord $M_0M_1$,
thus larger than the distance $MM_0$. Since $M$
is the midpoint of the circumcircle of the
isosceles triangle $PQR$, thus
$MN_0=MN_1$, such a configuration is not
possible unless $M,M_0,M_1$ coincide. Then
the triangle $PQR$ transforms into its
neighboring position by an infinitesimal rotation,
thus it remains unchanged during the shift along
the curve $c$, hence remaining a circle.

\begin{figure}
\includegraphics{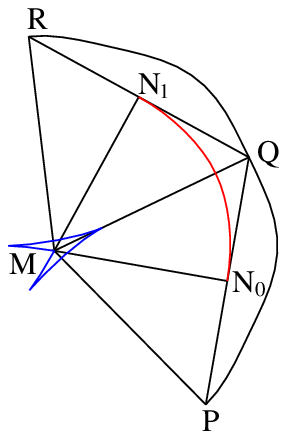}
\includegraphics{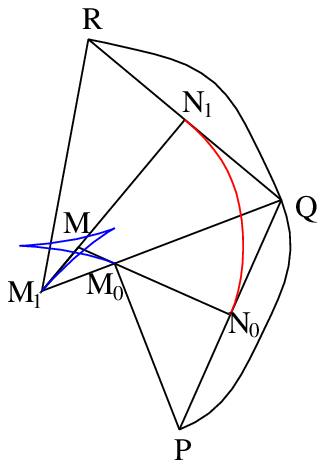}
\includegraphics{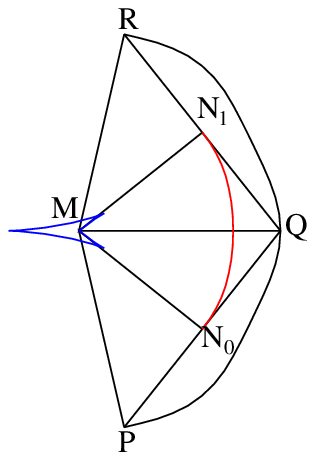}
\caption{The construction by Salkowski}
The curves $(P)$ are in black,
$(N)$ in red, and $(M)$ in blue
for $s_Q=\pi/7$, $\pi/14$, and $0$.
\label{fSal}
\end{figure}

I do not see a reason, why $M,M_0,M_1$ coincide in general.
They will coincide at points where the curvature of $(P)$
at $Q$ has an extreme. Then the neighboring
position need not be obtained by a mere rotation, since
the distance $PR$ is not constant.

Salkowski has based his argument on an obtuse angle at
$\angle PQR$. Such obtuse angles appear in several cases
considered in ref. \cite{We2}. The curves (\ref{rphi}) are 
convex for values $\epsilon$ up to
approximately $1/(2p^2)$. I choose the curve with
$p=7$ fold symmetry and $\delta_0=52.959^0$, that is
$\theta_0=37.041^0$. With $\epsilon=1/98$ the angle
$\angle PQR$ varies
between $102.99^0$ and $108.85^0$ and thus is always obtuse.
I use an approximation in linear order in $\epsilon$
for the curve (P),
\be
r(\phi) = 1+2\epsilon\cos(p\phi), \quad
\phi = s - \frac{\epsilon}p \sin(ps) \label{reps}
\ee
and choose
\be
s_P=s_Q-2\theta_0, \quad s_R=s_Q+2\theta_0.
\ee
The corresponding curves are shown in figure \ref{fSal} for
$s_Q=\pi/7$, $\pi/14$, and $0$. The curves $(P)$ are in black,
$(N)$ in red, and $(M)$ in blue.

For $s_Q=0$ and $s_Q=\pi/7$ the three points $M$, $M_0$, and $M_1$ coincide. But
in between, in particular for $s_Q=1/14$ the three points $M_0$, $M_1$, and $M$
differ. Note that curve
$(M)$ has cusps.
Thus the proof of his last theorem fails. This does not mean that
his theorem is disproved. Since in terms of eqs. (\ref{reps})
the range for $\epsilon$ for convex boundaries becomes smaller
with increasing $p$, it may be that there are not such ovals.

\subsubsection{Auerbach 1938} \label{1/2RemAuer}

Zindler had derived curves whose chords, which bisect the perimeter,
have constant length and bisect the area.
Auerbach\cite{Auerbach} 1938 rederived the solution by Zindler, but he
showed that these curves had also the property, that any chord acting
as water-line yields the same potential energy, and thus all orientations
are in equilibrium. The first five sections are for general densities
$\rho$. He obtains for the curvatures at $P$ and $Q$
\be
\kappa_Q - \kappa_P = 2\frac{\de\theta}{\de s}, \quad
\kappa_Q + \kappa_P = \frac 4d \sin\theta, \label{kapkap}
\ee
where $d/2$ is our $l$ and $\theta$ the angle between chord and tangent.
From section 6 on he considers the case $\rho=1/2$.
Auerbach derives the expressions for the
coordinates $x,y$ of the boundary, eqs. (\ref{Zin1}, \ref{Zin2}).
One has to replace $(d/2)\rm{ctg}(\theta)\rightarrow\hrho$ and
$d/2\rightarrow l$.

\subsubsection{Ruban, Zalgaller, Kostelianetz 1939}
\label{1/2RemRubZalKos}

Eugene Gutkin gives
a few remarks of personal and socio-historical character
at the end of his paper\cite{Gutkin13}.
He reports the cruel death of Herman Auerbach under the Nazi
regime.
He also reports that the archimedean floating problem was
popular among older mathematics students around 1939 in
Leningrad. The results of three of them, Ruban, Zalgaller and Kostelianetz were
published in the Proceedings (Doklady)
of the Soviet Academy of Sciences in a Russian and a shorter
French version.\cite{Ruban,SalKos} Only one of the authors,
Zalgaller (often written Salgaller), could continue his career after the war.
Kostelianetz did not return from the war.
Ruban returned as invalid from the war, no longer able to do
mathematics.

Although Ruban\cite{Ruban} and Salgaller and Kostelianetz\cite{SalKos}
published side by side, they obtained contradictory
results. Salgaller and Kostelianetz obtained solutions
for $\rho=1/2$ in agreement with Zindler, but Ruban
claimed that there are none besides the circle.
Ruban obtained the second equation (\ref{kapkap})
and correctly found that $sin\theta$ of the angle $\theta$
between chord and tangent are equal at both ends of the
tangent. Erroneously he concluded that both angles
are equal, but they add up to $\pi$.

In sect. 5 of his paper\cite{Ruban} Ruban introduces
the angle $\theta_0$ between chord and tangent
and curvature $\kappa_0$ (Ruban uses $k$ instead of $\kappa$)
of a circle of length $S$ and claims without proof
or explanation: If
\be
\theta_0 {\rm ctg} \theta_0 \not=
\frac{\pi nl}S {\rm ctg} \frac{\pi nl}S \label{Rubt}
\ee
for $n=1,2,...$, then there exists a number
$\epsilon>0$ so that the inequality $|\kappa(s)-\kappa_0|<\epsilon$
yields the equality $\kappa(s)=\kappa_0$, where $l$ is the length
of the arc.
Since $\theta_0=\pi l/S$, the inequality may be rewritten
$C_n\not=0$ with
\be
C_n = \cos(\theta_0)\sin(n\theta_0)
-n\cos(n\theta_0)\sin\theta_0. \label{Cn}
\ee
It is likely that Ruban, who has derived the relation
\be
b [\kappa(s+l)+\kappa(s)] = 4\sin\theta(s) \label{Rub5}
\ee
in eq. (5) ($s$ is the arc parameter) and used a Fourier expansion for 
$\kappa(s)$ considered
\be
\kappa(s) = \kappa_0 +a_n \cos \frac{2\pi ns}S
+b_n\sin \frac{2\pi ns}S.
\ee
Together with $2\theta(s) = \int_s^{s+l} \de t \kappa(t)$
and restricting to $a_n$ and $b_n$ in linear order
yields condition $C_n=0$ for nontrivial solutions
$a_n$, $b_n$. This is the starting point for non-circular
perturbative solutions, as given in \cite{We2}, where
$\theta_0=\pi/2-\delta_0$. Thus Ruban was close to a
solution, if he would have performed a perturbation expansion.
Obviously $C_n=0$ is always fulfilled for $n=1$. It
corresponds to a translation of the curve, compare sect. 4.2
of \cite{We2}. Thus Ruban's statement should not include
the case $n=1$.

In 1940 Geppert\cite{Geppert} gave the solutions for
$\rho=1/2$, but erroneously argued that there are no
solutions for $\rho\not=1/2$. He simply overlooked that
in this later case the points on the boundary are
end-points of two different chords, not one.

A general obstacle to find solutions for $\rho\not=1/2$
was that one expected that the circumference should be
divided in an integer or at least rational number $n$ of equal parts. This was
very
good for $n=2$, $\rho=1/2$, but it is not at all necessary
for $\rho\not=1/2$.

\section{Algebraic Curves by Greenhill} \label{Green}

In his 1899 paper\cite{Greenhill99} Greenhill gives special 
solutions expressed by pseudo-elliptic functions,
in which the cosine and sine of the angle $n\theta$ 
and $n\theta/2$, resp. are 
algebraic functions of the radius $r$.
Thus the curves are algebraic.

I do not attempt to go through the theory of the
pseudo-elliptic functions, but refer only to the main results.

Starting point is the expression for the polar angle $\theta$
as function of the radius $r$,
\be
\theta = \frac12 \int\de r^2 \frac{Ar^4+Br^2+C}{r^2\sqrt R},
\quad R=r^2-(Ar^4+Br^2+C)^2.
\ee
The integral is divided into two contributions,
\be
\theta = \theta' + (B- B') u,
\ee
with
\be
\theta' =\frac12 \int\de r^2 \frac{Ar^4+ B'r^2+C}{r^2\sqrt R},
\quad u = \frac12 \int\frac{\de r^2}{\sqrt R},
\ee
where $u$ is the arc length.

The shape of the curves depends on two independent
parameters. These may be the dimensionless $AC^3$ and $BC$,
or $\epsilon$ and $\mu$ in \cite{We4}, or $x$ and $y$ 
or $\beta$ and $\gamma$ by Greenhill.\cite{Greenhill99}

For a given periodicity in $\theta'$,
that is by an increase of $\theta'$ by $2\pi/n$ (class I) or
by $4\pi/n$, $n$ odd (class II), a certain relation between
these parameters is fixed. 
Since at Greenhill's time the integral for the arc length $u$,
expressed as elliptic function of the first kind, was
tabulated, one could easily calculate $\theta$ for these
cases.

If moreover one requires
$B=B'$, in order to obtain an algebraic curve, one has a
second condition, which allows only for single solutions.

\subsection{Class I}

For {\bf class I} one finds solutions of the form
\bea
\sin(n\theta') &=& \frac{H(q)\sqrt{P_1}}{Qq^{n/2}}, \label{eqH1} \\
\cos(n\theta') &=& \frac{L(q)\sqrt{P_2}}{Qq^{n/2}}, \label{eqL1} \\
H(q) &=& q^{n-1} + h_1 q^{n-2} + ... + h_{n-1}, \\
L(q) &=& q^{n-1} + l_1 q^{n-2} + ... + l_{n-1}, \\
P_1 &=& -q^2 +2(2\gamma+1) q -1, \\
P_2 &=& q^2 +2(2\beta-2\gamma-1) q +(2\beta-1)^2, \\
P &=& P_1P_2 = -(q^2+2(\beta-2\gamma-1) q -2\beta +1)^2 +16\beta^2\gamma q.
\eea
For $\gamma>0$ one uses $q=r^2$ and for $\gamma<-1$ one takes $q=-r^2$.
$P$ is related to $R$ by
\be
P = 16\beta^2|\gamma| R = 16\beta^2\gamma r^2
-\left[4\beta\sqrt{|\gamma|} (Aq^2 \pm Bq +C)\right]^2.
\ee
Hence $A,B,C$ are related to $\beta$ and $\gamma$ by
\be
4\beta\sqrt{|\gamma|} = A, \quad
\pm 8\beta\sqrt{|\gamma|} (\beta-2\gamma-1) = B, \quad
4\beta\sqrt{|\gamma|} (1-2\beta) = C.
\ee

The zeroes $q_{1,2}$ of $P_1$ are obtained as
\be
q_{1,2} = 2\gamma+1 \pm2\sqrt{\gamma(\gamma+1)}.
\ee
This yields the extreme values $r_{1,2}$ of $r$,
\be
r_{1,2} = \left\{\begin{array}{cc}
\sqrt{\gamma+1} \pm \sqrt{\gamma} & \gamma>0, \\
\sqrt{-\gamma} \pm \sqrt{-\gamma-1} & \gamma<-1.
\end{array}\right.
\ee
The scale of $q$ is chosen so that $q_1q_2=1$.
The other extreme values $q_{3,4}$ are the zeroes of $P_2$,
\be
q_{3,4} = -2\beta+2\gamma+1
\pm 2\sqrt{\gamma(2\beta-\gamma-1)}.
\ee
One may exchange $P_1$ and $P_2$ by simultaneously rescaling
$q$. This yields
\bea
P_1(\beta,\gamma,q) &=& -(1-2\beta)^2 P_2(\beta',\gamma',q'), \\
P_2(\beta,\gamma,q) &=& -(1-2\beta)^2 P_1(\beta',\gamma',q'), \\
P(\beta,\gamma,q) &=& (1-2\beta)^4 P(\beta',\gamma',q'), \\
H(\beta,\gamma,q) &=& -(1-2\beta)^{n-1} L(\beta',\gamma',q'), \\
L(\beta,\gamma,q) &=& -(1-2\beta)^{n-1} H(\beta',\gamma',q')
\eea
with
\be
\beta' = \frac{-\beta}{1-2\beta}, \quad
\gamma'=\frac{\gamma}{1-2\beta}, \quad
q'=\frac q{1-2\beta}.
\ee

The derivative of $\theta'$ calculated from eqs.
(\ref{eqH1}, \ref{eqL1}) yields
\bea
n\frac{\de\theta'}{\de q} &=&
\frac{2q\frac{\de H}{\de q}P_1+qH\frac{\de P_1}{\de q}-nHP_1}
{2qL\sqrt{P_1P_2}} \\
&=& -\frac{2q\frac{\de L}{\de q}P_2+qL\frac{\de P_2}{\de q}-nLP_2}
{2qH\sqrt{P_1P_2}}.
\eea
Thus one requires
\bea
2q\frac{\de H}{\de q}P_1+qH\frac{\de P_1}{\de q}-nHP_1
= nLP', \label{G1} \\
2q\frac{\de L}{\de q}P_2+qL\frac{\de P_2}{\de q}-nLP_2
= -nHP', \label{G2} \\
P' := q^2 +2(\beta-2\gamma' -1) q -2\beta+1, \label{G3}
\eea
so that
\bea
\frac{\de\theta'}{\de q} &=& \frac{P'}{2q \sqrt{P}}
=\frac{R'}{2q \sqrt{R}}, \\
P'=4\beta\sqrt{|\gamma|} R' &=& Ar^4+B'r^2+C, \\
B-B' &=& \pm 16\beta\sqrt{|\gamma|} (\gamma'-\gamma).
\eea
If one multiplies eq. (\ref{eqH1}) by $H$ and eq. (\ref{eqL1}) by
$L$, and adds both equations, then one obtains
\be
q^{n+1} \frac{\de}{\de q} \frac{H^2P_1+L^2P_2}{q^n} = 0,
\ee
which yields after integration
\be
\frac{H^2P_1+L^2P_2}{q^n} = {\rm const.}
\ee
Denoting this constant by $Q^2$, one obtains
\be
\cos^2(n\theta') + \sin^2(n\theta') = 1,
\ee
as required.

From eqs. (\ref{eqH1},\ref{eqL1}) one obtains
\be
\cos(2n\theta') = \frac{L^2(q)P_2-H^2(q)P_1}{Q^2q^n}.
\ee
Since
\be
q^n\cos(2n\theta') = \Re((x+\ie y)^{2n})
\ee
and $q=x^2+y^2$, these curves are algebraic curves.

To obtain solutions for eqs. (\ref{eqH1}, \ref{eqL1})
one has to solve eqs. (\ref{G1}-\ref{G3}).
The coefficients $\beta$ and $\gamma$ are solutions of
algebraic equations with integer coefficients.

\begin{table}
\begin{tabular}{|rrrcc|} \hline
 $n$ & $\beta$ \quad & $\gamma$ \quad &
 \S\ and fig. in \cite{Greenhill99} & fig. this paper \\ \hline
 2 & -1.36602540 & -1.57735027 & 23 8 & \ref{f21} \\
 3 & -0.37948166 & -1.14356483 & 24 9 & \ref{f31} \\
 4 & -0.19053133 & -1.06944356 & 26 10 & \ref{f41m} \\
 4 & 4.19179270 & 1.34344652 & - -& \ref{f41p} \\
 5 & -0.11633101 & -1.04168414 & 27 11 & \ref{f51a} \\
 5 & -4.26375725 & -2.97763686 & - - & \ref{f51b} \\
 6 & -0.07884381 & -1.02799337 & - - & \ref{f61m} \\
 6 & 2.21204454 & 0.41789155 & - - & \ref{f61p} \\ \hline
\end{tabular}
\caption{Constants of curves class I}\label{T1}
\end{table}

\begin{figure}
\parbox{5cm}
{\includegraphics{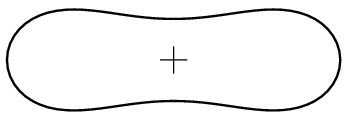}
%{\epsfig{file=green21.eps}
\caption{Curve with $n=2$}
\label{f21}}
\parbox{5cm}
{\includegraphics[angle=90]{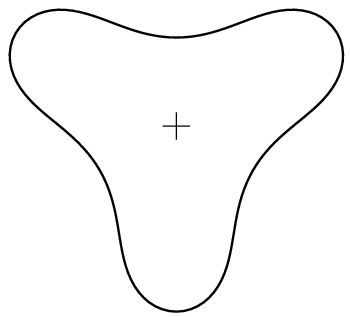}
%{\epsfig{file=green31.eps,angle=90}
\caption{Curve with $n=3$}
\label{f31}}
\end{figure}

\begin{figure}
\parbox{5cm}
{\includegraphics{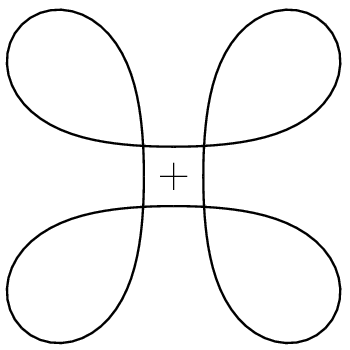}
%{\epsfig{file=green41pr.eps}
\caption{Curve with $n=4$}
\label{f41p}}
\parbox{5cm}
{\includegraphics[angle=90]{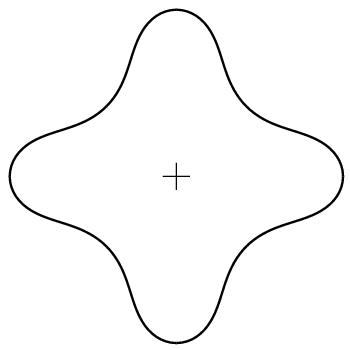}
%{\epsfig{file=green41mr.eps,angle=90}
\caption{Curve with $n=4$}
\label{f41m}}
\end{figure}

\begin{figure}
\parbox{5cm}
{\includegraphics{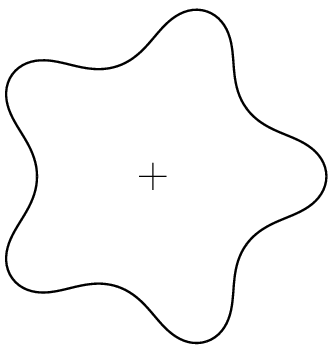}
%{\epsfig{file=green51a.eps}
\caption{Curve with $n=5$}
\label{f51a}}
\parbox{5cm}
{\includegraphics{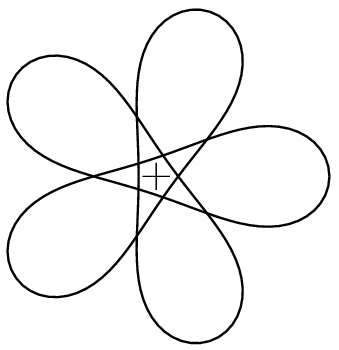}
%{\epsfig{file=green51b.eps}
\caption{Curve with $n=5$}
\label{f51b}}
\end{figure}

\begin{figure}
\parbox{5cm}
{\includegraphics{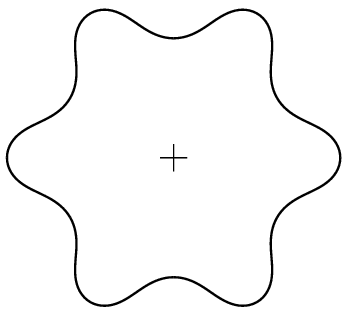}
%{\epsfig{file=green61m.eps}
\caption{Curve with $n=6$}
\label{f61m}}
\parbox{5cm}
{\includegraphics{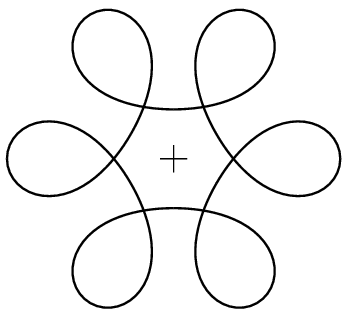}
%{\epsfig{file=green61p.eps}
\caption{Curve with $n=6$}
\label{f61p}}
\end{figure}

Several curves of class I are shown in figures \ref{f21} -- \ref{f61p}.
They are listed in table \ref{T1}.

\subsection{Class II}

For {\bf class II} the angle $\theta'$ can be written
for special values $A,B,C$ and odd $n$
\bea
\sin(\frac n2 \theta') &=& \frac{H_+ \sqrt{R_+}}{Qr^{n/2}},
\label{eqHp} \\
\cos(\frac n2 \theta') &=& \frac{H_- \sqrt{R_-}}{Qr^{n/2}},
\label{eqHm} \\
H_{\pm} &=& r^{n-2} \pm h_1 r^{n-3} + h_2 r^{n-4} \pm ...
  \pm h_{n-2}, \\
R_{\pm} &=& r \pm (Ar^4+Br^2+C), \\
R &=& R_+R_-.
\eea

Differentiating (\ref{eqHp}, \ref{eqHm}) one obtains
\be
\frac{n\de\theta'}{2\de r}
= \pm \frac{2r\frac{\de H_{\pm}}{\de r}R_{\pm}
+ H_{\pm}r\frac{\de R_{\pm}}{\de r}
-nH_{\pm}R_{\pm}}{2rH_{\mp}\sqrt{R_-R_+}}
\ee
This expression should yield
\be
\frac{\de\theta'}{\de r} = \frac{A'r^4+B'r^2+C'}{r\sqrt{R}}
\ee
This requires
\be
2r\frac{\de H_{\pm}}{\de r}R_{\pm}
+ H_{\pm}r\frac{\de R_{\pm}}{\de r}
-nH_{\pm}R_{\pm} = \pm nH_{\mp}(A'r^4+B'r^2+C'). \label{eqpm}
\ee
If we multiply the equation with the upper signs by $H_+$
and that with the lower signs by $H_-$, and add both then
we obtain
\be
r^{n+1}\frac{\de}{\de r} \frac{H_+^2R_+ + H_-^2R_-}{r^n} = 0,
\ee
which after integration yields
\be
\frac{H_+^2R_+ + H_-^2R_-}{r^n} = {\rm const.} \label{onepm}
\ee
which we set to $Q^2$. If we set
\be
H_1 = r^{n-2} + h_2 r^{n-4} + ..., \quad
H_2 = h_1r^{n-3} + h_3 r^{n-5} + ..., \quad
P = Ar^4+Br^2+C,
\ee
so that
\be
H_{\pm} = H_1 \pm H_2, \quad R_{\pm} = r \pm P,
\ee
then
\be
H_+^2R_+ + H_-^2R_- = 2(H_1^2+H_2^2)r + 4H_1H_2 P,
\ee
which is a polynomial of order $2n-1$ containing only terms
with odd powers of $r$. Thus eq. (\ref{onepm}) can only be
fulfilled for odd $n$.

From eqs. (\ref{eqHp}, \ref{eqHm}) one obtains
\be
\cos(n\theta') = \frac{H_-^2R_--H_+^2R_+}{Q^2r^n}.
\ee
Since
\be
r^n\cos(n\theta') = \Re((x+\ie y)^n)
\ee
and
\be
H_-^2R_--H_+^2R_+ = -2(H_1^2+H_2^2)P -4H_1H_2r
\ee
is a polynomial even in $r$, these curves are algebraic, too.

Eq. (\ref{eqpm}) is an equation for a polynomial of order $n+2$
in $r$. Equating the coefficients of the powers $n+2$ and $n+1$
yields $nA=nA'$ and $(n-2)Ah_1=-nA'h_1$. $A=0$ would yield
a constant curvature of the curve, thus only a circle or straight
line, we require $A\not=0$. Hence $A'=A$, $h_1=0$. The
coefficients of the zeroth power in $r$ yield
$-nCh_{n-2}=-nC'h_{n-2}$. Thus if $h_{n-2}\not=0$, then $C'=C$.

The simplest case is given by $n=3$.
For this case one obtains $A'=A$, $B'=B/3$, $C'=-C/3$.
Here one does not obtain necessarily $C'=C$, since
$h_1=0$. Requiring $B'=B$, $C'=C$ yields
\bea
\sin(3\theta/2) = \frac{r\sqrt{r+Ar^4}}{\sqrt{2r^3}}
=\sqrt{\frac{1+Ar^3}2}, \\
\cos(3\theta/2) = \frac{r\sqrt{r-Ar^4}}{\sqrt{2r^3}}
=\sqrt{\frac{1-Ar^3}2}.
\eea
Hence we obtain
\be
\cos(3\theta) = -Ar^3
\ee
and with $A=-1/a^3$
\be
r^3 = a^3 \cos(3\theta).
\ee
This curve is shown in figure \ref{fgp2p3c}.

For $n=5$ we find
\be
B= \frac 1{4C} - 4AC^2, \quad q = \sqrt 8 C
\ee
and
\be
B' = -\frac 1{20C} - \frac{12 AC^2}5.
\ee
Thus $B=B'$ is obtained for
\be
AC^3 = \frac 3{16}, \quad BC=-\frac 12.
\ee

A remark on $n=5$ of Greenhill\cite{Greenhill99}.
\S 14: The ratio of minimal radius and maximal radius is
$(\sqrt[3]{10}-1)/3=0.38481156$ as given in the paper.
\S 17 contains numerical errors: $c=(\sqrt[3]{10}-1)^2/3=0.44423982$,
and the ratio of maximal radius and minmal radius is
$(1+c)/(1-c)=2.59867451$, the inverse of the ratio of minimal
and maximal radius in \S 14.
The curves of \S 14 and \S 17 are not only of the same character,
but they are the identical.

\begin{figure}[h!]
\parbox{5cm}
{\includegraphics{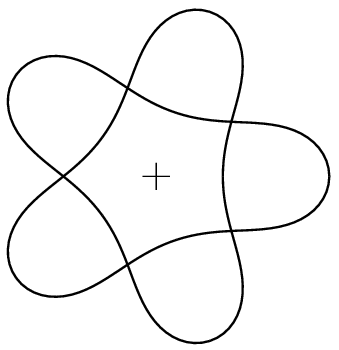}
%{\epsfig{file=green52.eps}
\caption{Curve with n=5}
\label{f52}}\hspace{2mm}
\parbox{5cm}
{\includegraphics{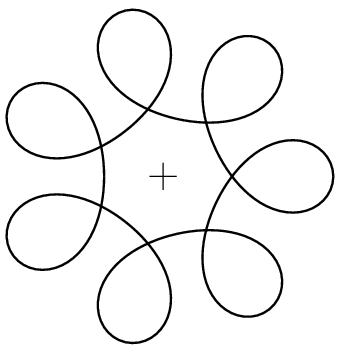}
%{\epsfig{file=green72m.eps}
\caption{Curve with n=7}
\label{f72m}}\vspace{2mm}

\parbox{5cm}
{\includegraphics{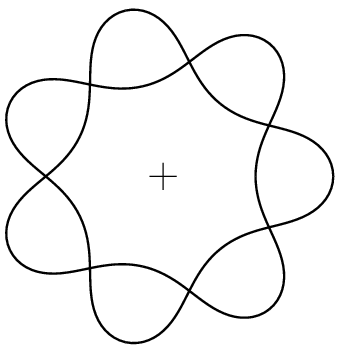}
%{\epsfig{file=green72p.eps}
\caption{Curve with n=7}
\label{f72p}}\hspace{2mm}
\parbox{5cm}
{\includegraphics{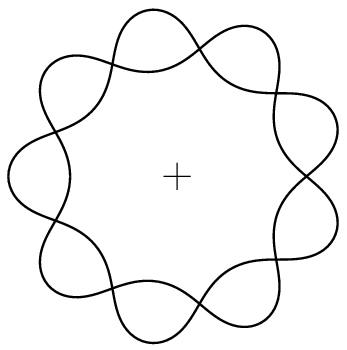}
%{\epsfig{file=green92.eps}
\caption{Curve with n=9}
\label{f92}}
\end{figure}

\begin{table}[h!]
\begin{tabular}{|rcc|} \hline
$n$ & \S and fig. in ref. \cite{Greenhill99} & fig. this paper \\ \hline
3 & 13 - & \ref{fgp2p3c} \\
5 & 14 3 & \ref{f52} \\
7 & 15 4 & \ref{f72m} \\
7 & 15 4 & \ref{f72p} \\
9 & 16 5 & \ref{f92} \\ \hline
\end{tabular}
\caption{Constants of curves class II}\label{T2}
\end{table}

If we follow the curve with a continuous change of the
direction of its tangent, then we have to circle twice around
the origin in the figures \ref{f52}, \ref{f72p}, \ref{f92},
until we return to the start of the curve, whereas in figure
\ref{f72m} we have to do this only once. The curves are listed in table
\ref{T2}.

\section{Bicycle Curves} \label{bicycle}

Most of the elastica with and without pressure and the Zindler curves
are bicycle curves. Finn\cite{Finn}
pointed out, that the class of bicycle curves is probably much
larger. We will first comment on his ideas, but also on an important
observation by Varkonyi\cite{Varkonyi09} before we give some
generalizations due to Mampel\cite{Mampel67} and some closed
curves with winding numbers different from one.

\subsection{Circle of centroids}

Tabachnikov\cite{Tabach06} and also Salkowski\cite{Salkow}
argue that one can give a segment of the front tire track.
One draws the tangent at one point of the track of the back wheel
in both directions and gives a smooth but arbitrary
track of the front wheel between these two points. Then one 
can continue the tracks in both directions as for example described
by Salkowski. Finn\cite{Finn} instead starts with a segment
of the back tire track, which is so long that the corresponding
pieces of the front tire tracks meet themselves.
In both cases certain conditions at the ends of the segments have to be met.

\begin{figure}[h!]
\parbox[b]{10cm}
{\includegraphics{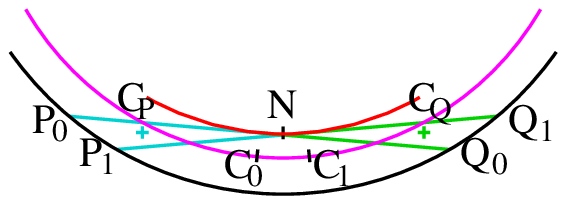}
%{\epsfig{file=watbic2.eps}
\caption{Bicycle tire tracks: Front track in black, rear track in red.
Centroids of the areas enclosed by the cyan-green chords and the black
track lie on the magenta circle}
\label{fwt2}}
\end{figure}

In section 2 and also section 6 of ref. \cite{Varkonyi09} Varkonyi
considered closed bicycle tracks. In fig. \ref{fwt2} two nearby locations
of the bicycle moving to the left is shown in cyan and 
those of the bicycle moving to the right in green.
They represent chords of length $2l$ of the black front track curve.
These chords and the black line enclose areas ${\cal A}_0$ and ${\cal A}_1$
of constant size $A_1$.
Their centroids are the points $C_0$ and $C_1$. Going from $P_0Q_0$
to $P_1Q_1$ the area $P_0P_1N$ is cut away and $Q_0Q_1N$ is added.
These areas are $l^2\alpha/2$ with (infinitesimal) angle $\alpha$
between the two chords. Their centroids $C_P$ and $C_Q$ are $2l/3$
apart from the point of intersection $N$ of the two chords. Thus
the centroid moves from $C_0$ to $C_1$ by the distance $r\alpha$
with $r_1=2l^3/(3A_1)$ parallel to the chords. Since $l$ and $A_1$ are
constant, the centroids lie on a circle of radius $r_1$.

Varkonyi now argues:
If the bicycle curve is closed, then the same arguments apply for 
the complementary area of size $A_2=A-A_1$, where $A$ is the area
enclosed by the black front track. The centroids of the complementary
area lie on a circle of radius $r_2=2l^3/(3A_2)$. 
In general it is not clear, whether these two circles are concentric.
Thus it is not clear whether closed bicycle curves correspond
to the boundaries of homogeneous floating bodies of equilibrium.
The centroid of the whole area lies on the connecting line between
the centers of the two circles. If one allows for an inhomogeneus
body, then the volume centroid and the mass centroid can differ.
Then one obtains floating bodies of equilibrium, if the mass centroid
lies in one of the circle centers.
For density $\rho=1/2$ both circles are identical. The known
boundaries for $\rho\not=1/2$. have dieder symmetry. Thus also
the centers of the circles are concentric. Whether there are other
solutions for closed bicycle tracks is not known.

\begin{figure}[h!]
\parbox[b]{5cm}
{\includegraphics[scale=0.5]{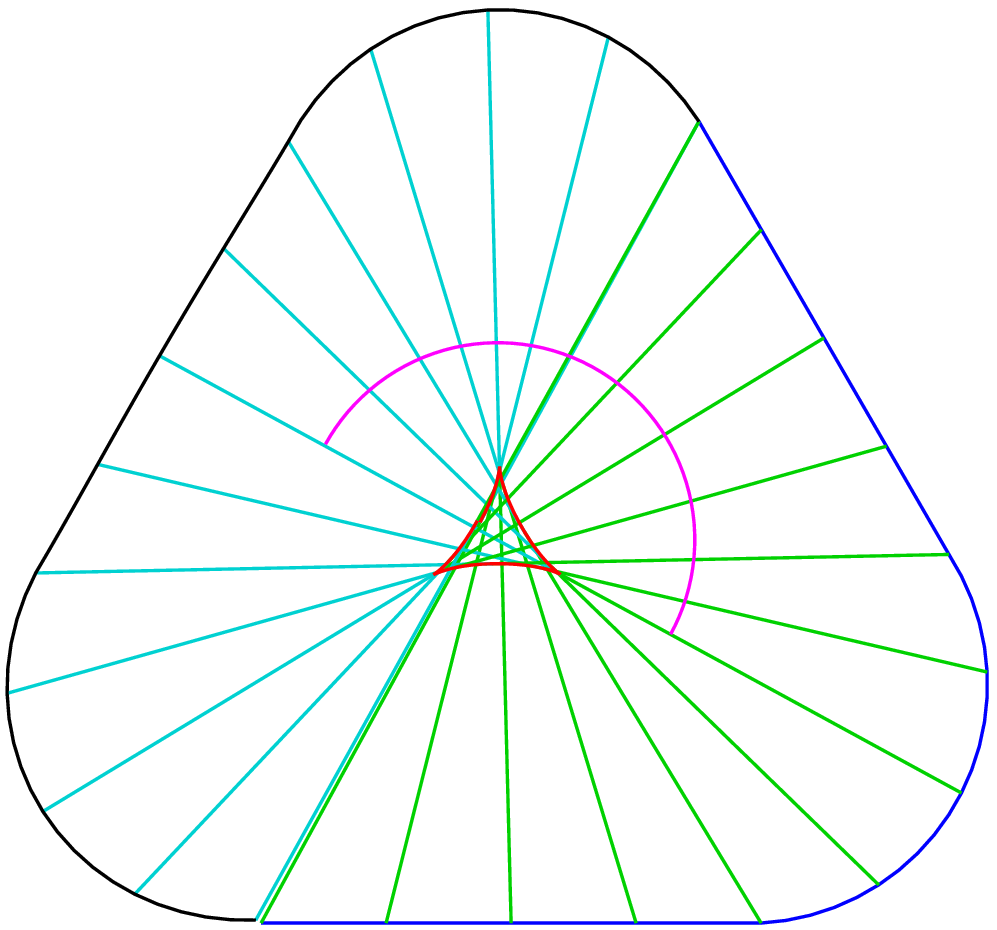}
%{\epsfig{file=fbogen3.eps,scale=0.5}
\caption{Nearly a triangle with rounded corners}
\label{fbg3}}\hspace{10mm}
\parbox[b]{6cm}
{\includegraphics[scale=0.5]{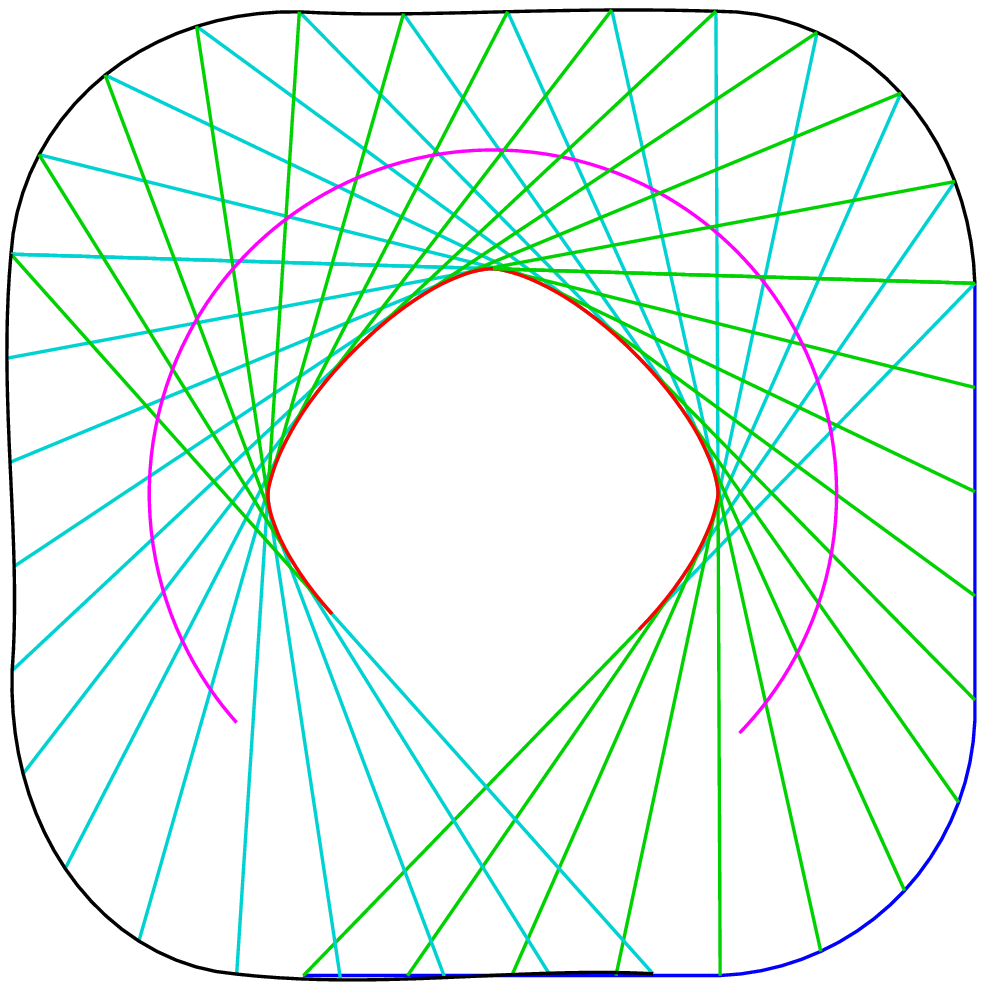}
%{\epsfig{file=fbogen4.eps,scale=0.5}
\caption{Nearly a square with rounded corners}
\label{fbg4}}
\end{figure}

\begin{figure}[h!]
\parbox[b]{5cm}
{\includegraphics[scale=0.4]{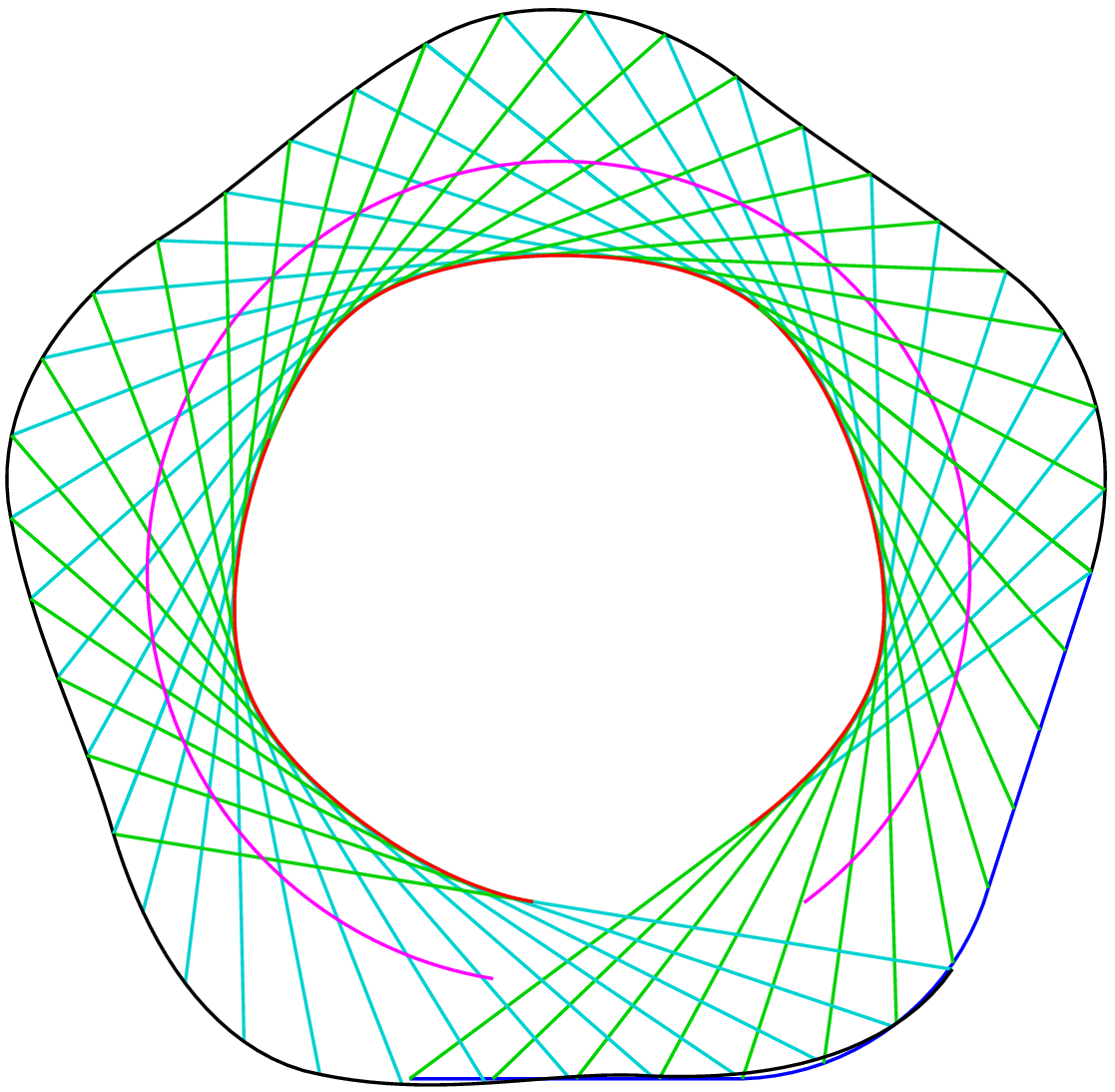}
%{\epsfig{file=fbogen5.eps,scale=0.4}
\caption{Nearly a pentagon with rounded corners}
\label{fbg5}}\hspace{10mm}
\parbox[b]{5cm}
{\includegraphics[scale=0.4]{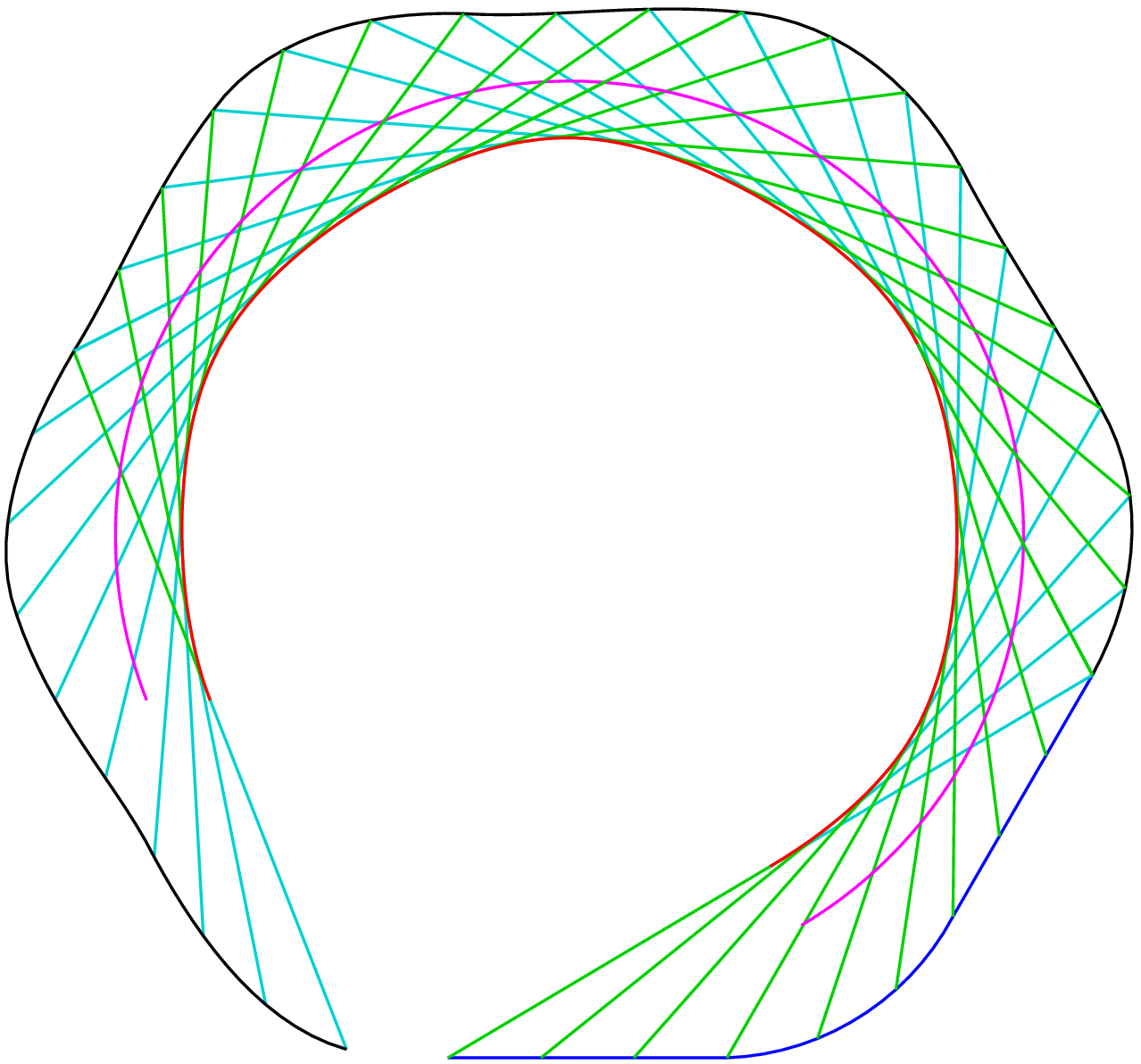}
%{\epsfig{file=fbogen6.eps,scale=0.4}
\caption{Nearly a hexagon with rounded corners}
\label{fbg6}}
\end{figure}

We give several examples of bicycle curves in the following.
The segment of the front tire track from which we start is shown in blue.
The following tire track is determined by adding Darboux butterflies and
shown in black.
(For Darboux butterflies see subsection \ref{Salk} and figure \ref{wb1}).
Chords are drawn after every fifth application of the
butterfly. The centroids of the enclosed area are connected and
shown in magenta. The appearence resembles part of a circle.
In figures \ref{fbg3} to \ref{fbg6} the initial segment consists of three
pieces, first a straight line of length $s/3$, then a circular arc
of the same length $s/3$ and then again a straight line of length
$s/3$. The circular arc changes the direction by an angle $2\pi/n$
with $n=3,4,5,6$ for figs. \ref{fbg3}, \ref{fbg4}, \ref{fbg5},
and \ref{fbg6}, respectively. The front tracks close nearly and 
show approximately polygons with $n$ rounded corners.
The butterfly procedure generates nearly circular arcs from
the (nearly) straight lines and nearly straight lines from the arcs.
These curves are very similar to the figures \ref{fhau3} and \ref{fhrc5},
the first and third of fig. \ref{fFB2}, the first of fig. \ref{fFB3},
and fig. \ref{fz13}.

\begin{figure}[h!]
\parbox[b]{5.5cm}
{\includegraphics[scale=0.5]{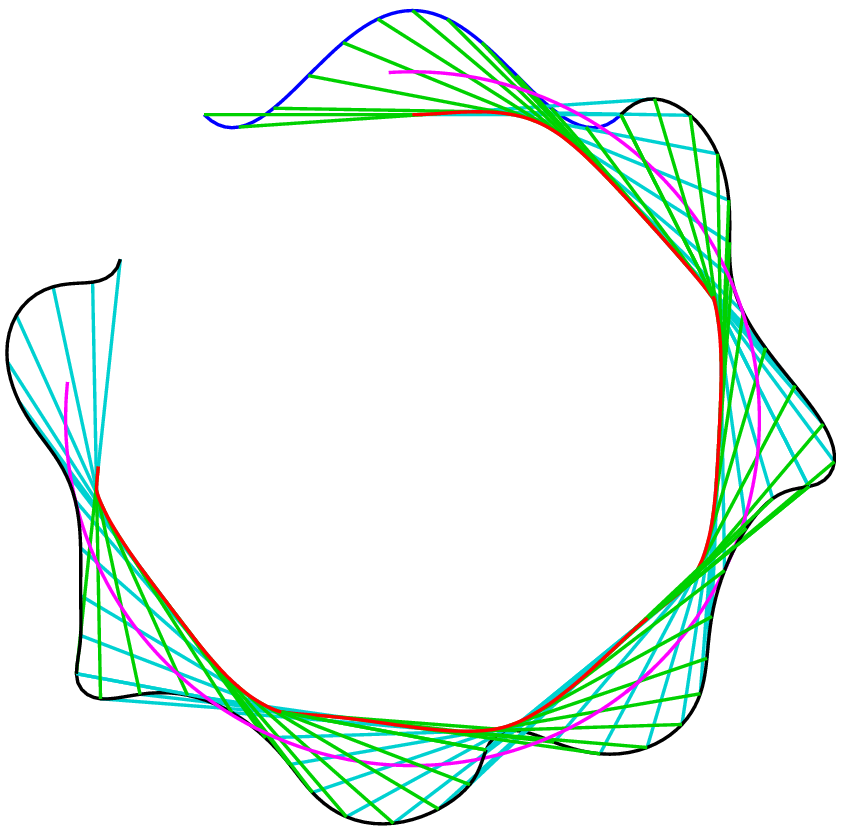}
%{\epsfig{file=fkurve2.eps,scale=0.5}
\caption{Based on eq. (\ref{eqfku}) with $c=2$}
\label{fku2}}\hspace{5mm}
\parbox[b]{5.5cm}
{\includegraphics[scale=0.25]{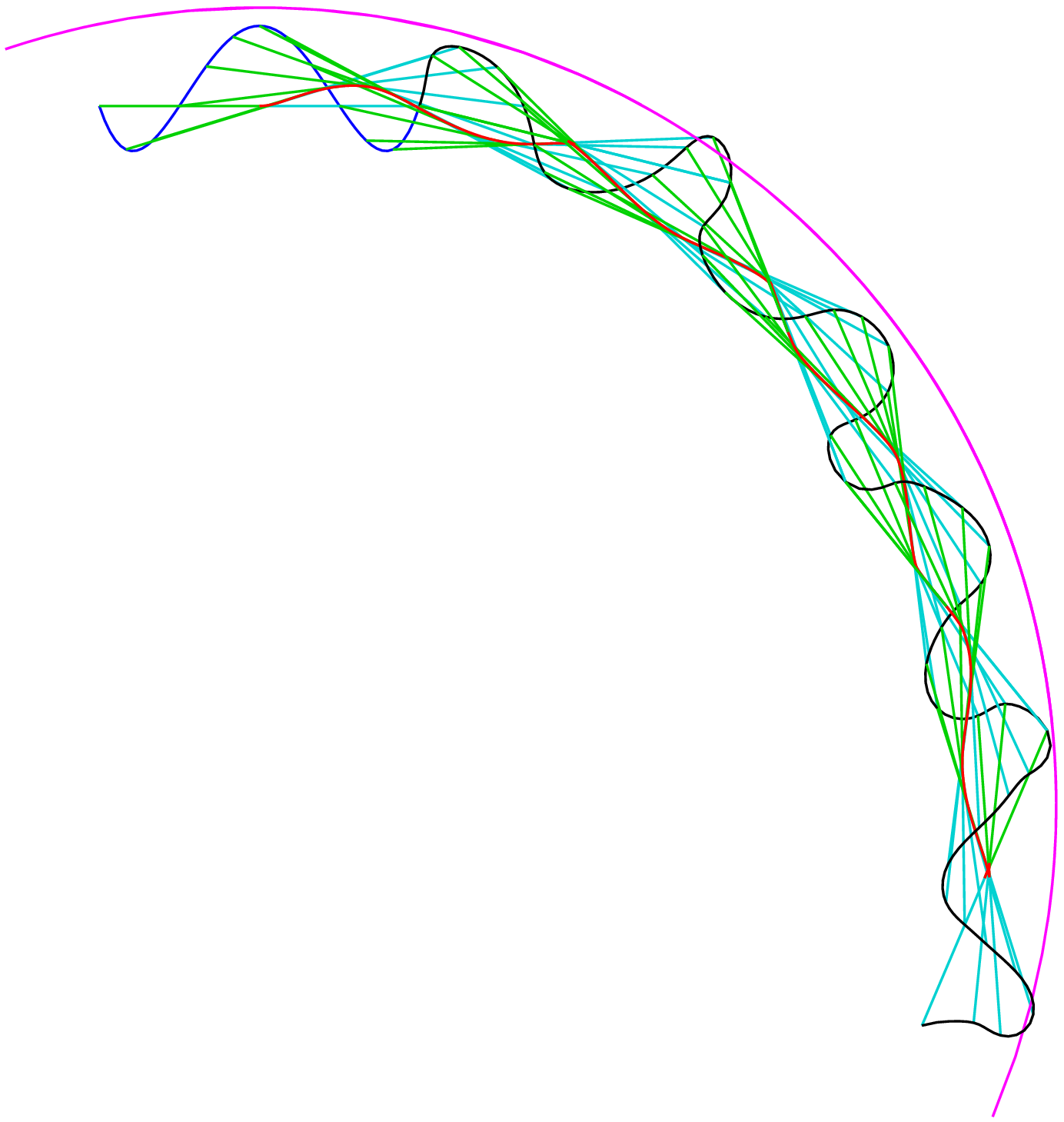}
%{\epsfig{file=fkurve4.eps,scale=0.25}
\caption{Based on eq. (\ref{eqfku}) with $c=4$}
\label{fku4}}
\end{figure}

\begin{figure}[h!]
\parbox[b]{5.5cm}
{\includegraphics[scale=0.2]{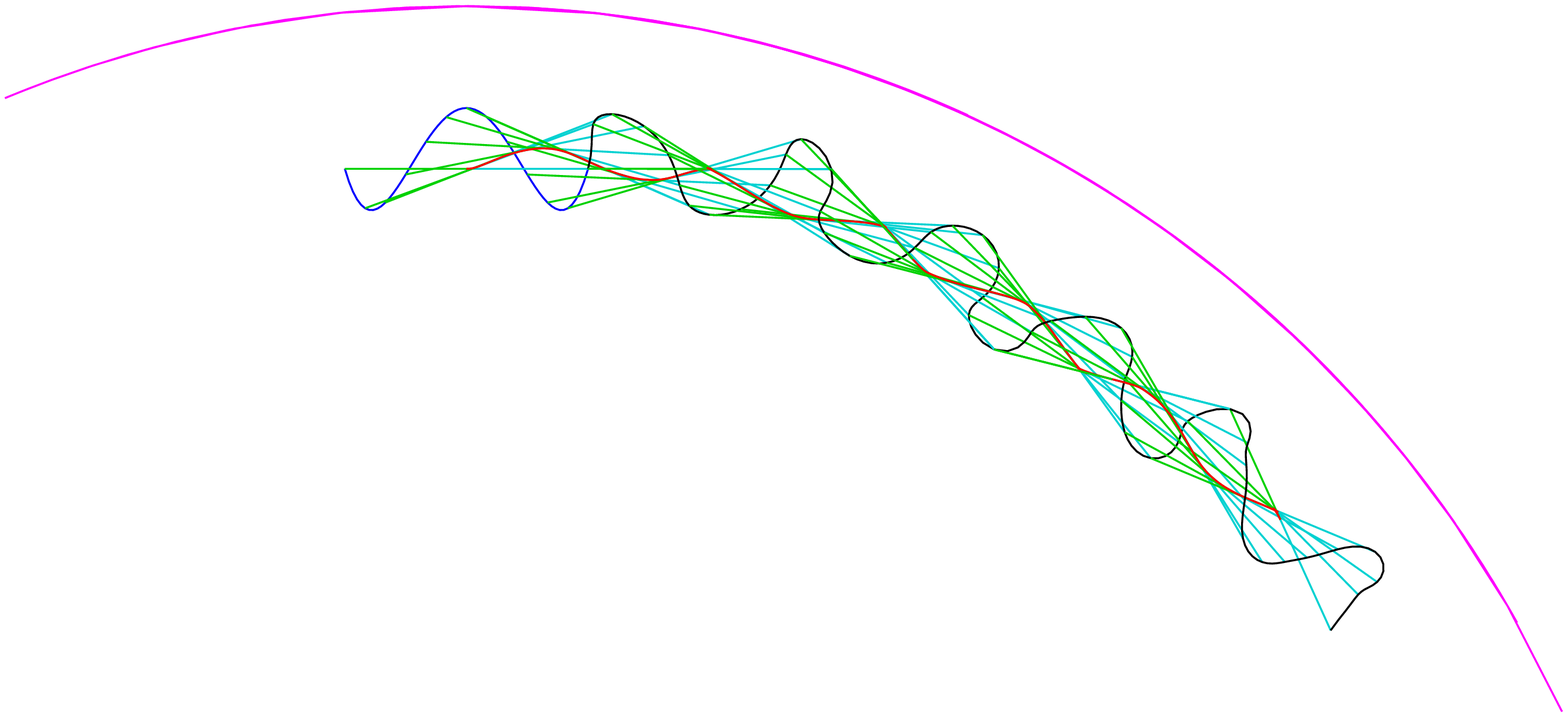}
%{\epsfig{file=fkurve45.eps,scale=0.2}
\caption{Based on eq. (\ref{eqfku}) with $c=4.5$}
\label{fku45}}\hspace{5mm}
\parbox[b]{5.5cm}
{\includegraphics[scale=0.2]{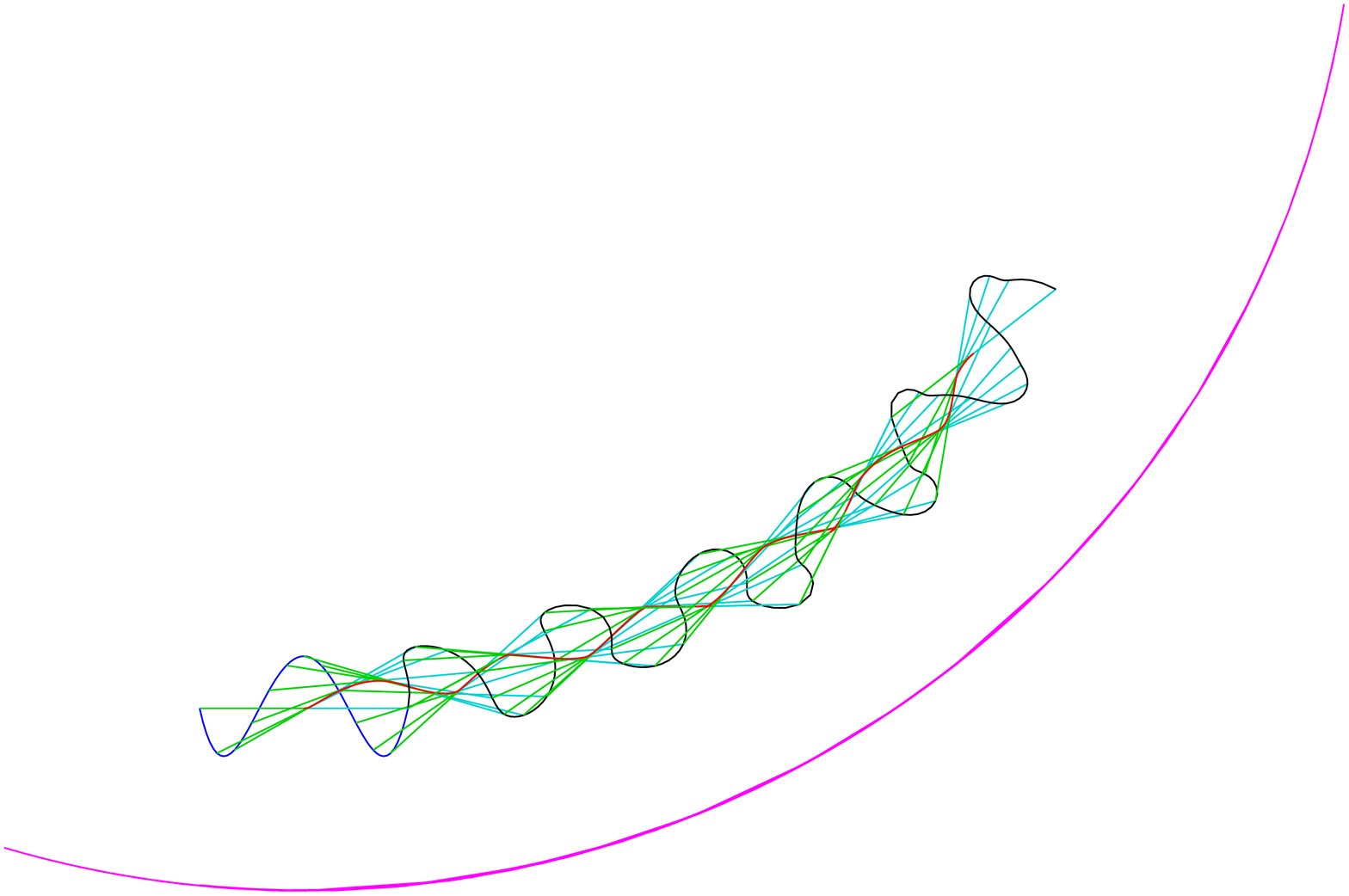}
%{\epsfig{file=fkurve55.eps,scale=0.2}
\caption{Based on eq. (\ref{eqfku}) with $c=5.5$}
\label{fku55}}
\end{figure}

\begin{figure}[h!]
\includegraphics[scale=0.5]{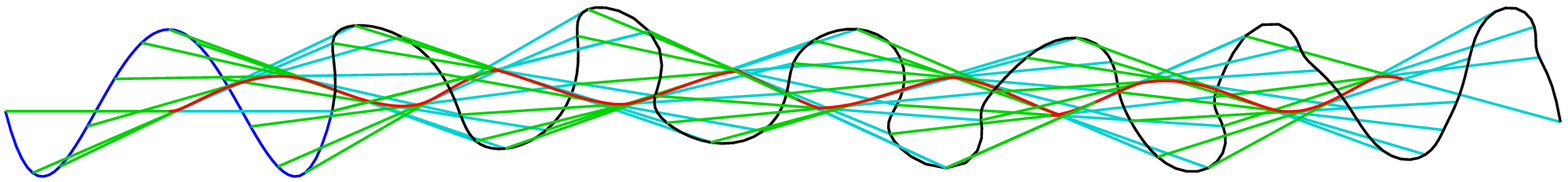}
\caption{Based on eq. (\ref{eqfku}) with $c=5$}
\label{fku5}
\end{figure}

\begin{figure}[h!]
\parbox[b]{5.5cm}
{\includegraphics[scale=0.25]{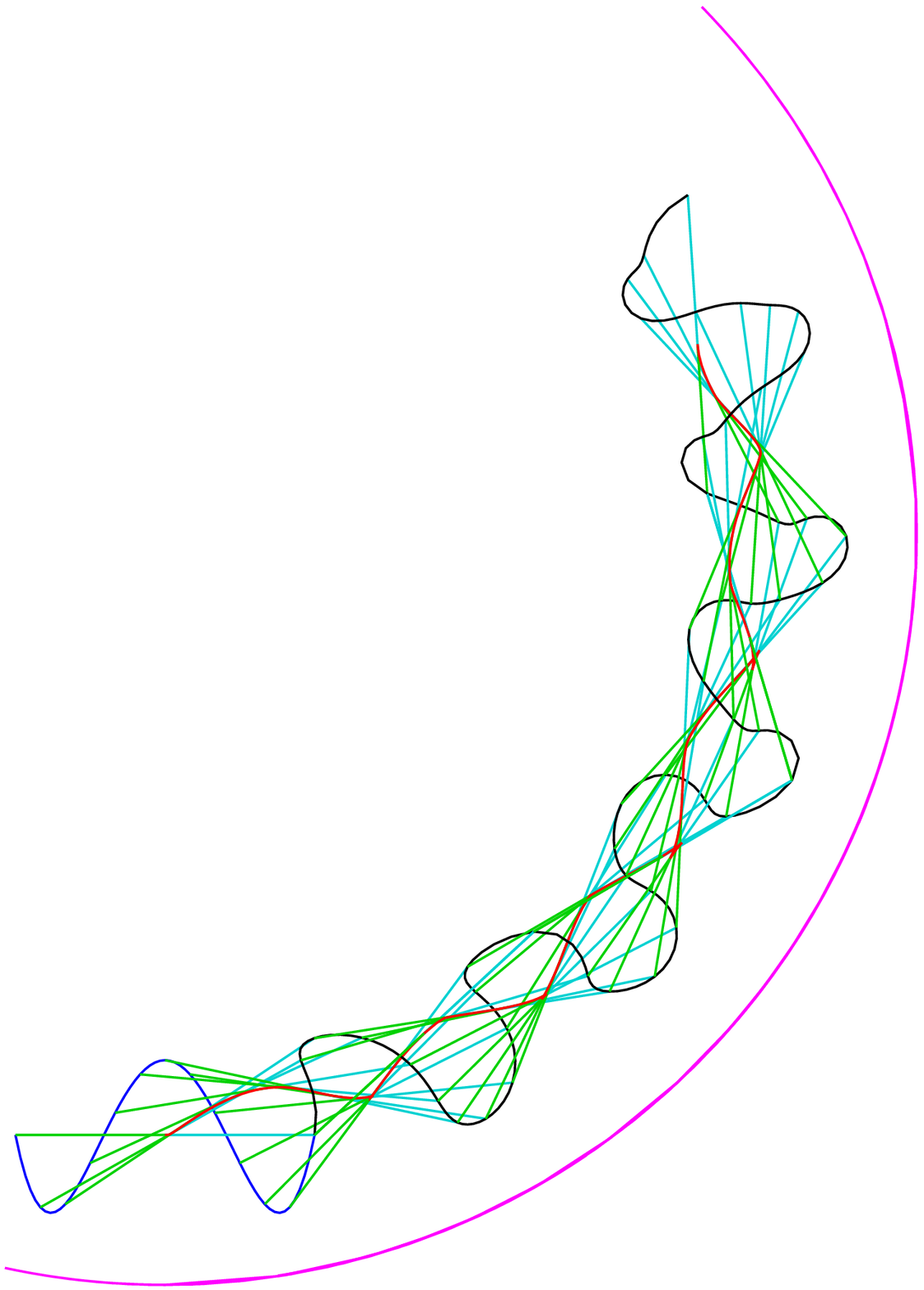}
%{\epsfig{file=fkurve6.eps,scale=0.25}
\caption{Based on eq. (\ref{eqfku}) with $c=6$}
\label{fku6}}\hspace{5mm}
\parbox[b]{5.5cm}
{\includegraphics[scale=0.5]{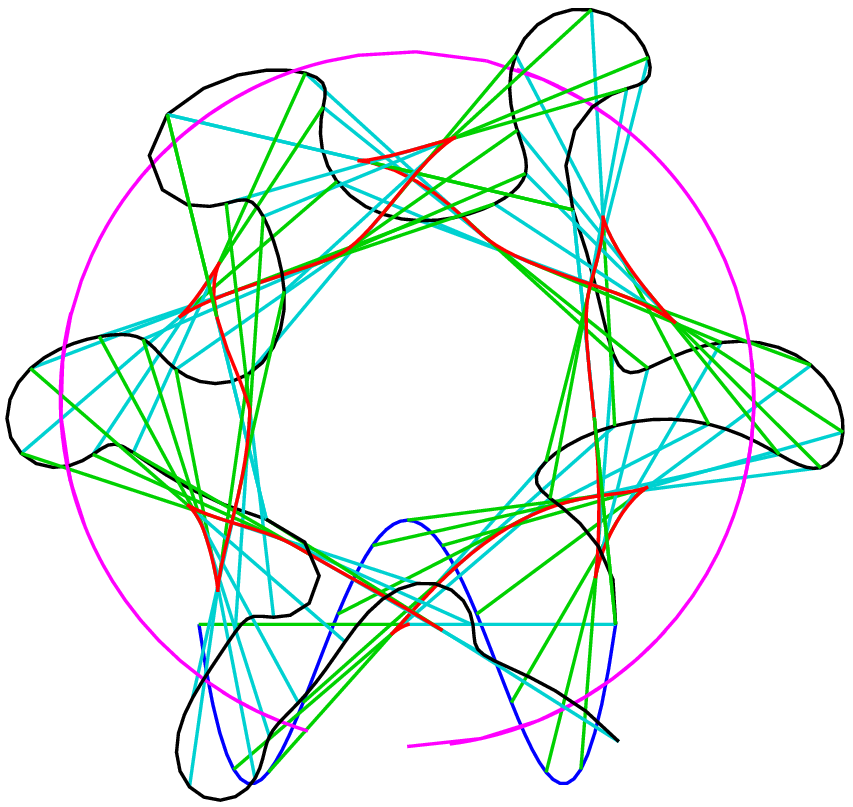}
%{\epsfig{file=fkurve8.eps,scale=0.5}
\caption{Based on eq. (\ref{eqfku}) with $c=8$}
\label{fku8}}
\end{figure}

A second class of front tire tracks are shown in figures \ref{fku2}
to \ref{fku8}. The initial front track segment is given by
\be
y=b(1-x^2)(1-cx^2). \label{eqfku}
\ee
We have chosen $b=1/2$ for all figures and $c=2,4,4.5,5.5,5,6,8$ for figures
\ref{fku2} to \ref{fku8}. The area on both sides of the chord are 
counted with opposite sign. Thus the 'centroid' may lie outside
the areas. For $c=5$ (fig. \ref{fku5}) the ares on both sides of the chord
are equal.
Thus a centroid is not defined in this case or lies at infinity.
From fig. \ref{fku8} we see that the tracks may become very wild.
This is also the case, if we use larger values of $b$ in eq. (\ref{eqfku}).

\subsection{Zindler multicurves}

Of course the curves we found for the floating bodies of
equilibrium
had to be closed after one revolution. This is not required
for bicycle curves. Also the Zindler curves
can be generalized to a larger class of bicycle curves, which
I call Zindler multicurves.

\begin{figure}[h!]
\parbox[b]{5cm}
{\includegraphics[scale=0.5]{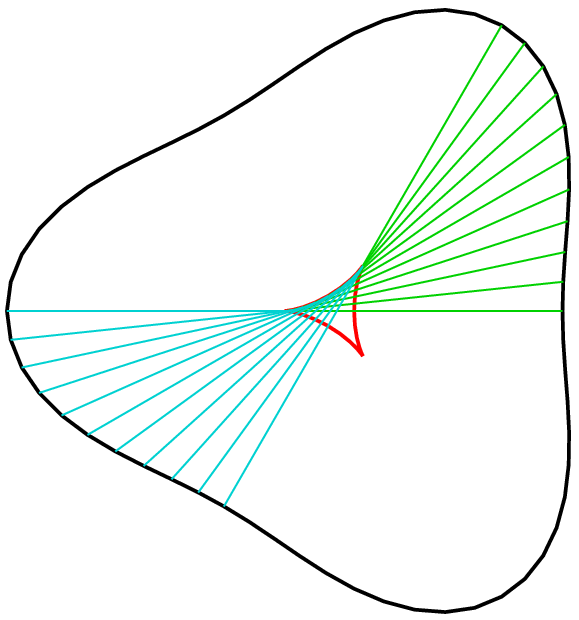}
%{\epsfig{file=fzms13.eps,scale=0.5}
\caption{Zindler curve with $m=1,n=3$}
\label{fz13}}\hspace{10mm}
\parbox[b]{5cm}
{\includegraphics[scale=0.5]{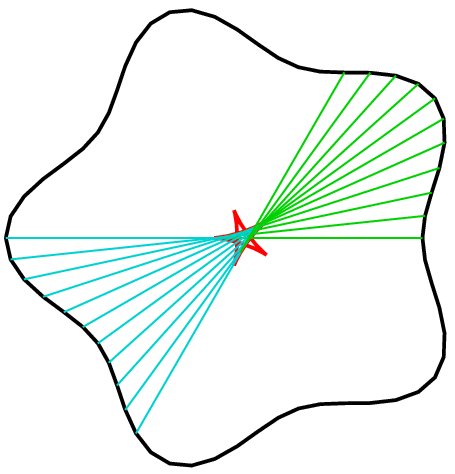}
%{\epsfig{file=fzms15.eps,scale=0.5}
\caption{Zindler curve with $m=1,n=5$}
\label{fz15}}
\end{figure}

\begin{figure}[h!]
\parbox[b]{5.5cm}
{\includegraphics[scale=0.5]{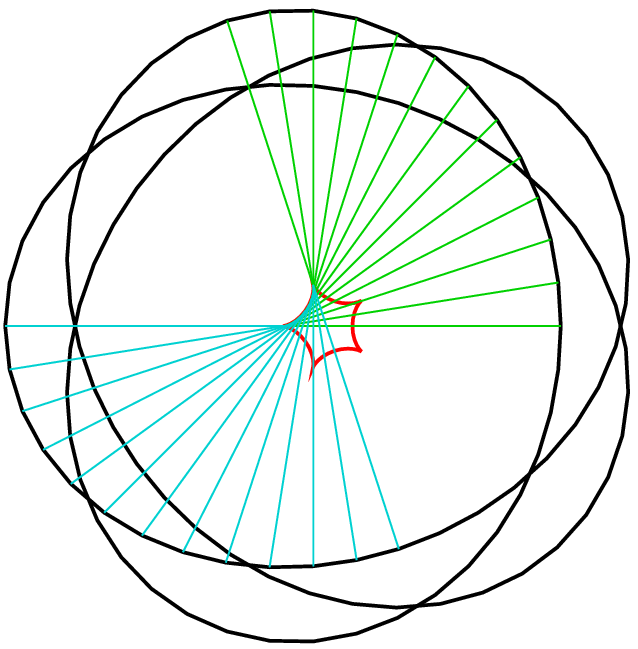}
%{\epsfig{file=fzms35.eps,scale=0.5}
\caption{Zindler multicurve with $m=3,n=5$}
\label{fz35}}\hspace{5mm}
\parbox[b]{5.5cm}
{\includegraphics[scale=0.5]{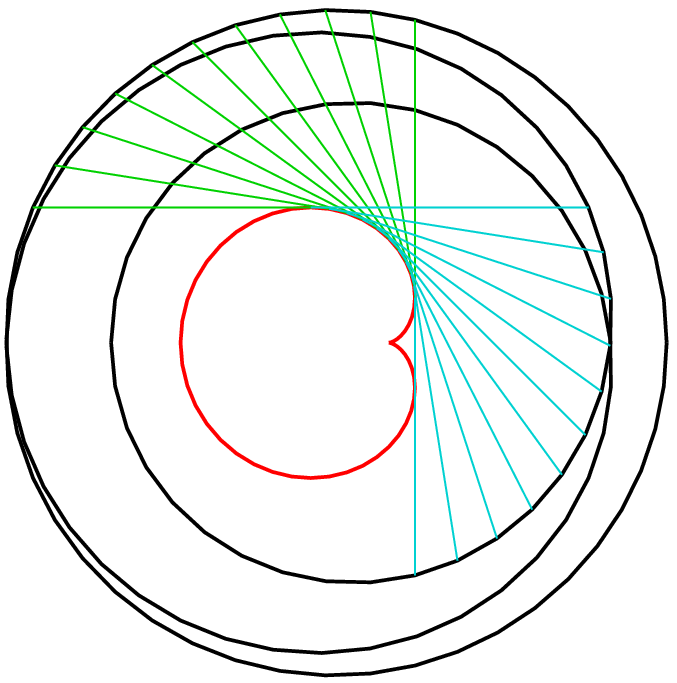}
%{\epsfig{file=fzms31.eps,scale=0.5}
\caption{Zindler multicurve with $m=3,n=1$}
\label{fz31}}
\end{figure}

\begin{figure}[h!]
\parbox[b]{5.5cm}
{\includegraphics[scale=0.5]{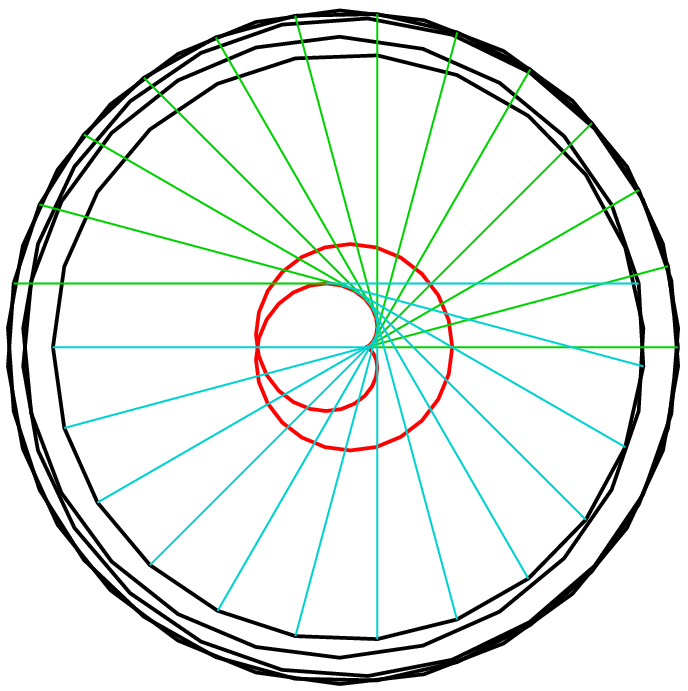}
%{\epsfig{file=fzms51.eps,scale=0.5}
\caption{Zindler multicurve with $m=5,n=1$}
\label{fz51}}\hspace{5mm}
\parbox[b]{5.5cm}
{\includegraphics[scale=0.5]{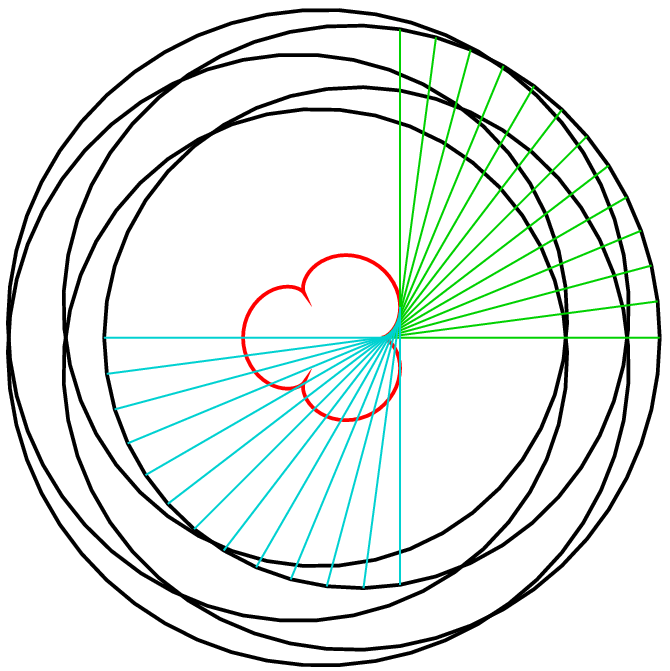}
%{\epsfig{file=fzms53.eps,scale=0.5}
\caption{Zindler multicurve with $m=5,n=3$}
\label{fz53}}
\end{figure}

The definition of the Zindler curves is generalized by replacing
eqs. (\ref{Zin1}, \ref{Zin2}) to
\bea
x(\alpha) = l \cos(m\alpha) + \xi(\alpha), &&
y(\alpha) = l \sin(m\alpha) + \eta(\alpha), \label{Zin5} \\
\xi(\alpha) = \int^{\alpha} \de\beta \cos(m\beta) \hrho(\beta), &&
\eta(\alpha) = \int^{\alpha} \de\beta \sin(m\beta) \hrho(\beta),
\label{Zin6}
\eea
where $m$ is an odd integer. 
The radius of curvature is now $|\hrho/m|$.
As in (\ref{Zin3}) we require
\be 
\xi(\alpha+\pi)=\xi(\alpha), \quad 
\eta(\alpha+\pi)=\eta(\alpha),
\ee
which again implies $\hrho(\beta+\pi)=-\hrho(\beta)$. The
curves for $m=1$ are Zindler curves. For larger $m$ the curves
are no longer double point free. Generally they repeat only
after $m$ revolutions. These curves are also bicycle curves,
since the argument around eq. (\ref{Zin4}) applies again.

Examples are
\be
\hrho(\beta) = (m^2-n^2) \sin(n\beta)
\ee
with odd $n$, $n\not=m$, and $n,m$ coprime. One obtains for
the envelopes (traces of the rear wheels)
\bea
\xi(\alpha) &=& \frac{n-m}2 \cos((n+m)\alpha)
+ \frac{n+m}2 \cos((n-m)\alpha), \\
\eta(\alpha) &=& \frac{n-m}2 \sin((n+m)\alpha)
- \frac{n+m}2 \sin((n-m)\alpha).
\eea
These envelopes are known as hypocycloids for $n>m$ and as epicycloids
for $n<m$. They wind $m$ times around the origin and
have $n$ cusps. These cusps point outward for hypocycloids  and 
inward for epicycloids.
With $\alpha=u/2$ the Zindler multicurve is parametrized by
\be
z=x+\ie y = \frac{n-m}2 \ex{\ie(n+m)u/2}
+ \frac{n+m}2 \ex{\ie(m-n)u/2} + l \ex{\ie mu/2}. \label{cycloid}
\ee
In particular for $m=1,n=3$ one obtains
\be
z = \ex{2\ie u} + 2\ex{-\ie u} + l \ex{\ie u/2}. \label{cycloid12}
\ee
Examples of such Zindler curves, $m=1$, are shown in figures
\ref{fz13} and \ref{fz15}.
We show four examples of Zindler multicurves with $m=3$ and $m=5$
in figures \ref{fz35} to \ref{fz53}.

\subsection{Mampel's generalized Zindler curves}

\begin{figure}[h!]
\parbox[b]{3.5cm}
{\includegraphics[scale=0.5]{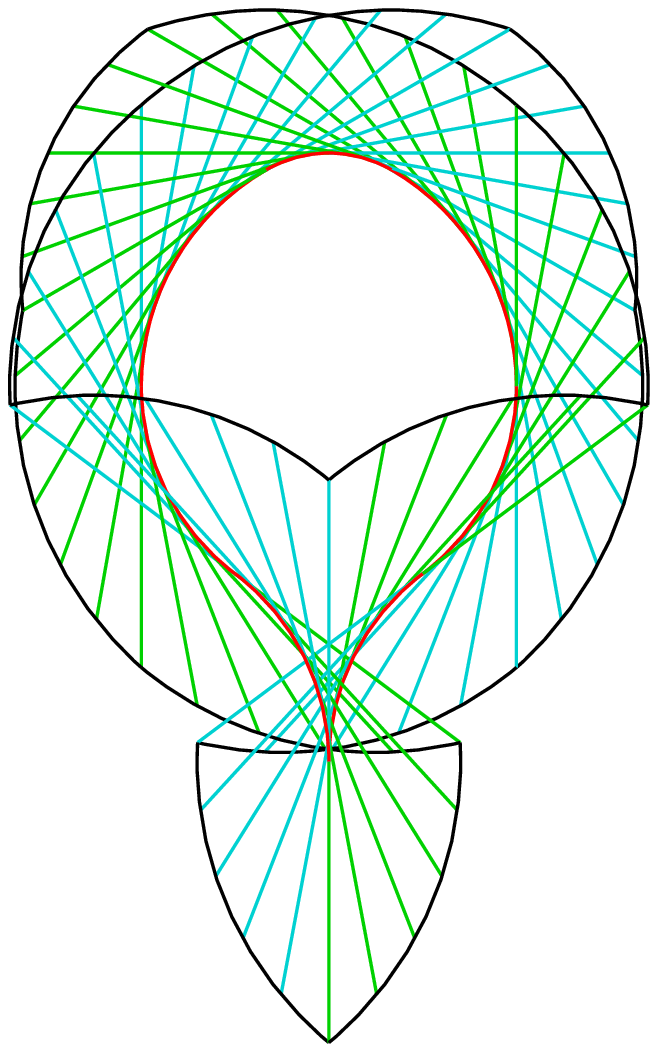}
%{\epsfig{file=Mampelc.eps,scale=0.5}
\caption{Envelope with one cusp pointing outside}
\label{fMc}} \hspace{2mm}
\parbox[b]{3.5cm}
{\includegraphics[scale=0.5,angle=90]{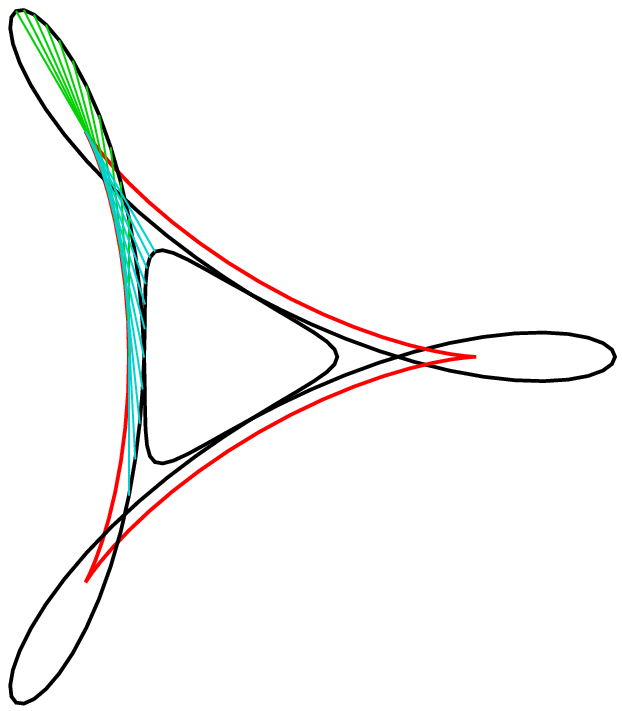}
%{\epsfig{file=fzmc13a.eps,scale=0.5,angle=90}
\caption{Curve according to (\ref{cycloid12}) with $l=1.6$.}
\label{fzmc13a}} \hspace{2mm}
\parbox[b]{3.5cm}
{\includegraphics[scale=0.5]{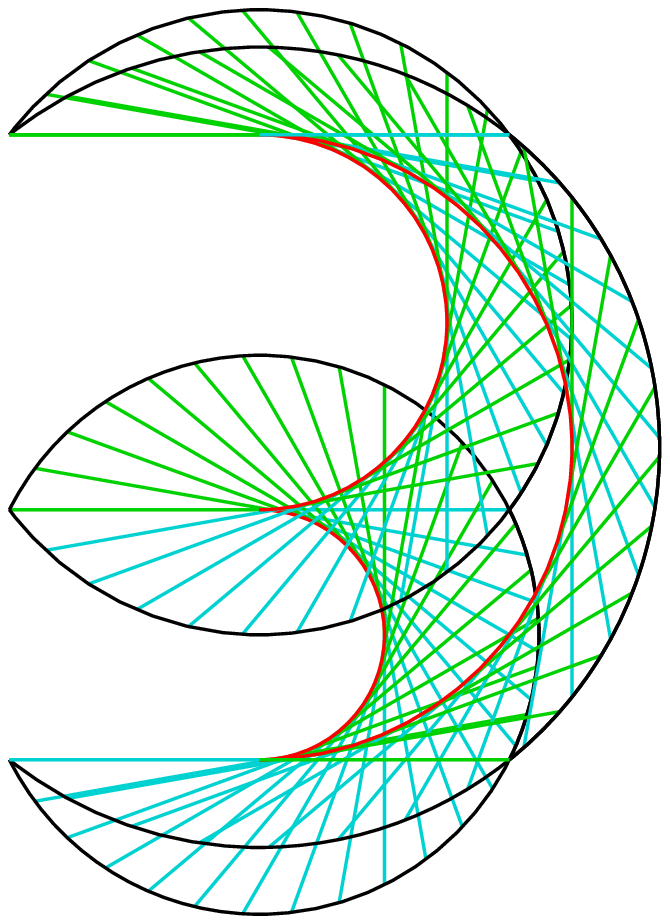}
%{\epsfig{file=Mampelg.eps,scale=0.5}
\caption{Three cusps pointing to the left.}
\label{fMg}}
\end{figure}

Mampel\cite{Mampel67} considers generalized Zindler curves.
He introduces envelopes (denoted as 'Kern' $\cal{K}$)
with an odd number of cusps.
He attaches tangents with constant length $l$
in both directions. The endpoints of the tangents form
his generalized Zindler curves $\cal{Z}$, irrespective of any
convexity. Examples similar to Mampel's figures 8a, 8b, and 9,
are shown in figures \ref{fMc} to \ref{fMg}. The curves
\ref{fMc} and \ref{fMg} consist of circular arcs.
The envelope of fig. \ref{fzmc13a} is a hypocycloid.

\subsection{Other bicycle curves}

\begin{figure}[h!]
\parbox[b]{6cm}
{\includegraphics[scale=0.4,angle=180]{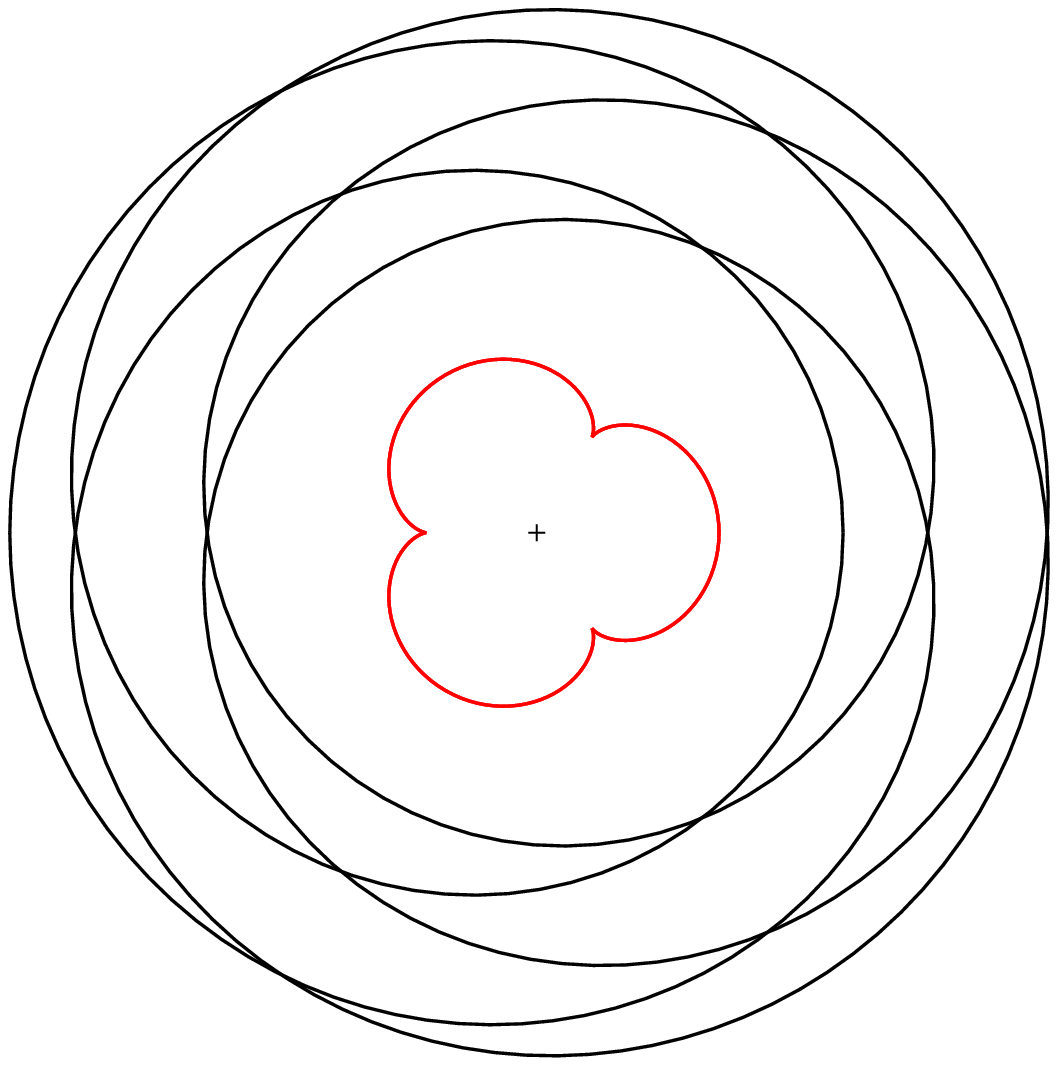}
%{\epsfig{file=fgp1p3s.eps,scale=0.4,angle=180}
\caption{Curves with ratio = 1.72}
\label{f13s}} \hspace{2mm}
\parbox[b]{6cm}
{\includegraphics[scale=0.4]{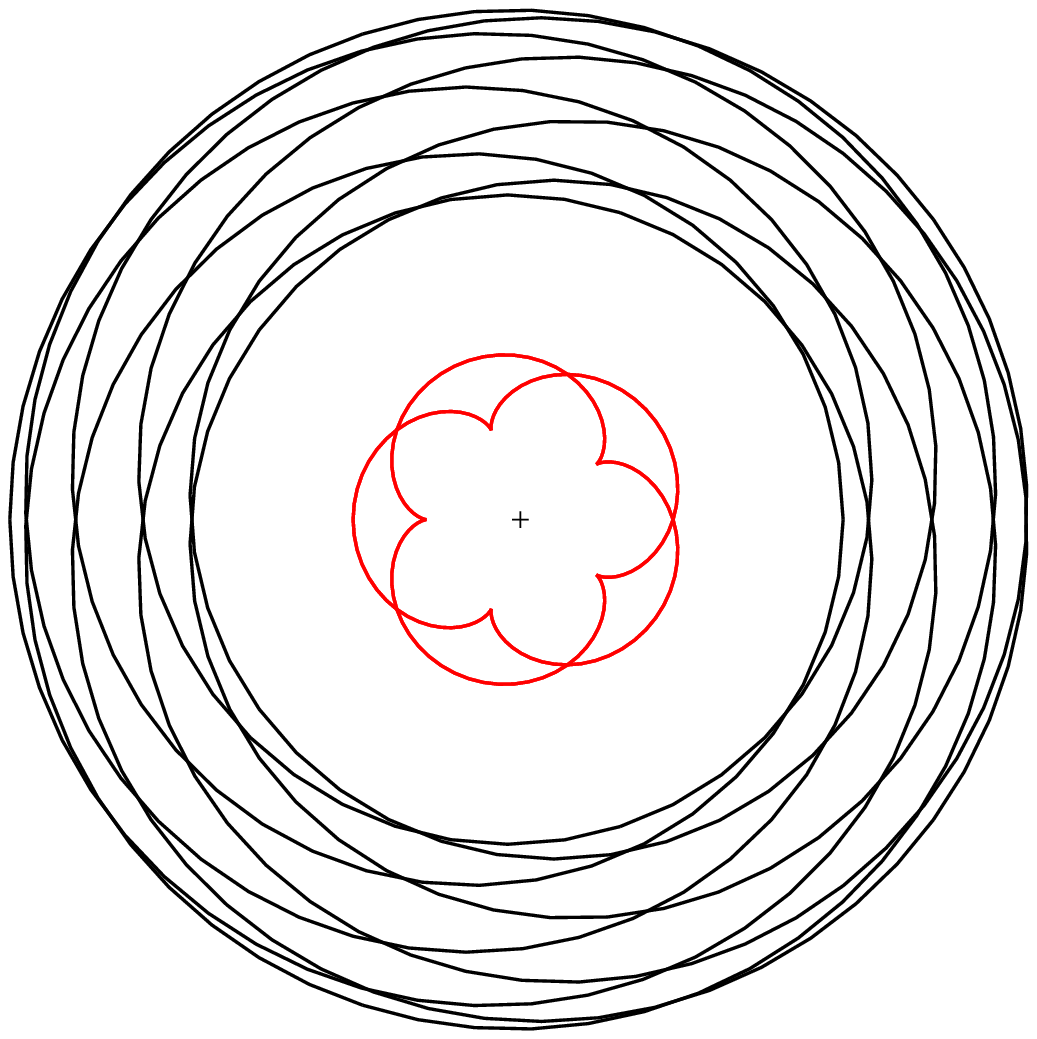}
%{\epsfig{file=fgp1p5u.eps,scale=0.4}
\caption{Curves with ratio = 1.582}
\label{f15u}}
\end{figure}

We show some buckled rings, which turn around the
center several times. The ratio of the maximal radius and
the minimal radius is given.
The buckled ring figure \ref{f13s} is very
similar to the Zindler multicurve figure \ref{fz53}.

\begin{figure}[h!]
\parbox[b]{6cm}
{\includegraphics[scale=0.4]{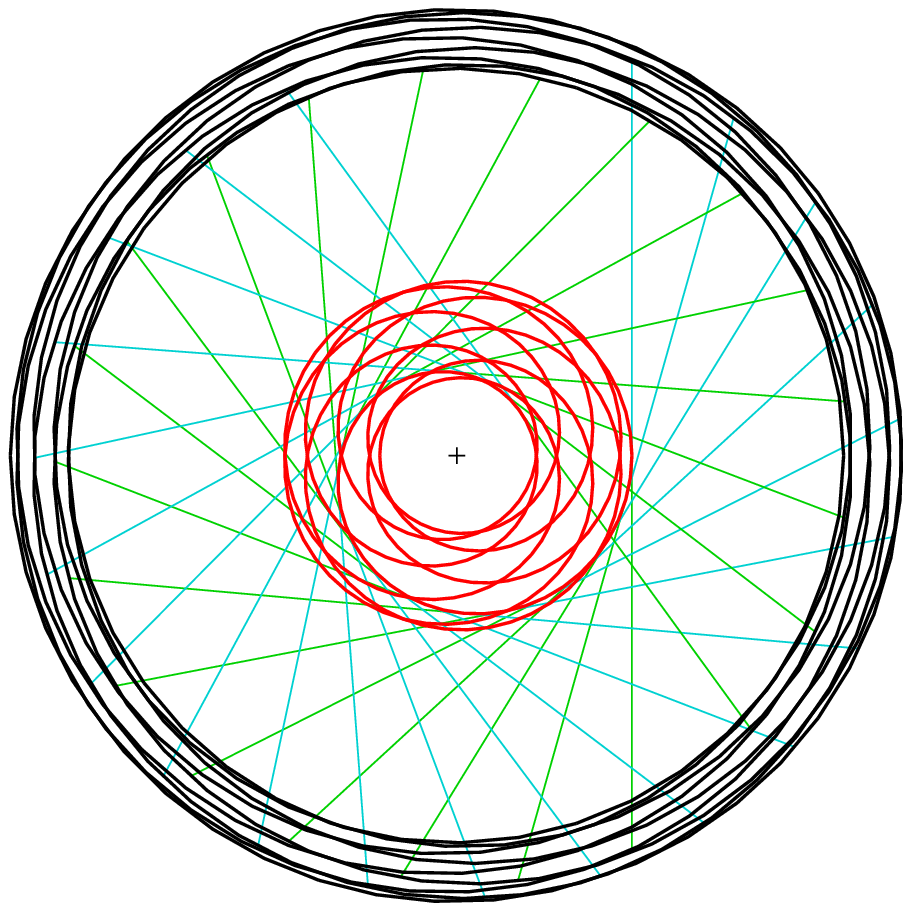}
%{\epsfig{file=fgp1p5m.eps,scale=0.4}
\caption{Curves with ratio = 1.1546}
\label{f15m}}
\parbox[b]{6cm}
{\includegraphics[scale=0.4]{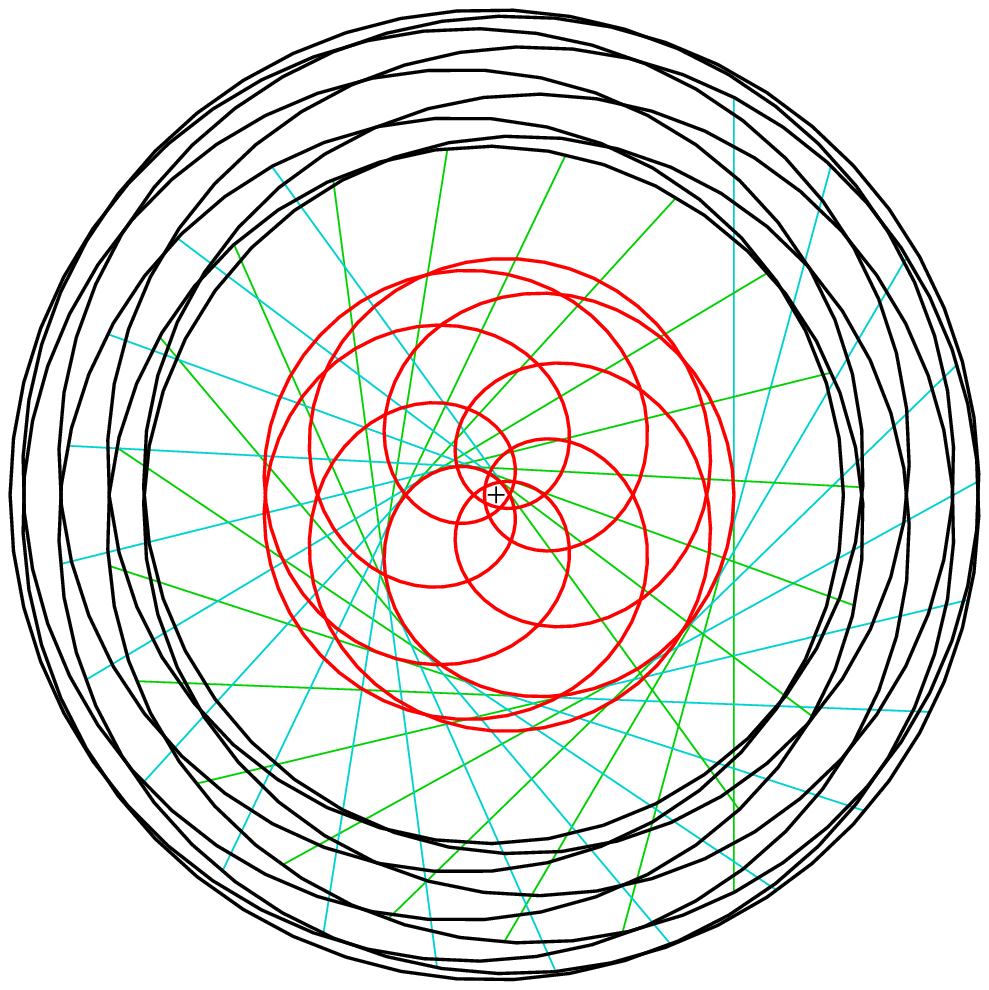}
%{\epsfig{file=fgp1p5n.eps,scale=0.4}
\caption{Curves with ratio = 1.399}
\label{f15n}}
\end{figure}

\begin{figure}[h!]
\parbox[b]{6cm}
{\includegraphics[scale=0.4]{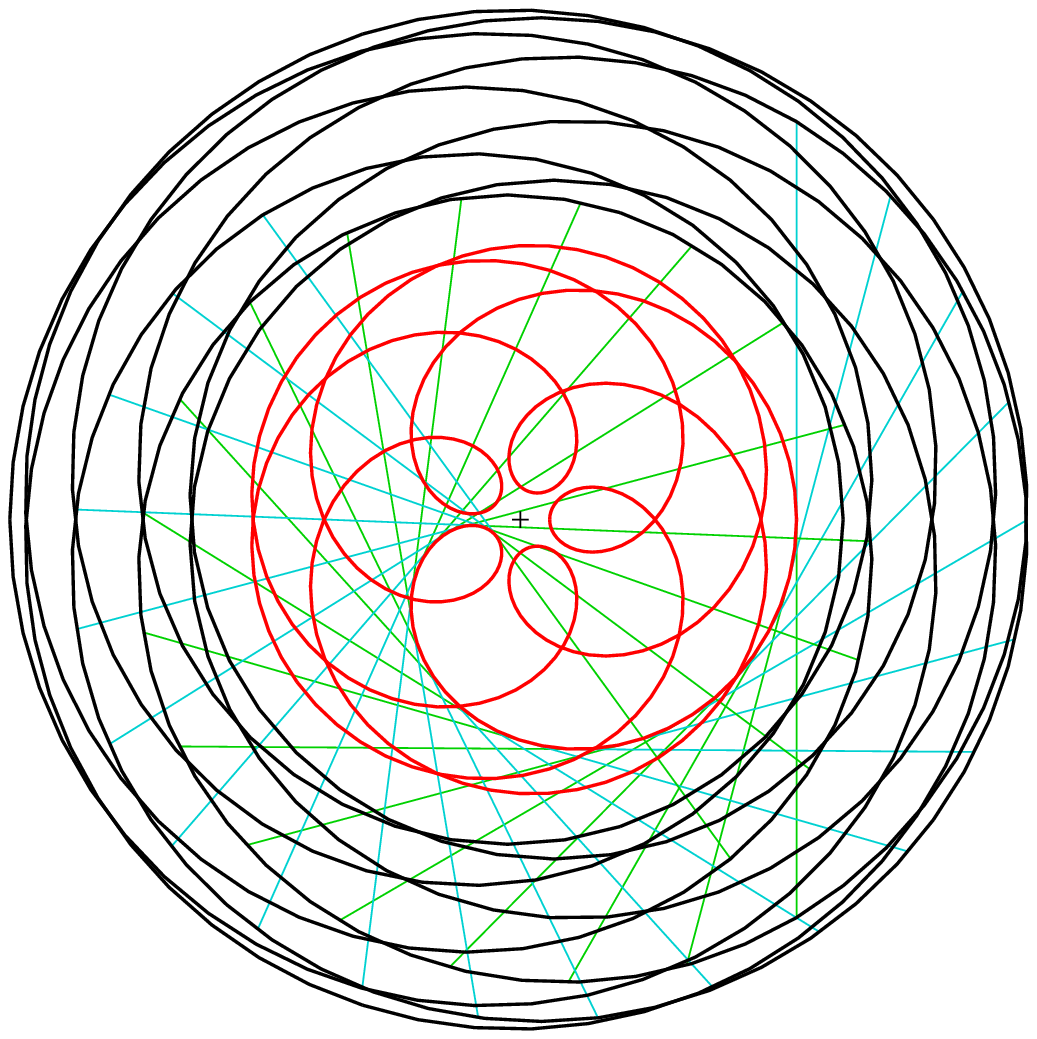}
%{\epsfig{file=fgp1p5o.eps,scale=0.4}
\caption{Curves with ratio = 1.582}
\label{f15o}}
\parbox[b]{6cm}
{\includegraphics[scale=0.4]{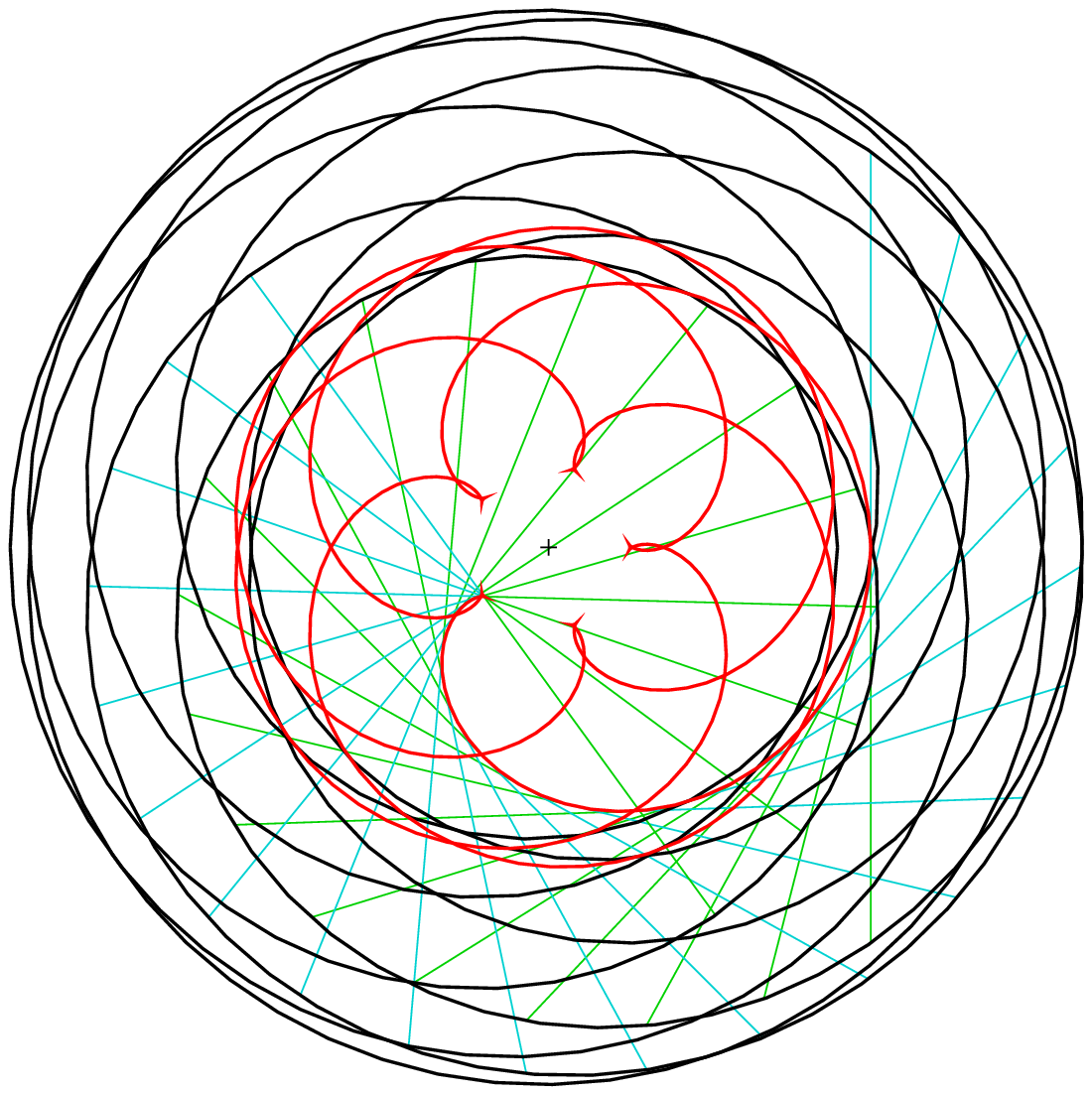}
%{\epsfig{file=fgp1p5p.eps,scale=0.4}
\caption{Curves with ratio = 1.8637}
\label{f15p}}
\end{figure}

\begin{figure}[h!]
\parbox[b]{6cm}
{\includegraphics[scale=0.4]{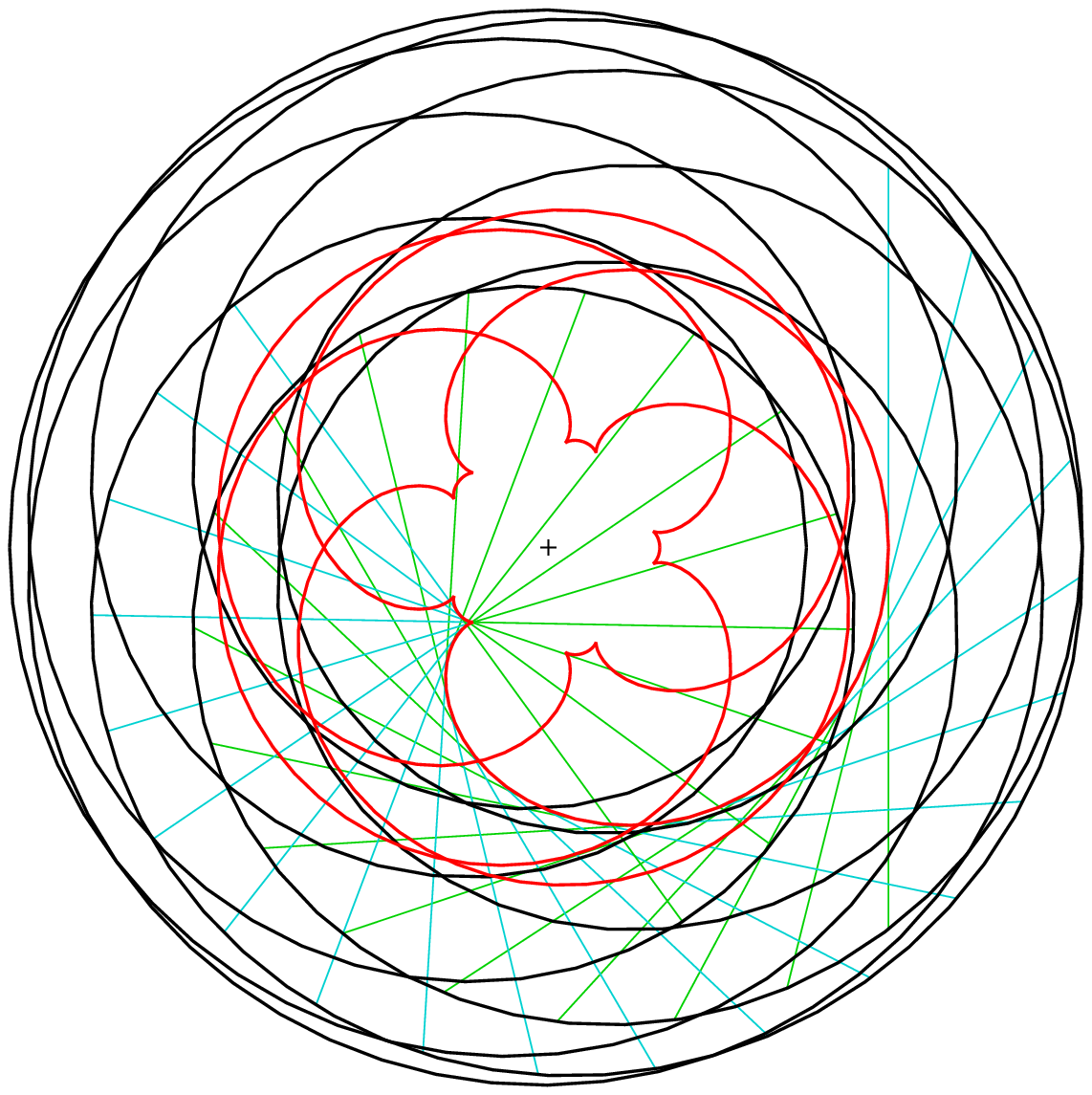}
%{\epsfig{file=fgp1p5q.eps,scale=0.4}
\caption{Curves with ratio = 2.0869}
\label{f15q}}
\parbox[b]{6cm}
{\includegraphics[scale=0.4]{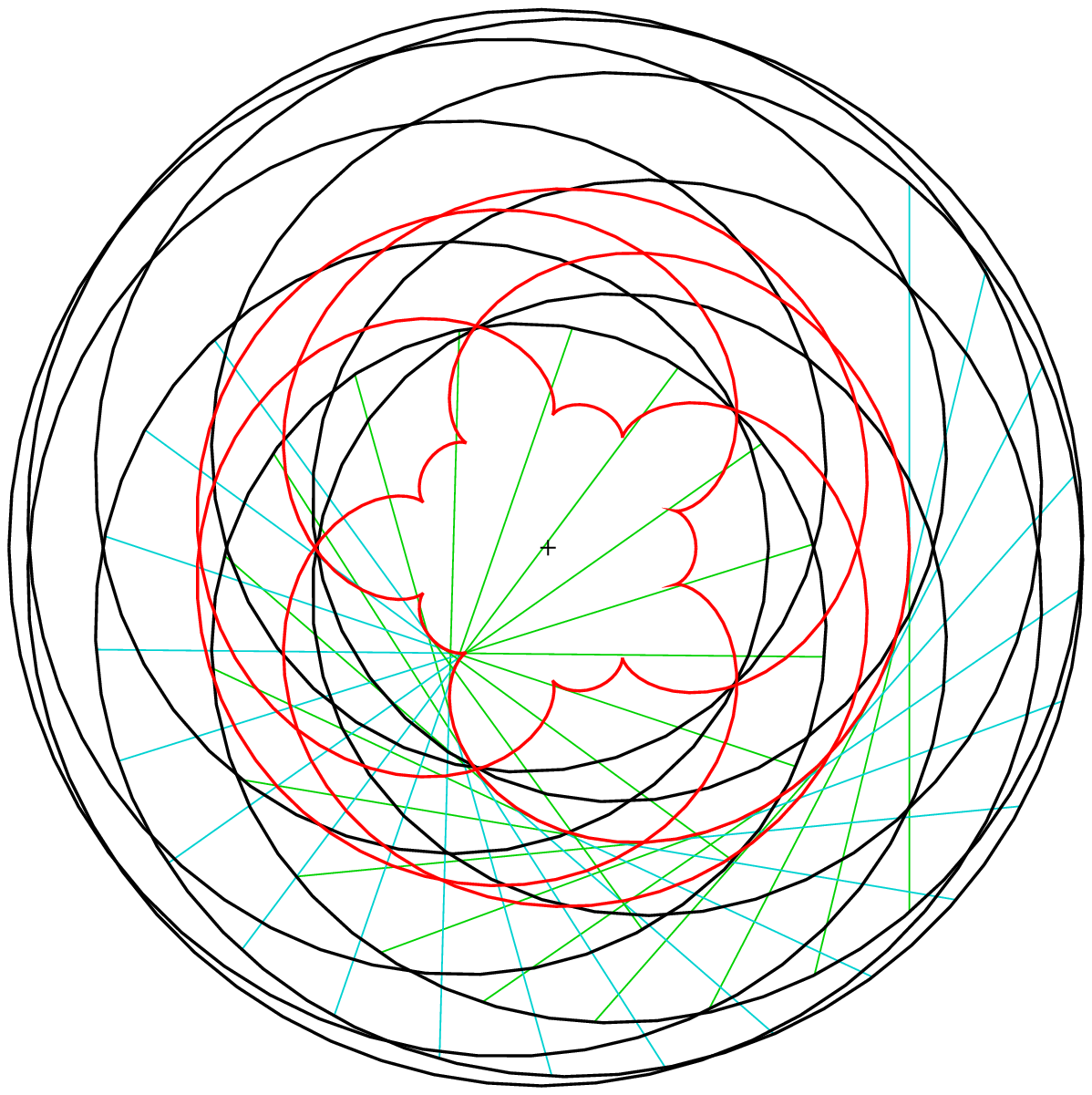}
%{\epsfig{file=fgp1p5r.eps,scale=0.4}
\caption{Curves with ratio = 2.4451}
\label{f15r}}
\end{figure}

Figures \ref{f15u} to \ref{f15r} show buckled rings which turn
around the center nine times. These buckled rings have two
different envelopes. The smaller one has five cusps pointing
inward, figure \ref{f15u}. The outer envelope, figure \ref{f15o}
has no cusps.
If the ratio of the largest distance to the
smallest distance from the center is not too large,
then the outer trace for the rear tire is without cusps.
This is the case for the figures \ref{f15m} to \ref{f15o}.
If the ratio becomes larger, then cusps appear as seen in
figures \ref{f15p} to \ref{f15r}.

\section{Conclusion}

A short historical account of the curves related
to the two-di\-men\-sio\-nal floating bodies of equilibrium
and the bicycle problem is given in this paper.
Bor, Levi, Perline and Tabachnikov found that quite a
number of the boundary curves had already been described as {\it Elastica} and
{\it Elastica under Pressure} or {\it Buckled Rings}.
Auerbach already realized that curves described by Zindler
are solutions for the floating bodies problem of density 1/2.
An even larger class of curves solves the bicycle problem.

The subsequent sections deal with some supplemental details:
Several derivations of the equations for the
elastica and elastica under pressure are given.
The properties of Zindler curves and some
work on the problem of floating bodies
of equilibrium by other mathematicians is discussed.
Special cases of elastica under pressure lead to
algebraic curves as shown by Greenhill.
Since most of the curves considered here are bicycle curves,
we added some remarks on them.

{\bf Acknowledgment} I am indebted to Sergei Tabachnikov for
many useful discussions. I am grateful to J.A. Hanna and M. Bialy
for useful information.

\end{document}